# ON HUMAN ROBOT INTERACTION USING MULTIPLE MODES

*DISSERTATION*

*Submitted by*

**Neha Baranwal**

In Partial Fulfillment of the Requirements

For the Degree of

**DOCTOR OF PHILOSOPHY**

*Under the Supervision of*
**Prof. G.C. Nandi**

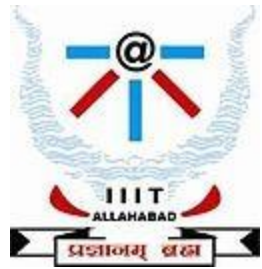

DEPARTMENT OF INFORMATION TECHNOLOGY

भारतीय सूचना प्रौद्योगिकी संस्थान, इलाहाबाद

**INDIAN INSTITUTE OF INFORMATION TECHNOLOGY, ALLAHABAD**

*(A centre of excellence in IT, established by Ministry of HRD, Govt. of India)*

April 25, 2016

| 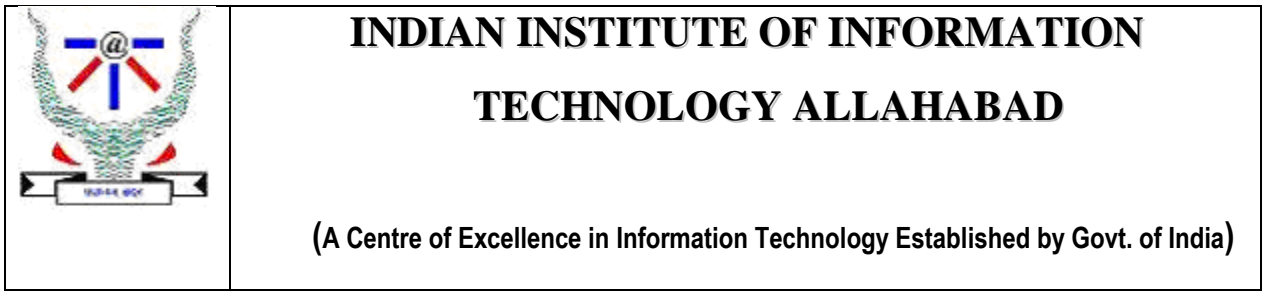 | **INDIAN INSTITUTE OF INFORMATION TECHNOLOGY ALLAHABAD**<br><br>(A Centre of Excellence in Information Technology Established by Govt. of India) |
|---|---|

# *CANDIDATE DECLARATION*

I, Neha Baranwal**,** Roll No. **RS-115** certify that this thesis work entitled "*On Human Robot Interaction Using Multiple Modes"* is submitted by me in partial fulfillment of the requirement of the Degree of **Doctor of Philosophy** in Department of **Information Technology**, **Indian Institute of Information Technology, Allahabad.**

I understand that plagiarism includes:

1. Reproducing someone else's work (fully or partially) or ideas and claiming it as one's own.
2. Reproducing someone else's work (Verbatim copying or paraphrasing) without crediting
3. Committing literary theft (copying some unique literary construct).

I have given due credit to the original authors/ sources through proper citation for all the words, ideas, diagrams, graphics, computer programs, experiments, results, websites, that are not my original contribution. I have used quotation marks to identify verbatim sentences and given credit to the original authors/sources.

I affirm that no portion of my work is plagiarized. In the event of a complaint of plagiarism, I shall be fully responsible. I understand that my Supervisor may not be in a position to verify that this work is not plagiarized.

**Name Neha Baranwal**                                                                                           **Date: 25/04/2016**

*Enrolment No: RS-115*

*Department of Information Technology,*

*IIIT-Allahabad (U.P.)*

*IIIT-A, Robotics and AI Lab*                                                                                                     *I*

| 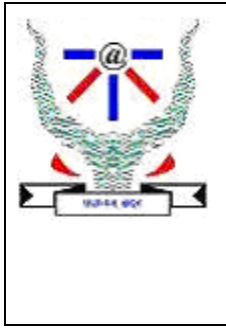 | **INDIAN INSTITUTE OF INFORMATION TECHNOLOGY ALLAHABAD**<br><br>(A Centre of Excellence in Information Technology Established by Govt. of India) |
|---|---|

# *CERTIFICATE FROM SUPERVISOR*

I do hereby recommend that the thesis work done under my supervision by **Ms. Neha Baranwal (RS 115)** entitled "*On Human Robot Interaction Using Multiple Modes*" be accepted in partial fulfillment of the requirement for the Degree of **Doctor of Philosophy** in Department of **Information Technology, Indian Institute of Information Technology, Allahabad.**

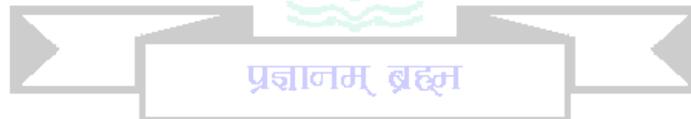

_______________________

**Prof. G.C. Nandi**

**Supervisor, IIIT-Allahabad**

# ABSTRACT

Today robotics is a vibrant field of research and it has tremendous application potentials not only in the area of industrial environment, battle field, construction industry and deep sea exploration but also in the household domain as a humanoid social robot. To be accepted in the household, the robots must have a higher level of intelligence and they must be capable of interacting people socially around it who is not supposed to be robot specialist. All these come under the field of human robot interaction (HRI). Our hypothesis is- "*It is possible to design a multimodal human robot interaction framework, to effectively communicate with Humanoid Robots*". In order to establish the above hypothesis speech and gesture have been used as a mode of interaction and throughout the thesis we validate our hypothesis by theoretical design and experimental verifications. Individually, none of the modes has been sufficient for a good communication because each of them has its own potentials and drawbacks, such as gesture works well when light condition is sufficient and the performance goes down as the light becomes poorer. Similarly for speech, as the noise level increases the recognition accuracy of speech signal decreases. To solve such challenges, a multi-modal framework for HRI has been addressed where the results obtained from gesture and speech have been combined. Two humanoid robots, NAO and Humanoid Open Architecture Platform (HOAP-2) have been utilized as testbeds for experimentations and to test our hypothesis. This thesis has three major portions. The first portion deals with the gesture based interaction where we worked on both Isolated as well as continuous Indian Sign Language (ISL) gestures. Although ISL has similarity with other sign languages like American Sign Language (ASL), Korean Sign Language (KSL) etc. in terms of analysis, it differs in terms of their context. Most of the ISL gestures have been made up of two hands and are dynamic in nature which induces the ambiguity issue. This challenge has been solved by proposing a vision based gesture recognition method where discrete wavelet transform (DWT) and Mel Frequency Cepstral Coefficient (MFCC) have been applied. Since we believe DWT's multiresolution (time frequency) capability coupled with MFCC's spectral envelop (showing vividly the spikes of the signal) ability could produce better unambiguous features which might improve recognition rate. The database of ISL gestures have been created using a webcam and Canon EOS camera. Sign language has been normally used by hearing impaired society. To make the communication bonding between deaf and normal persons, a novel NAO based continuous ISL gesture recognition framework has been developed. Continuous Sign Language is a sequence of gestures that generates a meaningful sentence. The major challenges of any gesture recognition system are proper segmentation, background elimination, person invariance, overlapping frame abstraction and extraction of appropriate features. The silhouette images of each video frames have been created using background modelling technique where background subtraction, morphological filtering, color based segmentation and face extraction have been used as a part of the process. Moreover, the gradient based method has been applied on these silhouette images for finding the overlapping frames. These key frames have been used for removing uninformative gestures and also discovering the start and end point of the individual gesture in a complete sentence. Geometrical shape of the hand has been used as a feature, with the combination of other two features orientation, and hand motion (velocity or speed). These features have been extracted using discrete wavelet transform, orientation histogram, wavelet descriptor, chain code and spatio-temporal changes. We have applied and compared various machine learning techniques like Hidden Markov Model, Euclidean distance, Mahalanobis distance, K-nearest neighbor for gesture classification. A novel machine learning technique known as possibility theory based hidden markov model (PTBHMM) has been proposed. This technique has been used as a classifier for classifying gestures. Reduction of time complexity is a big challenge in any real time applications. With the help of PTBHMM, we minimize this time complexity issue. A concept of possibility theory has been applied to redesign and solve all the three fundamental problems (Evaluation, Decoding and Learning) of conventional HMM. PTBHMM deals with both uncertainty as well as imprecision whereas probability theory based HMM only handles uncertainty. We have proved both theoretically and experimentally that the time complexity of PTBHMM (NT) is N time less than the time complexity of classical HMM ($N^2T$) without compromising the classification accuracy. It has been very efficient for real time gesture based human robot communication system. After classification robot performs a lexical analysis to generate a normal language which has been understandable to normal persons.

The second contribution of the thesis is the development of speech based HRI. A Hindi speech has been recognized using the HTK toolkit and discrete wavelet transform (DWT) with Human Factor Cepstral Coefficient (HFCC). HTK toolkit is an HMM based toolkit developed by Cambridge University and is very efficient for English speech recognition. Thus we have applied it for Hindi speech recognition and found satisfactory results. Due to some implementation issues like handling of speech signal in real time environment, installing of HTK toolkit with windows etc. related to the HTK toolkit we go for another approach which is DWT with HFCC. Here the cepstral coefficients of speech signal has been extracted using MFCC and HFCC both, which has been applied after DWT decomposition. DWT removes the noise occurred during speech recording by dividing the whole signal into high frequency and low frequency zones. Then cepstral coefficients of these low frequency part is calculated. This is how we extracted features which are substantially used for classifications.

All the experiments have been performed on isolated as well as continuous speech. Proposed technique has been compared with the existing technique by calculating the word error rate and signal to noise ratio (SNR) value. ISL gesture and their speech dataset have been created in robotics and artificial intelligence laboratory at Indian Institute of Information Technology, Allahabad, India. Also we have tested our proposed algorithms on a benchmark dataset available for the researcher (Sheffield Kinect gesture dataset [177] and English speech dataset [36]) and found a respectable amount of accuracy in real time scenario. Finally, we propose a fusion model between speech and gesture for HRI, which provides unambiguous communication between human and robot. Decision level fusion has been proposed for combining two different modes (speech and gesture). Fusion has been performed on the basis of likelihood values obtained after classification. Finally, it has been tested on two humanoid robots NAO and HOAP-2.

# Table of Contents

# List of TABLES

# List of Figures

# List of Abbreviations

HRI            Human Robot Interaction

ASL            American Sign Language

KSL            Korean Sign Language

ISL             Indian Sign Language

DWT          Discrete Wavelet Transform

MFCC         Mel-Frequency Cepstral Coefficient

HFCC         Human Factor Cepstral Coefficient

HOAP-2      Humanoid Open Architecture Platform-2

HMM          Hidden Markov Model

KNN           K-Nearest Neighbor

SVM           Support Vector machine

PTBHMM    Possibility Theory based Hidden Markov Model

WD             Wavelet Descriptor

PCA            Principal Component Analysis

ICA             Independent Component Analysis

MFD           Modified Fourier Descriptor

OH             Orientation Histogram

# Dissertation Publications

## *List of Journal Publications*

J1. Neha Baranwal and G.C.Nandi." **An Efficient Gesture based Humanoid Learning using Wavelet Descriptor and MFCC Techniques**". Published in International Journal of Machine Learning and Cybernetics (Springer) (SCI) (2.692 impact factor) (Vol. 8, issue 4, pp. 1369-1388, 17-april-2016)( DOI:10.1007/s13042-016-0512-4).

J2. Neha Baranwal and G.C.Nandi." **Real Time Gesture based Communication using Possibility Theory based Hidden Markov Model**" Published in International journal of computational Intelligence (Wiley) (SCI) (0.964 impact factor) (Volume 44, No. 4)(SN.1467-8640)(28-04-2017)( **DOI:**10.1111/coin.12116)

J3. Neha Baranwal and G.C.Nandi." **A Mathematical Framework for Possibility Theory Based Hidden Markov Model"** Published in International journal of Bio-Inspired Computation (Inderscience) (SCI) (1.935 impact factor) (Vol. 10, No. 4, 2017 (http://www.inderscience.com/info/ingeneral/forthcoming.php?jcode=ijbic)

J4. Neha Baranwal and G.C.Nandi **"Development of a Framework for Human-Robot interactions with Indian Sign Language Using Possibility theory" published** International journal of social robotics (springer) (SCI) (2.009 impact factor)(30-05-2017)(DOI 10.1007/s12369-017-0412-0).

J5. Neha Baranwal, Shweta Tripathi and G.C.Nandi. "**A Speaker Invariant Speech Recognition Technique Using HFCC Features in Isolated Hindi Words**" published in Inderscience journal (International Journal of computational Intelligence Studies) (Volume 3, issue 4, PP. 277-291, January 2014)(https://doi.org/10.1504/IJCISTUDIES.2014.067031).

J6. Avinash Kumar Singh, Neha Baranwal and G.C.Nandi." **Development of Self Reliant Humanoid Robot for Sketch Drawing**". Published in International Journal of Multimedia tools and applications (Springer) (SCI) (1.541 impact factor) (pp. 1-24, 09-February-2017) ( **DOI**:10.1007/s11042-017-4358-x).

J7. Avinash Kumar Singh, Neha Baranwal, and G. C. Nandi. "**A rough set based reasoning approach for criminal identification.**" International Journal of Machine Learning and Cybernetics (2017): 1-19.

## *List of Conference Publications*

C1. Neha Baranwal and G.C. Nandi. "Possibility Theory based Continuous Indian Sign Language Gesture Recognition" published in **35th IEEE TENCON 2015- 2015 IEEE Region 10 Conference**, pp. 1-5, 1-4 Nov. 2015.

C2. Neha Baranwal, Neha Singh and G.C. Nandi. "Implementation of MFCC based Hand Gesture Recognition on HOAP-2 using WEBOTs Platform" published in **3rd IEEE International Conference on Advances in Computing, Communications and Informatics (ICACCI 2014)**, pp. 1897-1092, September 2014 (**Best paper award**).

C3. Neha Baranwal, Neha Singh and G.C. Nandi. "Indian Sign Language Gesture Recognition Using Discrete Wavelet Packet Transform" published in **IEEE International Conference on Signal Propagation and Computer Technology (ICSPCT 2014)**, pp. 573-577, July 12-13, 2014.

C4. Neha Baranwal, Ganesh Jaiswal and G.C. Nandi. "A Speech Recognition Technique Using MFCC with DWT in Isolated Hindi Words" published in **Springer International Conference on Advanced Computing, Networking, and Informatics (ICACNI 2013)**, pp. 697-703, June 12-14, 2013.

C5. Avinash Kumar Singh, Neha Baranwal and G.C. Nandi, "Human Perception based Criminal Identification through Human Robot

# ACKNOWLEDGEMENTS

This thesis is the end of my four years of Ph.D. journey with so many ups and downs. Sometime I got frustrated and sometimes enjoy the achievements which is necessary for a good researcher. During initial phase of Ph.D. i have explored various research topic of my interest and become stuck, eventually, after so many discussions with my supervisor and other colluges, I settled for an interesting research problem which had piqued my interest for quite some time.

      I would like to thank all those peoples who encourage me when my hope down, motivates me to achieve my goal. First of all, I would like to express my sincere gratitude to my supervisor **Prof. G. C. Nandi**, not only for his incredible academic support, but also for giving me so many wonderful opportunities. This thesis would not have been possible without his constant technical advice, innovative thinking and endless enthusiasm. I am grateful for not only providing guidance and but also for the avid interest that he showed in my work. I would also thank my supervisor for introducing me to robotics, for opening my eyes to this wonderful world, and for his energy, enthusiasm and dedication which were a constant source of inspiration. For these and for many other reasons I will be always indebted to him.

My thanks extend to **Dr. Pavan Chakraborty**, in-charge of robotics lab and head of Ph.D. cell, **Dr. Rahul Kala**, assistant professor, IIIT-Allahabad, who has generously provided invaluable knowledge and advice and with whom I have greatly enjoyed discussing aspects of my thesis.

My special thanks goes to my special friends **Avinash, Anup and Seema**, without them I cannot imagine my Ph.D. Their invaluable friendship provides me an enjoyable working environment.

My heartfelt thanks goes to **Ms. Manisha Tiwari**, who has helped me in experiments, assignments, printouts etc. throughout my Ph.D. research. I am also profoundly grateful to Shweta Tripathi, Neha Singh and Kumud Tripathi for assisting me in my research work.

Finally, I would be eternally grateful to my parents **Mr. Gopal Lal** and **Mrs. Rita Lal**, and my younger brother Naveen, who have given me no end of love, support and understanding during the sometimes trying days, weeks and months that it took to complete this thesis.

# Chapter 1
# Introduction

*This chapter gives a brief idea about the world of robotics, the different modes in which one can interact with the robot. Among all modes of communication, speech and gesture are gaining importance, which in turn increases the importance of gesture recognition, speech recognition and the combination of both (multimodal). Basic concepts about speech and gesture and the techniques related to their recognition are then defined, which can be utilized for the implementation of these two. Work done previously in these two prominent areas of research has been then described. Ultimately, the problem definition and the motivation for the formulation have been defined. Finally, the layout of the whole thesis has been depicted.*

## 1.1 Human Robot Interaction (HRI)

Today robotics is a vibrant field of research and it has tremendous application potentials not only in the area of industrial environment, battle field, construction industry and deep sea exploration but also in the household domain as a humanoid robot. Human's physiological and behavioral identities, like face, behavior, gesture, posture and speech are necessary for the efficient and safe interaction. It permits the robot to understand what users want, and to bring forth an appropriate reply. Human robot interaction (HRI) [1, 2, 3] is the study of interactions (how people respond to robot and vice-versa) between the human and robots. With the advances in manufacturing and artificial intelligence, more human like and sophisticated robots have been developed like ASIMO [4], NAO [5], etc. to share the same workspace as human does. HRI is the blend of different technologies like Human Computer Interaction, Robotics, Natural Language Understanding, Social Sciences and Artificial Intelligence[222][223]. In 1941, Isaac Asimov stated the three Laws of Robotics as, there should be no adverse effects of the Robot on humans. (i) They should never injure themselves. (ii) The robot should obey the human beings for all the orders not causing confliction with the First Law. (iii) Keeping the First and Second Law in mind, Robot should take care of the security of its own self [6]. There are various fields where robots have been used to achieve the target with human beings. This interaction needs the generalized medium used by the normal people. The general framework of human robot interaction is shown in Figure 1.1.

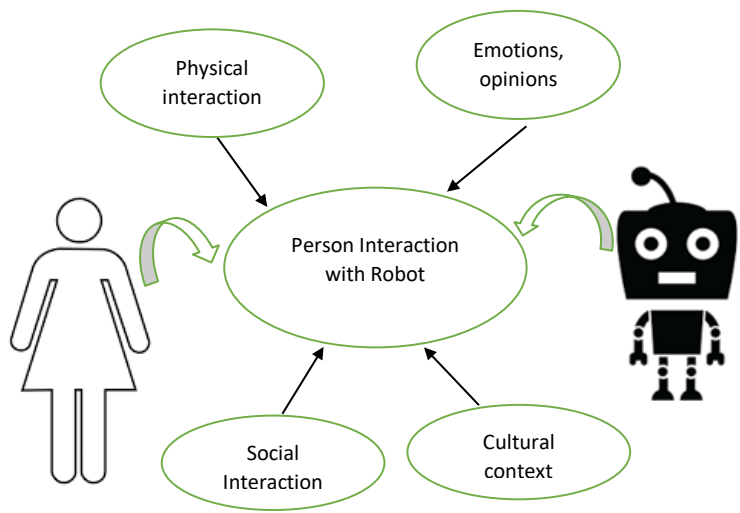

Figure 1.1 General framework of human robot interaction

### 1.1.1 Types of Modes

There are different techniques through which a human can interact with the Robots which is presented in Figure 1.2.

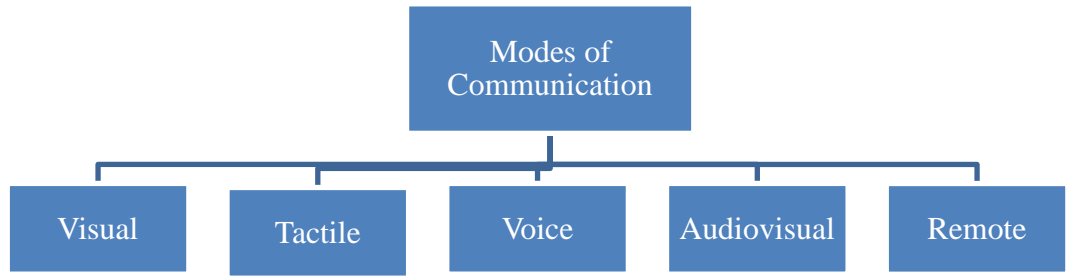

Figure 1.2 Modes of Communication

Some of them are as follows:

- **Visual:** It is a non-verbal mode of communication known as gesture where we use the movement of body parts to interact. Gestures can be of any type, like it can be head gesture, hand gesture, anybody gesture or even facial expressions [7][224][225][226]. Different types of gestures have been shown in Figure 1.3.

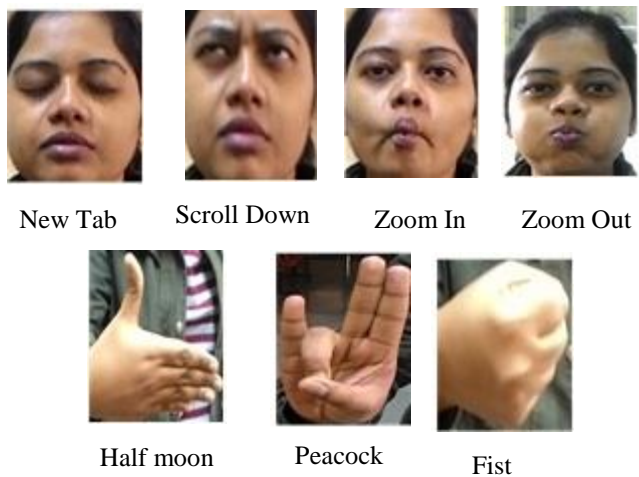

New Tab  Scroll Down  Zoom In  Zoom Out

Half moon  Peacock  Fist

Figure 1.3 Types of gestures

- **Tactile:** There are two types in tactile mode: tactile screen sensing and tactile skin sensing. Tactile skin sensing [8] as the name suggests, is skin sensing, whereas in the field of Robotics, tactile screen sensing is used.

The tactile screen analyzes where the force is being given, as well as in what amount. A typical tactile sensor is shown in Figure 1.4.

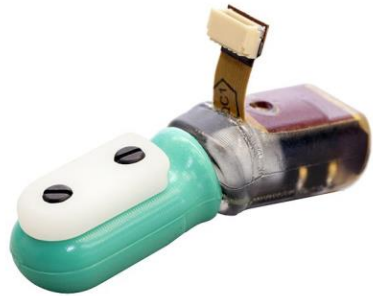

Figure 1.4 Tactile sensor

- **Voice:** Voice is the sound that comes out of the lungs and the vocal folds and when that sound is molded to produce a sound which is in some decidable form, it is called speech. In this mode of communication, speech is used for which the speech relevant technologies like automatic speech recognition and speech synthesis are used. Speech is the simplest and easiest medium which has been used to communicate [9]. Vocal Tract System, which produces voice is shown in Figure 1.5.

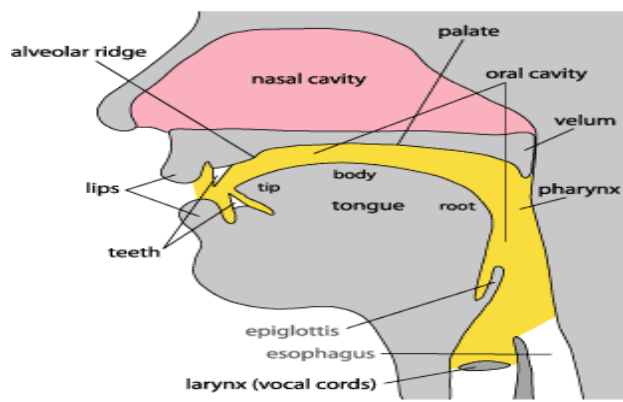

Figure 1.5 Vocal Tract System [10]

- **Audiovisual (Multimodal):** One or more ways of communication can be integrated. Gesture based and Speech based communications are the most important among them, which is the concentration of the researchers these days. Even we can combine these two modes of communication to remove some sort of ambiguity [11, 12]. Ambiguities, in the sense like, if we say "hi" to a Robot, using only gesture based communication; it is difficult for the robot to identify whether it is hi or "bye". But if we combine the gesture based communication with voice, it becomes much more convenient for the robot to determine what exactly the thing that we want to communicate. The multimodal human robot interaction is shown in Figure 1.6.

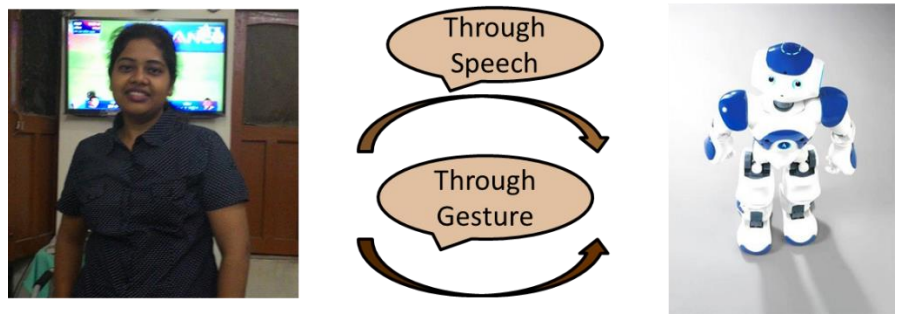

Figure 1.6 Human Robot Interaction

- **Remote:** Last but not the least is remote, where the robot can access the remote information using internet through its processing unit which is its computer [13]. In this way it can access the weather information, news, bus timetables, e-mails etc. Internet access through Web-2.0 is the best part of any robot.

Most of the persons use two modes to convey information to others: one is a gesture and the other one is speech. In this thesis both the mediums are used for communication and finally a generalized fusion model (speech and gesture) is developed to exchange the information between human and robot.

### 1.1.2 Gesture Based Interaction

*"Gestures are expressive meaningful body motions, -i.e., The physical movements of the fingers, hands, arms, heads, face, or body with the intent to convey information or interact with the environment."[14]*

Gestures have relevance in many fields, and some of the application domains of gestures are as follows:

- *Virtual reality:* These applications use gestures for manipulations of objects in real time exclusively through the hands. Model or avatar is controlled and most suitable for game playing [15].
- *Augmented reality:* Mainly uses tracking of markers or specific objects can be traced and tracked though machine vision and extensively used for the visualization of virtual objects [16].
- *Robotics & tele presence:* For controlling and commanding the robot's movement and behavior have significant application in the area of health care, military, entertainment etc.
- *Desktop system application:* For many simple applications which facilitates the user with a lesser amount of use of keyboard and mouse based interaction [15, 16]. In the same way pen gestures and mouse gestures are used for many applications as described in [17, 18].
- *Graphics applications:* Sutherland used gestures for graphics applications in 1963 [19], found the first application of gesture in HCI domain. The gestures of strokes, lines, circle and many more shapes captured and used for controlling the applications in [19, 20].
- *Communication applications:* For human computer interaction, interfaces are the greatest concern of the user and developer as given in [21, 22]. From 1980, research started for natural human computer interaction with graphics and other desktop applications as reported in [23]. The emphasis of the gesture research was mainly on hand gestures as reported in [22].

Here we concentrate on hand gestures specially on sign language gestures. Sign language are the gestures performed either using one hand or both the hands. It represents a particular symbol like hello, you, two, go etc. When the signer perform the sign at a particular stand of time and at that time a hand is not moving the pose is called the posture and when the signer perform the sign and the hand is continuously moves then that is called the gesture. Sign language is the only means of communication among deaf and dumb community. Thus, in today's scenario almost every country has tried to make their own sign language for the ease of their people. Because the sign language does not imply as a universal language. Likewise, every country is having their own sign language containing their own syntactical and grammatical variations. American Sign Language [24], British Sign Language [25], Japanese Sign Language [26], Arabian Sign Language [27] and many more are the example of the efforts that have been made for the ease of impairs. In similar way India has its own sign language known as the Indian Sign Language (ISL) [28] consists of both dynamic and static hand movements.

Gesture recognition has been performed in two ways, one is a sensor based [29] and another one is a vision based [30]. Nowadays vision based gesture capturing and recognition becomes the best option due to the advancement of technologies coming in this area. In this thesis, we focus on vision based ISL gesture recognition and their applications in humanoid robots.

### 1.1.3 Speech Based Interaction

Another medium of interaction which has been frequently used for communication is speech. Speech Recognition (SR) [31, 32] is about the analysis of some unique characteristics of the speaker's voice to convert it into text form. In other words, we can say, using speech recognition, we can identify what is being said. Speech recognition has been classified into 4 parts shown in Figure 1.7.

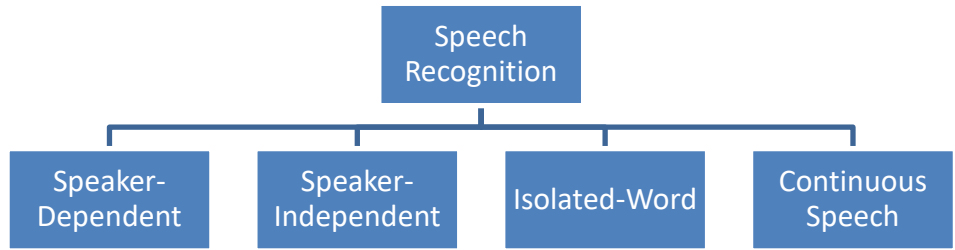

Figure 1.7 Classification of Speech Recognition

- *Speaker-Dependent v/s Speaker-Independent:* Speaker-Dependent [33, 34] systems recognize speech from only one speaker, while speaker-independent [33, 35] system can recognize anyone's speech.
- *Isolated Word SR v/s Continuous Speech SR:* There is a brief pause between the spoken words in Isolated Word SR [36, 37], while in Continuous Speech SR [38, 39] no pause is required.

There are various application areas where speech has been applied few of them explained as:

- *Dictation*: In various fields of work documentation of the work is a big need [40, 41]. Doing this documentation work through speech makes the task quite easy as well as fast.
- *Application in Fighter Aircrafts, Helicopters, and Training Air Traffic Controllers:* These also take the advantage of speech recognition, as through speech, commands can be given, which reduces the need of extra personnel [42, 43]. As well as, at the same time one can have control of several tasks [44], as at the time when the person's hands are busy doing some task; she/he can have control of the other ones through voice.

To establish the communication using speech, we first identify speech what he or she tries to say that is what speech recognition is. Here we focus on speaker independent speech recognition technique where Hindi as well as English speech is recognized. Then these speeches are performed by a humanoid open architecture platform HOAP-2 [45, 46] robot in the form of a gesture.

### 1.1.4 Multimodal Interaction

Next mode of interaction is the multimodal interaction [47, 48, 49] where different modes like gesture, speech, behavior etc. have been combined to make interaction effective. In this thesis speech and gesture have been combined as a multimodal. Fusion has been performed in two ways, one is feature level fusion and another one is a decision level fusion. In this type of fusion a conclusion has been done on behalf of decisions obtained from the individual mode (speech and gesture) by applying some conditions.

## 1.2 Motivation of the Research

As per the data provided by Census 2013, the differently-abled population in India is 26.8 million. In percentage terms, this stands at 2.21 percent [50, 51]. There has been a marginal increase in the differently-abled population in India, with the figure rising from 21.9 million in 2005 to 26.8 million in 8 years where 5.07 million people have been affected with hearing impairment and 5.03 million people having problem of vision [52, 53]. Apart from these, today the life of every person is fast and busy since mostly men and women both are working. These people needs a helper to do a daily life tasks, but no one has time to help others. Therefore, we need a device that follows our instructions. For this a humanoid robot has a good tool for solving all these problems. Currently various types of robots like mobile robot, social robot, etc. have been developed. Social robots need a higher level of intelligence than industrial robots and they must be socially capable of interacting with the people around who are not robot specialist. But today humanoid robots have not very interactive; this is the biggest drawback of using a humanoid robot. Therefore, scientists have been motivated to make robots human friendly, so that people can easily interact with the robots.

## 1.3 Research Objectives

Our research objective is to design a multimodal communication interface for humanoid robot so that people can perfectly interact with it in real time. It has four basic objectives:

- To build a person invariant gesture recognition framework which identifies an Indian Sign Language (ISL) gestures in real time environment.
- To explore and validate a machine learning technique which reduces the time complexity of the HRI system.
- To design a speaker invariant speech recognition framework which recognizes a Hindi speech word in real time scenarios.
- To fuse speech and gesture mode in such a way that human robot interaction can be done effectively.

## 1.4 Statement of the Problem

The main aim of the thesis is to design and develop a multimodal communication framework with humanoid robots using gesture and speech. Various challenges arise to make human robot interaction effective in real time scenario. Here we use two modes of interaction, gesture and speech. Each mode has its own challenges. The challenges related to each modes are discussed below:

- Following are the challenges which have been addressed corresponding to hand gesture:
  1. To localize the hand from the full human body. This leads to the problem of hand segmentation.
  2. Extraction of the start and end point of individual gesture (key frame extraction) in between the sequence of gestures which distinguishes informative and uninformative gestures.
  3. Abstraction of appropriate features with respect to the hand.

- 4. To recognize a gesture in real time. This leads to the research challenge of minimizing the computation cost of gesture recognition systems.
- Challenges corresponding to the speech signals:
  1. Reduction of noise from recorded speech.
  2. Abstraction of appropriate features with respect to the speech.
  3. Identification of the correct words from the variation of pitches of a human voice (slow, medium, high).
- Challenges associated with multimodal fusion:
  1. Construction of a fusion technique to fuse different mediums which makes the interaction effective.
  2. To define the conditions for fusion means on which basis the two different modes are combined.

## 1.5 Deliverables of the Thesis

There are various ways through which human can interact with robots. In this thesis, we used three ways (gesture, speech and a combination of both) of communication. In gestures mode of communication we proposed three frameworks one is for Isolated and two are for continuous ISL gesture recognition. In any gesture recognition system segmentation, key frame extraction, feature extraction and classification are the major issues which we have solved using different techniques like gradient method, DWT, MFCC. Another mode of interaction we have used is speech mode. In any speech recognition system noise reduction and feature extraction are the two major problems which are being addressed in this thesis using discrete wavelet transform and Mel Frequency Cepstral Coefficient and Human Factor Cepstral Coefficients. The last work of the thesis is the multimodal interaction, this work we have done because a single mode of communication is not sufficient for active interaction. Therefore we have proposed a combine modal known as multimodal framework for human robot interaction and finally it is simulated on HOAP-2 and NAO humanoid robots. The layout of the thesis has been shown in Figure 1.8.

- We have developed a novel framework for isolated gesture recognition where MFCC features have been extracted by processing of transforming images by discrete wavelet transform. The proposed framework for ISL gesture recognition is divided into 5 modules: Data Acquition where data collection has been done through webcam and Canon EOS camera then each video has been divided into a sequence of frames. These frames has been further gone to the pre-processing phase in which hand extraction, silhouette image formation, and boundary point extraction are performed. In this module DWT has been applied for finding boundary points of each silhouette image of gestures reduces the size and analyses the images w.r.t time as well as frequency. These features have been further processed for finding MFCC coefficients which has been done in feature extraction module. The MFCC feature extraction technique has been widely used in speech processing here we explored MFCC in hand gesture recognition in combination with DWT and see the performance on image frames. Due to spectral envelop property of MFCC, the proposed method gives a high recognition rate with less processing time. Further probes have been classified using SVM and KNN in the classification phase. In the last module these classified gestures are performed by a HOAP-2 robot using Webots platform.
- Further research has been done on continuous ISL gesture recognition where wavelet descriptor and possibility theory based hidden markov model (PTBHMM) is applied for feature extraction and classification. Any gesture recognition system has five modules data acquisition, preprocessing, key frame extraction, feature extraction and classification. Here PTBHMM is used for classifying a probe gesture. After data collection background modelling and segmentation has been done for hand subtraction and extraction of the start and end point of informative gestures. Gesture segmentation is done using gradient based key frame extraction method. This helps us to break each sentence into a sequence of words (isolated gestures) and also obliging for extracting frames of meaningful gestures. Finally feature extraction and classification has been done. Features extracted from these gestures are classified using PTBHMM.

Here all the three problems (evaluation, decoding and learning) of HMM are solved using possibility theory. By applying this possibility theory, the time complexity of real time gesture recognition system gets reduced by N number of times of the complexity of classical HMM. Experimentally, we have also proven that the computational complexity of PTBHMM is very less in comparison to classical HMM.

- In the third module we have proposed a framework in which NAO humanoid robot recognizes continuous ISL gestures in real time environment and then translates it into text format. These texts are further matched with the knowledge database of NAO robot using the shortest distance method. This matching will also increases the classification accuracy. Here the database has been created using NAO vision sensors which are accessed through an NAO MATLAB API. This database contains various sentences of ISL, commands etc. which is helpful for normal persons to understand commands through the robot. These are very much helpful in the communication established between human and robot or deaf and dumb persons and human. In recognizing any gesture preprocessing and coarticulation detection are the major issues which has been solved in this module. After recognizing a sentence or commands, NAO converts it into a speech format or answer to deaf persons into gesture format. In this way, the difference between the normal people and the hearing impaired community has been minimized which improves the communication ratio between Deaf and Dumb community and normal persons. Here continuous gestures has been tested in real time environment together with dark color, full sleeve dress using an NAO humanoid robot.

- Subsequently, we have described a speech recognition technique where HTK toolkit is used. HTK toolkit has been based on the bigram model developed by speech vision and robotics group of the Cambridge University Engineering Department. For each module having one special function like for feature extraction HCopy command, for recognition HVite command is used. All the experiments are performed on Hindi speech recorded using audacity software in robotics and artificial intelligence laboratory. Performance of the toolkit has been evaluated in two different environment one is speaker dependent and another one is speaker independent. MFCC and LPCC are used as a feature for these speech signals and found that MFCC works better than the LPCC features.

- We have proposed a new concept for speaker invariant speech recognition where DWT with HFCC is used. DWT is used for reducing the noise present in the speech signal after that, HFCC features of the speech signal are extracted. We have applied HFCC with DWT for isolated Hindi as well as English word recognition. Words are recognized using Bayesian classifier and HMM technique. HMM is a best tool for recognizing any speech signal because signals are time dependent and HMM predicts the result on next time instances. Performances of various techniques like HFCC with DWT, DWT with PCA, MFCC with PCA, only PCA and MFCC DWT with PCA etc. are analyzed and compared with a proposed approach (DWT with HFCC). Among all we found that the performance of HFCC DWT with Bayesian and HMM method is better than the other existing technique.
- Fusion is generally used for making the system more reliable. The major issues of fusing two different styles (gesture and speech) are a type of fusion technique, what the normalization parameter of the dataset are and how to fuse two different modes. Therefore, in this thesis we used decision level fusion for fusing gesture and speech mode. It is a technique used for combining the decisions obtained from different modes. We define a five different types of cases for both the modes on behalf of the results obtained after classification. Then we merge the decisions obtained from those cases and obtained the final results. The results are based on the dominance factor means which mode is more dominating. If both the modes are equally dominating then we include a weighting factor which is calculated using the weighted average method and weight calculation method. This method gives the very prominent results when conditions of one mode is not very favorable.

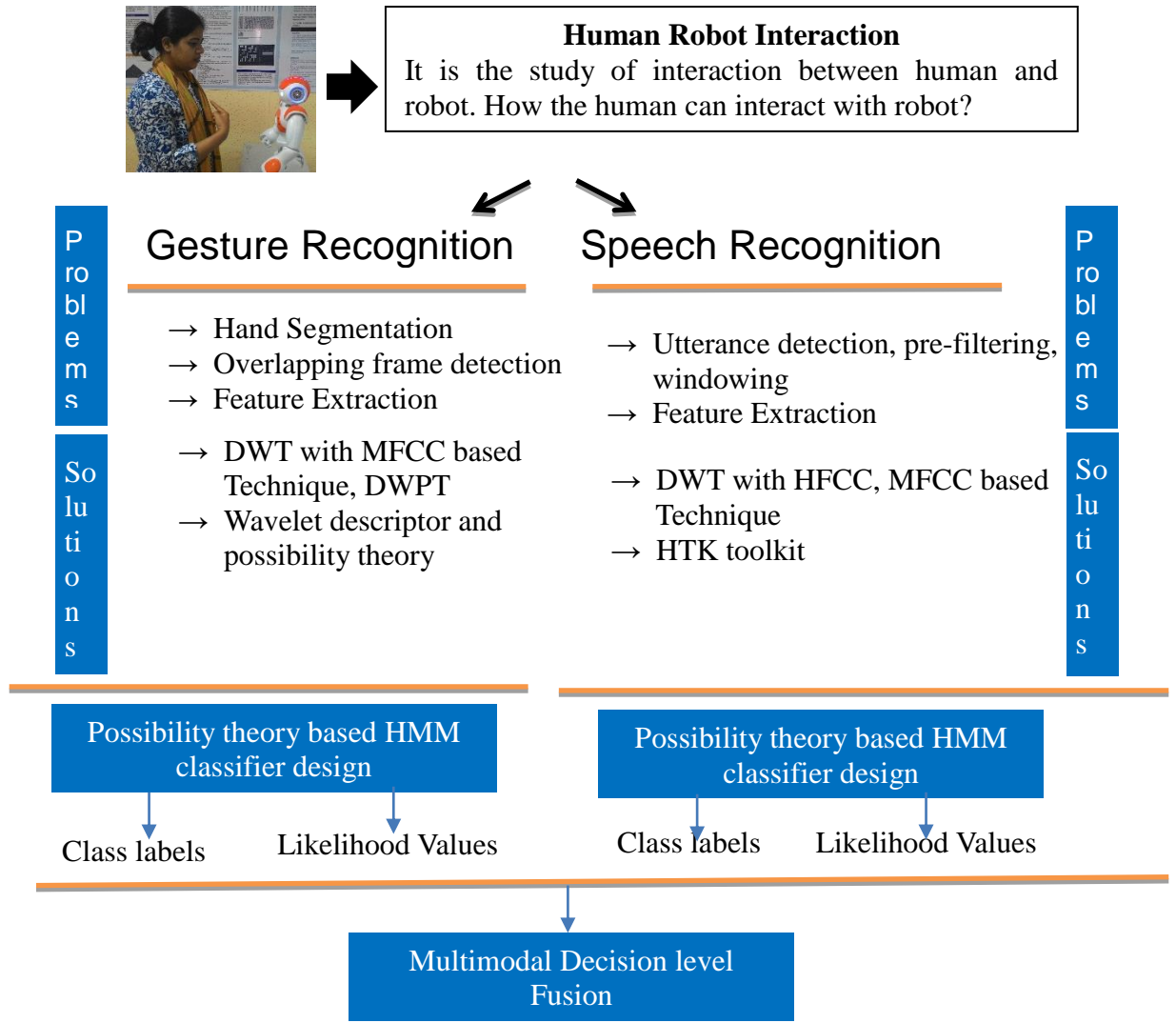

Figure 1.8 Layout of the thesis

## 1.6   Structure of Thesis

This thesis is divided into seven chapters. An overview of human robot interaction and their different modes and the objective of introducing this research are defined in **Chapter 1**. This chapter addresses the background motivation for human robot interaction through multimodal. The research challenges and original contributions of the thesis are discussed in this chapter. Finally, the orientations of the entire thesis are mentioned at the end of this chapter.

*Chapter 2 –* **Analysis of Previous Research** provides a comprehensive analysis of literature related to different modes like gesture, speech and a combination of both for solving human robot interaction problem. In this chapter, we categorize the literature into three portions- first portion covers the works related to gesture recognition. The complete work is divided into two parts, one covers the isolated gesture recognition and another covers the continuous gesture recognition. Any gesture is recognized into two ways sensor based approach and vision based approach. Here we focus on the vision based approach. Therefore, maximum literature focuses on the domain of vision, but we give the details of few work related to other sensor based approach. The second part gives the detail about the previous work done in the field of speech recognition. Where we focus on the literature related to speaker invariant speech recognition technique. Most of the work focuses on the sampling and features of the speech signal because these two are major components of any speech recognition system. End of the thesis deals with the multimodal decision level fusion where we explain the literatures related to the fusion techniques. We mainly focus on the works which is used for combining the results obtained from two different sources.

*Chapter 3 –* **Isolated Gesture Recognition Techniques:** Here we explain the work which recognizes the various types of dynamic and static gestures in different background and light conditions. We apply Mel frequency cepstral coefficients (MFCC) technique for extracting gesture features. MFCC is generally used in speech recognition domain, but we applied it on gestures and present the performance. After rigorous analysis, we found that discrete wavelet transform (DWT) with MFCC technique give good accuracy on images also. Finally, few of the gestures have been implemented on the HOAP-2 humanoid robot for communication purpose.

*IIIT-A, Robotics and AI Lab* 16

*Chapter 4 –* **Development of Strategies for Continuous Gesture Recognition Using Possibility Theory Based Hidden Markov Model** describes the methodology used for recognizing continuous gestures. We have designed a possibility based classifier which solves all the three problems of classical HMM. The continuous gesture dataset is vary ambiguous in nature due to minor change occurred in their shape. These ambiguities have been handled here by applying possibility theory. Possibility theory handles both uncertainty as well as imprecision. Theoretically as well as experimentally we proved in this chapter that it also reduces the time complexity N times less than the time complexity provided by the classical HMM ($N^2T$).

*Chapter 5 -* **Development of an NAO based Experimental Framework for Human-Robot interactions with Continuous Indian Sign Language** illustrates the process flow of humanoid robot based continuous gesture recognition. This chapter deals with the four main points which are preprocessing, overlapping frame detection, feature extraction and classification and text generation. All the data have been taken through the NAO humanoid robot. NAO has a vision camera of 1.22mp. The detail configuration we have discussed in this chapter. After video capturing of various gestures, color based hand segmentation is done. Overlapping frames are detected for separating one gesture from another gesture. Then three feature's shape, speed and orientation of the hand are extracted and combined using feature mapping technique. Finally, it is classified using possibility theory and then indexing is performed for text generation.

*Chapter 6 -* **Multimodal Human Robot Interaction** framework incorporates a speech recognition system as a part of this framework. A HMM based toolkit is applied for classification of Hindi speech data. Here we apply this toolkit for speech recognition where MFCC and LPCC features are extracted. Experiments are performed on different types of environment like speaker dependent and speaker independent and their results are compared. We have also discussed a limitation of HTK toolkit. To minimize the limitations of HTK toolkit we proposed a DWT with HFCC based speaker invariant speech recognition system, which provides an intuitive insight into speech recognition framework for Hindi speech. The major problems we deal is an audio recording, utterance detection, pre-filtering, windowing, feature extraction and classification. Here MFCC and HFCC features of human voices are extracted and classified using HMM and Bayesian classifier.

Subsequently speech and gesture have been combined to get better interactions. For integrating both the modes we used decision level fusion because these two modes are completely different in nature therefore it is very difficult to integrate. These two modes have different features, different properties because one is signal and another one is image therefore it is better to integrate at the level of decision. We define five types of test cases on behalf of the decisions obtained after classification. In these test cases we define certain types of condition and then we combine it.

*Chapter 7* – **Conclusion, Recommendation for Future Work** attempts to conclude the thesis with possible ideas and enhancements for future work. It also offers an insight into the limitations of different modes of human robot interaction framework. It suggests exploring feasible aspects with untapped potential for the enhancement of the identification process in the gesture and speech domain. Finally, it gives the idea about how to integrate different modes.

# Chapter 2

# Analysis of Previous Research

*This chapter presents a review and analysis of multimodal based human robot communication and interaction research. In this review, various methods related to gesture, speech and multimodal have been explored, which could enhance the natural interaction among people and robots. After going through the literature we found that no single medium is enough for a communication establishment between human and robot. This is the fact to inspire and consider a multimodal based human robot communication system.*

## 2.1 Human Robot Interaction

In HRI, learning is an important task, when the environment is structured and the robot has to work in the real world domain, behavior based learning is preferred [54]. Behavior based system [54, 55, 56] can execute modules continuously and in parallel. Another approach which used is robot learning by demonstration [57, 58]. Learning is the ability by which system acquires new concepts, skills to adapt the environment changes and accordingly tune its performance. The teacher gives the instruction; iteratively these instructions may be refined. This type of work is expected to be done in the learning process. Thus, the learning process is further categorized as skill learning and task learning. Learning is performed by observation, imitation, experience and learning from multiple demonstrations.

### 2.1.1 Humanoid Robots

The robot is defined as "An Intelligent robot is a mechanical creature which can function autonomously". Industrial robots are used to performing repeated tasks in **highly structured environments**. A humanoid robot is an autonomous robot with human like appearance, it can adapt to changes in its environment or itself and perform tasks like human being. Examples of humanoid robots are: Asimo by Honda [59], HOAP by Fuzitsu [60], NAO by Aldebaran Robotics [61], Robonaut by NASA [62], Hubo by KAIST [63] etc.. Humanoid robots have revolute joints and the movement of the body parts obtained by rotation. For programming the humanoid robot, through demonstration [57, 64, 65, 66], observation [67], imitation [69, 70], prototype building [71], the kinematics demonstration [66], tele operation [72, 227,228] and genetic programming [73] approaches are in use. Learning of the new behavior and reproduction of behavior in unknown situation can be taught through artificial intelligence approaches.

### 2.2 Gesture based Human Robot Interaction

Two schools of thoughts for gesture recognition are – (i) Use vision based instruments and approaches (ii) Use specific tracking devices and data gloves. Both the approaches have significant advantages – Vision based approaches gave the liberty to user, to act independently, they are usually less costlier where as specific tracking devices provide accurate data for further use. But these methods also suffer from several disadvantages- vision based technique heavily depend upon the environment conditions, lighting, distance from the camera, obstacle, occlusion etc. and need to solve machine vision problems. The associated disadvantages with specific data capture devices are: Users dependability on the device, expensive, needs specific hardware arrangement hence not suitable for simple applications.

### 2.2.1 Data Glove based Gesture Recognition

For hand gestures dataset the interaction of gloves wearing hands with the computer is necessary. And in a report the instrumented gloves provide joint angle values according to the movement of the fingers. Virtual environment interface technology [74], a review by Youngblut, Encarnac¸ao's survey on input technology [75] and survey on human movement technology [76] by Mulder presented a deeper analysis of over 25 different tracking devices on the market now a day. Magnetic tracker, acoustic tracker, and inertial tracker are three basic tracking devices have been practiced. Instrumented gloves are more often than not employed in non-vision technique to measure finger movement and it utilizes several kinds of detector. In the recent era, many cost effective gloves are available in the market. The 5DT date glove is one of the examples. According to the specifications [77], for overall flexion of the four fingers and the thumb, it uses five fiber optic sensors. Statistical analysis for gesture recognition using the data-glove is presented by the Rung- Huei Liang et al. [78]. Dynamic gesture consists of many things like frame at a single time stamp called the posture that is a static information of a gesture in a particular time and orientation how much degree hand is moving weather the hand is straight or vertical etc. The position of the hand is also a critical issue. Hand position detection where the hand is situated is necessary and important to determine. The hand may be staged in front of the body or near or far from the physical structure. For the training and testing of the system words of a Taiwanese Sign language is used. The database consists of the posture, motion element and orientations. Concept of Range Of Motion is used by Parvini and shahabi [79]. Where data of several hand gestures are collected using wearable sensor. Several features like position, orientation, and finger bending by hand, are captured using different sensors attached in a light weight hand glove. For finger bending analog flex, for hand position and orientation ultrasonic or magnetic flux having six degrees of freedom is used. These types of gloves are used for generating the 3 dimensional model of hand gestures. This technology has been first used by Thomas G. Zimmerman [80] for establishing the interaction between human and computer. An uwave based sign language gesture recognition system has been proposed by Jiayang Liu et al. [81]. uwave requires a single training sample of each gesture and give satisfactory results on 4000 samples having eight types of gesture. A three axis accelerometer is used for database collection. The KHU-1 data glove [82] based approach has been used for 3D hand motion and tracking and recognition of hand gesture. The data glove is consisted with three tri axis accelerometer types of sensors, one controller along with one Bluetooth. It transmits motion signals to system through wireless medium via Bluetooth. The kinematic theory is applied to construct 3D hand gesture model in digital space. The rule based algorithm has been utilized in order to recognize hand gestures using KHU-1 data gloves and 3D hand gesture based digital model. Laura Dipietro, Angelo M. Sabatini, and Paolo Dario have suggested a survey paper based on glove based system [83, 84,229] and their applications. This paper provides an insight about the understandings of the system and the salient characteristics in application point of view. It explores several areas of applications like biomedical and engineering sciences. The recent technology has been associated with glove based system. One research paper is entirely constituted on recognition of Chinese Sign Language [85] using a data glove based system. This paper tells about the recognition process by considering the data glove as an input device for extracting features from hands. The words in Chinese sign language have been customized by posture, orientation of the palm and trajectory planning while movement of hands. In this paper Cyber glove with 18 sensors has been opted for an input device. Ishikawa M, Matsumura H, has proposed a recognition system based on the self - organization method [86]. The entire recognition process has been followed by acute measurement of finger joints using data glove, extracting out the gestures from a sequence of hand shapes and proper aligning of the data length. Finally, input vector has to be generated for self-organization towards clustering of hand gestures. The recognition of hand gesture explores the applications towards virtual reality using glove based system [87, 88]. This concept is used in human computer interaction by developing a neural network model using back propagation algorithm and radial basis function. Here data glove is to be considered as an input device for measuring of 18 angle values of different finger joints. Kadous [89] and Sturman [87] have listed some issues related to both the techniques (data glove and vision). The cost factor is bit effective in case of robust and complex posture and gesture recognition by instrumental glove and tracking devices. And a vision based technique is bit inexpensive. User comfort level is high in vision based technique as in this technique user may not have to wear that glove and tracking devices processing time can also be saved.

### 2.2.2 Vision based Gesture Recognition

In this approach a video of hand gestures is taken using web cam, HD camera, Kinect etc. There are two approaches in vision based systems, one is 3D model based approach and the second one is an appearance based approach. In 3D approach volumetric [90] or skeleton [91] image of hand gesture has been created and classified using various classifiers like HMM, SVM etc. Yale Song et al. [92] proposed a 3D model based hand gesture recognition system where 3D model has been generated for a body frame captured using the camera. Both the static and dynamic information makes the system more reliable in capturing and extracting the body model. From this 3D model a shape of hand has been extracted and classified. A combination of 3D appearance and motion dynamics of hand are used as a feature vector by Ye. Guangqi et al. [93]. These features have been extracted using an object centered approach where hand volume is calculated using region based coarse stereo matching algorithm. HMM and neural network are used for classifying a controlling and manipulative gesture. Mahmoud Elmezain et al. [94] proposed a gesture recognition system where the 3D depth map is used for hand segmentation. Then 3D dynamic feature of hand is generated using mean shift and the Kalman filter algorithm. HMM is used for classifying these dynamic features and achieved 90 percent classification accuracy. Another approach is appearance based, in this method a 2D image of the hand is extracted from a video sequence. Various features like shape, area, contour, velocity etc. has been extracted from

*IIIT-A, Robotics and AI Lab*                                                                                                           23

these 2D images and further it is learned using any mathematical model. Appearance based technique is proposed by Davis and shah [95]. They have used a finite state machine to model four different stages of a gesture and then it is classified using HMM. Lee Jaemin et al. [96] proposed the Kinect based gesture recognition system where depth images of hand are extracted. Kinect contains two types of sensors, one is for capturing the color data and another one is for capturing the depth information. Arm movement and hand shapes are extracted from this hand region. These features are then classified using Gaussian based HMM. Another Kinect based robot control methodology is proposed by Biao MA et al. [97]. Several types of gestures is recorded for various applications. When the organization picks out the correct gesture, a command referred to that gesture is sent to the robot controller through the wireless medium and then that command is performed by the automaton. The author used various techniques like Histogram, OTSU's technique, edge detection and Hough transform for hand segmentation, feature extraction. For reducing the aliasing effect, low pass filter is applied. Deng-Yuan Huang et al. [98] proposed a hand gesture recognition system where preprocessing, noise reduction and feature extraction are handled using convolution, Gabor filtering and PCA. Gabor filter is a combination of wavelet and kernel function. These two functions are helpful for noise reduction and feature extraction. Here the angle of orientation is used as a feature vector, then the dimension of these features are reduced using PCA. For classifying a probe gesture, SVM is used and achieved a 95.2% classification accuracy. Fingertip based static hand gesture recognition system is proposed by the Ra'eesah Mangera [99, 100]. Colour, depth and anthropometric measurements of hand are used for hand segmentation and detection. Parallel boarder based, depth based and also K-curvature based methods are used for fingertip detection. K-mean and neural network are used for classifying these fingertips. All the experiments are performed on Visapp 2013 dataset which is a CMU military dataset having 8 gestures of 29 persons. The author Martin [101] discussed about a really important topic that is, how many cameras are sufficient for distinguishing a special motion. First, he experimented with one camera and then go for more than one camera.

After analysis he shows that one camera is sufficient for recognizing any gesture. Stereo camera is only needed for depth analysis. Here contour, edges of hands are used as features for hand gestures. Heap and Samaria proposed an active shape analysis technique (smart snake model) for hand gesture recognition [102]. This project is developed for multimedia application at the Olivetti Research laboratory. Gestures are classified using genetic algorithm where data is captured using a networked camera. MFD (modified Fourier Descriptors) is used for hand shape extraction is proposed by Licsar and Sziranyi [103,229,230]. The edges and contour of open hands are extracted and evaluated properly that makes these techniques reliable for recognizing any gestures. Principal Component analysis is used for calculating principal components from these feature vectors, which reduces redundancy. Birk et al. [104, 105] proposed a vision based gesture recognition system where the dataset of hand gestures is captured using a monocular camera on a moving vehicle. The major issues that arises is how to handle the ego motion of the vehicle which is handled using the spectral registration method. This method is generally used for underwater scenarios. Then the motion vector is used as feature which is being classified using various classifiers. Wxelbat [106] developed a layer wise hierarchical architecture. In which higher level used some information from lower level and a feature for fist, flat hand posture or a waving gesture is generated. Vision based point gesture recognition system is developed by M. Czupryna [107]. These point gestures are used as an application in the human computer interaction domain. It similarly worked as a mouse. Experiments are performed on various types of hand poses like zoom-in, zoom-out, left click, right click etc. are captured using a webcam. Senthil Kumar [108, 109] developed a user control gesture recognition system running on the computer. It classifies various types of gestures like point gesture, click gesture, reach gesture etc. in the form of image signal. All these gestures are used as a controlling command in any real world applications like video games, graphical editor, virtual flight simulation etc. A probabilistic framework [110] has been applied for the recognition and reconstruction of gestures for humanoid robot which has been composed by PCA, ICA and HMM techniques. The key features of the gesture are extracted into preprocessing stage and the comparison of decomposition between PCA and ICA techniques have been carried out to reduce the dimensionality of the feature space. A continuous Hidden Markov Model approach is applied to train the gesture[231].

## 2.3 Related research on Isolated Sign Language Recognition

There are various types of gestures like pantomimes, language like gesture, gesticulation, emblems, iconic and sign language gestures. In this thesis, we have chosen sign language gesture because sign languages are the most promising way of expressing feelings, commands and so on. Also, it is helpful for hearing impaired society. This is the only medium that they have used to share their feelings, needs or whatever they try to express. Every nation possesses its own sign language similarly Indian people has their own sign language call Indian Sign language (ISL). This section explains the work related to SL as well as ISL recognition. Shape, orientation, moment and velocity are the main features in any sign language recognition system. Heap and Samaria used active shape analysis technique for hand gesture recognition [102]. MFD (modified Fourier Descriptors) has been used for classification of hand shapes by Licsar and Sziranyi [103]. The edges and contour of open hands are extracted and evaluated properly that makes these techniques reliable for complicated gestures. In general, one camera is used by Starner [90] and Martin [101] and they suggested that one camera can provide appropriate information. More than one camera can be used in case of getting depth or stereo data. Jafreezal and Fatimah [111] jointly suggested applying MFCC as a feature extraction technique for recognition of ASL database. They evaluated this technique with 10 gestures and get a satisfactory recognition rate. R.R. Igorevich et al. [112, 113] proposed a gray scaled histogram based hand gesture recognition technique where the stereo camera is used for collecting hand motion in 3D. Gray scale histogram method is used for finding the depth of hand images using a disparity map. This is useful for recognizing motions at different distances. South African Sign language recognition system has been proposed by the S. Naidoo et al. [114] where HSV color based segmentation method is used for hand segmentation. Colour based histogram is used for feature extraction and classification is done using SVM classifier and achieved a satisfactory results on various backgrounds and illumination conditions. Joyeeta and Karen [115] jointly used Karhunen-Loeve transform as a feature vector for hand gesture recognition. Histograms of local orientation used as feature vectors by William T. Freeman and Michal Roth [116]. They added advantages of using Orientation as robustness towards lighting condition and Histogram as translational invariance. ANN based sign recognition system is proposed by the Engy R. Rady et al. [117] in the paper. The dataset which he used consists of five digits from zero to five and nine symbols performed by the four individual signers at the different time stamp. Feature vectors extracted using Daubechies 4 tap filter. The length of the feature vector provided as the input in the feed forward, back propagation neural network is 44. An appearance based model is proposed Vaishali S. Kulkarni et al. [118] for American Sign Language recognition. The edges of the hand are used as a feature vector for classifying an unknown ASL. Hand gesture recognition using orientation histogram is presented by Anup Nandy et al. [119, 120, 121]. In his paper the author used the dataset of Indian sign language for robotic communication. Calculated orientation angle values are kept in one vector. Orientation histogram is the robust technique under rotation and invariant to lightening condition. Bhattacharya distance is used for calculating the distance between the training and testing feature vector of gesture. HMM based classification is also done for recognition. Finally a WEBOTS simulation has been presented in imitation learning. P.V.V Kishore et al. [122] proposed a gesture recognition technique in which a video dataset of Indian sign language is converted into text or audio file. Here the image differencing method is used for extracting the change done in the moment of hand, from one frame to another frame. This gets rid of the redundancy present in the dataset. In this paper Fourier descriptors are used as a feature extractor for extracting shape of the hand. A work related to ISL is proposed by Jay Shanker Prasad et al. [123] in which various clustering method like K-Means, Mean shift clustering, hierarchical clustering and fuzzy c-means clustering are used for cluttering the data having similar properties.

A codebook of each gesture is generated and classified using HMM. The fuzzy logic based technique is employed by P.V.V. Kishore et al. [124] where ISL dataset is used for demonstration of the result. The size of the feature vector is minimized by applying the PCA on the extracted features. After the feature extraction for the recognition step Sugeno fuzzy classifier is used. After that corresponding speech is played. In paper [125, 126] M.K. Bhuyan et al. proposed an ISL recognition system where three features (shape, trajectory and motion) of hand are combined and classified using dynamic time wrapping algorithm. The finite state machine is used for finding the informative gestures present in the complete video sequence. Experiments are performed on 24 ISL words having fixed background and get satisfactory results. Another work on ISL is proposed by Bhuyan et al. [127]. This work is done in two parts, first part recognizes an ISL gesture and the second part animate it. For gesture animation, various parameters by hand like finger position, orientation and metatarsophalangeal joints has been calculated. With the help of these parameters a 3D hand model is created for gesture animation which looks similar to the original one.

## 2.4     Related research on Continuous Sign Language Recognition

Most of the work is done for isolated sign language recognition, but very few literatures are available in the field of Continuous sign language recognition because of its complex nature which we discussed here.

A real time, the continuous ASL recognition system is proposed by T. Starner et al. [128] where he proposed two types of systems. One system in which the camera is mounted on the desk and get 92 percent classification accuracy. In second system where the camera is put at the head of the person. Here the author gets the 98 percent classification. In both the systems HMM is used as a classifier. Japanese sign language recognition system is proposed by Murakami et al. [129]. Recurrent neural network is used for learning and classification. Experiments are performed on three types of data one is static gestures called postures, second one is dynamic gestures and last is the continuous JSL. M.K. Bhuyan et al. [130, 131] proposed a continuous ISL gesture recognition system where three dimensional features orientation, hand position and length of an ellipse fitted on hand trajectory are used and classified using the cumulative random field. Key frame extraction has been done by considering 3 hand motions, preparation, stroke and retraction. With the help of these three stages, gesture spotting has been done. In this paper results are analyzed on both isolated gestures as well as continuous gestures and got 96% and 88.9% accuracy. He also compares his result with existing motion chain code (MCC) method and achieved 26.1% better accuracy. Here only one hand has been used for performing any gesture. Also, each gesture has been recorded in constant light. These are the downsides of this framework. Rung-Huei Liang et al. [132] proposed a Data Glove based real time, continuous gesture recognition. For this they have created a large vocabulary Taiwanese sign language translator. They solved the problem of key frame extraction using time-varying parameter detection. They detected the discontinuities in frames and do statistical analysis on four features posture, position, orientation, and motion. For gesture recognition they have used Hidden Markov Model. The average recognition rate of this system is 80.4%. The limitation of this framework is that it is person dependent and using gloves, which is very expensive and need physical connection between user and computer. Britta Bauer [133, 134] proposed suitable Features for Video-Based Continuous Sign Language Recognition. Here they are using hand location, hand shape and hand orientation as features and recognition using HMM. But limitation is the user has to wear simple colored cotton gloves to make this system to run in real time. Britta Bauer proposed another system that is Video-Based Continuous Sign Language Recognition Using Statistical Methods. Here they are using Hidden Markov Model for recognition using a single video camera. For a lexicon of 97 signs without a language model this system achieves an accuracy of 91.8%. But this system is user dependent and user has to wear cotton gloves to run this system in real time. Dharani Mazumdar [135, 136] suggested a method of Gloved and Free Hand Tracking based Hand Gesture Recognition. They show the result of the difference between gloves based system and free hand system. Glove based system is illumination independent, background independent, but free hand tracking suffers these problems. In this work they highlight the importance of glove based system for real time computation in any environment. But the main problem is glove based system is very expensive and not feasible for everyone. Jung-Bae Kim proposed a Fuzzy Logic and Hidden Markov Model based system for recognition of Korean Sign Language [137]. Using these methods, they recognize 15 KSL sentences and Obtain 94% recognition ratio. Using fuzzy partitioning and state automata, they reject unintentional gesture motions such as preparatory motion and meaningless movement between sign words. They concentrate on two features like speed and change of speed for hand motion. Here the author has done a sentence based recognition so no need to pause between sign words. This system has the high computational burden with illumination dependency. Sunita Nayak et al. [138] proposed an unsupervised approach of extracting any sign from continuous sentence, these signs are the part of the sentence. Here recurrent pattern of gestures has been estimated using relational distribution. This method is used for extracting frames containing some information from a complete video sequence. Then, iterative conditional modes (ICM) are used for finding the start point and width of a particular gesture. Monte Carlo method is used for calculating the probability of an unknown sign using conditional probability density measurement. All the experiments are performed on ASL dataset. Another study on continuous sign language has been proposed by W.W. Kong et al. [139] where sign language sentences are recognized using a probabilistic approach. This procedure is performed in two parts using conditional random field. Firstly the sentences are breaking into a word, then the output coming from these words using support vector machine and conditional random field is fused using Bayesian networks. Here the movement epenthesis problem is solved by proposing semi Markov CRF decoding scheme and get 96% accuracy with known signers and 87% accuracy with unknown signers. Another work which handles, hand segmentation and the movement epenthesis problem of continuous sign language is proposed by Ruiduo Yang et al. [140, 141]. These two problems has been solved by proposing enhanced level building algorithm. The performance of proposed algorithm has been compared with conditional random field and latent dynamic CRF based approach.

## 2.5 Related Research on Speech based Human Robot Interaction

Another medium which has been frequently used for human robot interaction is speech. It is a most common medium used by every person for communication establishment. Lots of work has already been done in this field here we have discussed a few of them. Y. Lee et al. [142] tried to compare different speech features in terms of the class-relevant and class-irrelevant information based on Shannon Algorithm. The features that have been compared are melscaled FFT, cepstrum, Mel-cepstrum, and wavelet features. The experiment has been done on the TIMIT corpus and the best feature of speech to uniquely identify is given as Mel-Scaled FFT. Mel frequency based discrete wavelet transform has been used by the Z. Tufekci et al. [143, 144]. First the speech signals are Mel scaled and then the DWT applies to the log filter bank to compute the coefficients. The system is tested against the noise and simple speech signals without noise. Sub band based features are similar to the mel scaled Discrete wavelet transform. A best basis algorithm has been applied by the C.J Long et al. [145]. They try to represent the speech signal efficiently before the classification. The novel and unique concept of super wavelet is represented. Coding is done using the Adaptive wavelets and best basis algorithm for transforming the on the signal. Neural net classifier is applied to sorting and the input to the neural classifier is the wavelet coefficients. Laszlo Toth et al. [146] shows why HMM gives good recognition accuracy with speech dataset. Because HMM is based on the Naïve bays and it calculates the condition probability with incorrect bias. There are two main parts of any speech recognition system one is outlier detection and another one is phoneme classification. These two parts are perfectly handled by the Naïve bays. As the size of the frame increases the probability also increases, which gives an indication towards small segment. Here the probability of an unknown speech is calculated using Bays rule, but in a different manner and achieved a good amount of accuracy similar as HMM. A. N Kandpal et al. [147] implemented PCA and ICA (Independent Component Analysis) for voice recognition and separation of speech. He concluded with the concept that PCA extracts the features of the voice without any data loss, while after implementing ICA the recovered signal is not similar to the original one. Later on, M. A. Anusuya et al. [148, 149] developed a SR system for Kannada speech recognition. After computing the discrete wavelet transform of the speech, MFCC coefficients have been forecast, over which Principal Component analysis has been applied for identification purposes. He has also done a comparative analysis of different wavelet families with PCA applied for recognition and found that Daubechies 4, 5-level decomposition and the Discrete Meyer wavelets have given comparable performance. PCA in that approach can be replaced by any other technique like HMM to increase the accuracy of the system. Using the approach used in the paper SR system for any other language could be developed. The system could be made much more speaker independent by taking the number of utterances of number of speakers with different age, gender and accent. A lot has been done in the area of speech recognition, in the last decade. Simply not much work is there for Hindi Language in this sphere. In paper [150, 151] Robert Wielgat et al. proposed a HFCC based bird species recognition technique. Here the voices of polish bird species are recorded in very noisy environment. Than HFCC features of these voices are taken out and classified using dynamic time wrapping (DTW) classifier. Experiments are performed using two features MFCC and HFCC. The result has shown that the HFCC based recognition technique provides higher recognition rate in comparison to MFCC based technique. S. Ranjan [152] worked well for Hindi Language recognition. DWT over the signal has been applied and then he applied LPC for recognition. Again, here, the best DWT category is being tested among the five DWT families, db2, db3, db4, db5 and db8. The illustration depicts the sound performance of db8 level 5. By increasing the number of voice samples the accuracy of the system could be increased. As well as, the accuracy could be increased by using HMM for recognition purposes. The approach could be used to develop the recognition system for any other language. D. Spiliotopoulos et al. [153] developed a Hygeiorobot's dialogue manager, for the interaction purposes with the robots which do not carry keyboards and mouse with them and are very helpful where people are either unaware or less aware of computer fundamentals. This Hygeiorobot's dialogue manager has been designed for the assistance of the robot in hospitals. It is just a simulation which could be implemented in a real robot. A. A. M. Abushariah et al. [154] came up with a system that recognized the English digits using MATLAB. They used MFCC for feature extraction and HMM for the recognition. They performed their experiment in two types of environments, clean and noisy. As well as, their system gave a performance both as an isolated word recognizer and as a continuous speech recognizer in both of the environments. They obtained very poor results with noisy speech and get a sufficient amount of accuracy with isolated words having no noise. Tarun Pruthi et al. [155] in 2000, described a Hindi speech recognizer, for the Hindi digits for two male speakers. He used LPC for feature extraction and HMM for recognition purposes. The system is giving a good performance, but as it is speaker specific its performance needed to be enhanced up to some extend. The vocabulary of only 10 digits is quite small. Next Gupta [156] gave an isolated word SR for Hindi language in 2006. They also used continuous HMM for recognition and the vocabulary is again the ten Hindi digits. The system this time again is giving satisfactory performances for speaker-independent environments, only the major shortcoming is the small vocabulary of the arrangement. K. Kumar et al. [157] tried to make the system much more efficient as they increased the vocabulary upto thirty Hindi words. The scheme also demonstrates good performance for speaker independent environments. He used the Hidden Markov Toolkit to develop the system and trained the system for 30 Hindi words with the data collected from eight speakers. The system's overall performance is 94.63%. Perspective talking based human robot interaction framework has been proposed in this paper [158]. Here author used a cognitive architecture called polyscheme for interaction establishment. The author used a similar frame of reference that the astronaut used for communication. His technology has been inspired from the technology that the astronaut used. Through polyscheme a TCP-IP protocol has been used for robot communication. After executing all the project and lots of analysis the author reached an efficient solution for human robot interaction.

## 2.6 Related Research on Possibility Theory

In any real time application time complexity is the major concern because it shows the efficiency of the system. In real time scenario, it is necessary that the system responds very quickly when we give some command or performed some action. Various algorithms have already been developed to solve the time complexity issue. Here we tried to reduce the time complexity of classical Hidden Markov Model by introducing a new concept of possibility theory. The literature related to possibility theory has been discussed below.

D. Dubois and H. Prade et al. [189] have further introduced an axiomatic approach of possibility theory (combination of necessity measure and plausibility measure) to handle uncertainty. This theory measures the degree to which an event can occur means up to what range a particular event can happen. Authors [190] also compared possibility theory with probability measures and showed what the differences are. Here the authors consider an example of pens and explain the concept of probability and possibility. W is the number of pens that Rohit can use day after tomorrow where the value of W is 2, 4, 6, 7 and 8. The possibility $\Pi(W)$ shows the degree with which a pen can be used by Rohit. $p(W)$ is the probability of a pen being used by Rohit. Probability and possibility values for a number of pens are shown in Table 2.1. Table 2.1 shows that the possibility of using 2 pen is 0.6 but the probability of drawing two pen is 0.5. Again the possibility of using 8 pens is 1 but the probability of drawing 8 pens is 0. All these values of probability and possibility show that it is not necessary, when the possibility is high, then the probability is also high and vice versa.

Table 2.1 Comparison of probability and possibility dataset

| W | 2 | 4 | 6 | 7 | 8 |
|---|---|---|---|---|---|
| $\Pi(W)$ | 0.6 | 1 | 1 | 0.5 | 1 |

| p(W) | 0.5 | 0.5 | 0 | 0 | 0 |
|---|---|---|---|---|---|

In paper [188] proposed a possibility based model for handling investment problems. Here monitory and nonmonetary both types of aspects are considered. Author considers an example of construction project where the investment will be done in such a manner that the cost will be minimized and construction will be more. It is very difficult for finding the accurate knowledge about the price, material, etc. because all this information is vague in nature and also they don't have any prior information. For handling these two situations the author has applied a concept of possibility theory to deal with uncertainty in construction project.

The author [191] also explained the concept of subjective possibility theory where the Bayesian network has been used for solving uncertainty problems. The possibility distribution function is applied for calculating likelihood values. If x ϵ X then Πϵ [0, 1] represents the degree of possibility of element x.

Andrew A. Alola et al. [192] proposed an analysis of possibility theory for handling uncertainty. He also explains the difference between possibility theory and probability theory, means how possibility will differ from probability. Comparative analysis of qualitative and quantitative approach has also been done. In this paper author also explains the fields like health and the environment, sciences, IT and engineering, economics and real life problems where possibility theory can handle uncertainty more effectively than probability theory

Radja et al. [193] proposed a possibility based classifier for classifying a remotely sensed images. Here classifier is designed by combining various operations (continuative and disjunctive) of possibility theory. Possibility theory handles both uncertainty as well as imprecision using possibility and necessity functions. Functionalities and details about the possibility theory has been explained and applied for classifying the given dataset. The author measures, possibility in various ways like using Klir transformation, variable transformation and Dubois and Prade transform and finally he calculates the possibility in his own way by applying a Gaussian probability measure using Klir transformation. After that basic operations like conjunction and disjunction are used for finding the classification accuracy. Classification accuracy has been measured by calculating the maximum possible value. Author also compares the results obtained from various transformations and probabilistic classifier and found that the accuracy of proposed approach increases from 72.72 percent to 90.23 percent.

In this paper [194] the author proposed a Naive Bayesian classifier based on the possibility theory. This theory works well on continuous data set. Here two concepts of possibility theory have been applied, one is based on the probability, possibility concept and the second one is a purely possibilistic concept. In first one possibility has been calculated with the help of Gaussian distribution function and in the second one the data values have been represented in the format of possibility distribution. The second approach is much more effective in detecting an unknown illustrations in which the instances are ambiguous. In this approach possibility classifier has been designed using hybrid aspects and also using K-nearest neighbor concept. Here author describes about a possibility Bayesian rule in place of probabilistic Bayesian concept. It is generally used for their simplicity, less space (storage space) and efficiency. This approach is tested on 9 types of UCI repository dataset (wine, iris, etc.) and found that the proposed method works well on all the dataset and gives 94 percent accuracy.

The author [195] compared the views of various researchers about the conditional possibility and its drawbacks. Then she proposed her own theory to overcome all the drawbacks present in the existing literature which is defined as: Zadeh's approach: In this approach the author defines the possibility as:

$$(\forall X_1)\left(\pi_{v_1}(x_1)\right) = sup_{x_2} \epsilon\ X_2 \pi_{v_1,v_2}(x_1 x_2) \qquad (2.1)$$

Now the conditional possibility distribution [17] is defined as:

$$\Pi v1/v2(*/y2) : X1 \rightarrow [0,1] : x1 \rightarrow \Pi v1, v2(x1 y2) \qquad (2.2)$$

$$\Pi v1/v2(*/y1) : X2 \to [0,1] : x2 \to \Pi v1, v2(y1x2) \qquad (2.3)$$

$X_1$ and $X_2$ are the universe of discourse, $v_1$ and $v_2$ are the variable lies within $X_1$ and $X_2$: $x_1$ and $x_2 \in X_1$ and $X_2$.

Hisdal's approach: She used the axiomatic approach as well as a probability, possibility concept for defining conditional possibility. Here conditional possibility distribution has been calculated using minimum function in place of conditional probability distribution where product is used.

$$\forall x_1, x_2 \in X_1 \times X_2 \left(\pi_{\vartheta_1 \vartheta_2}(x_1 x_2)\right) = \min(\pi_{\vartheta_1/\vartheta_2}(x_1/x_2), \pi_{\vartheta_2}(x_2)) \qquad (2.4)$$

Here the operation min is similar to the operation product in the probability theory. This technique overcomes the problems present in the Zadeh's approach. The Hisdal's approach also has some drawbacks known as Hisdal's problem which is solved in Coolman's approach. In her proposed approach fuzzy type 1 and type 2 function is used for solving the problems presented in Hisdal's and Sugano's approach.

From existing literature we have found that the possibility theory is much more effective than the probability theory when data is vague or ambiguous. Therefore, we have replaced the probability in Hidden Markov Model (HMM) [196] with the possibility theory. The main objective of replacing the probability theory with the possibility theory is to reduce the time complexity as well as space complexity with a good amount of accuracy. HMM is a stochastic process depends on Markov theory. It has many application areas like speech recognition [196], speaker identification, fingerprint identification, gesture recognition, etc.

This theory is very effective in the production of intermediate states. Let we have a video consisting of a sequence of frames $I_1, I_2, I_3$..... In this sequence of frames every next frame will depend on previous frame, i.e. $I_3$ will depend on $I_2$ $I_2$ will depend on $I_1$ etc. With the help of HMM algorithm, we find out the probability of being in state $I_{n+1}$ when the observation sequence is $O_1, O_2, O_3$..... But this algorithm fails to predict the next state when intermediate frames are very vague.

Therefore, we used a possibility theory to solve all three fundamental problems (Evaluation, Learning, and Decoding) of HMM.

## 2.7     Related Research on Multimodal Human Robot Interaction

From past literature, it has been observed that any single way of communication is not sufficient for a good interaction framework. Therefore the researcher moves towards another approach called multimodal. In this section the literature related to multimodal interaction and their fusion strategies has been discussed. V. Y. Budkov [159] proposed a multimodal framework for human computer interaction and it is extended to human robot interaction. Emotions, speech and gesture are used as a modality of interaction which are also helpful for handicapped individuals. Four ways manual entries, dialogue, gesture and emotions are employed for making interaction between them. All the experiments are done on mobile robot equipped with loud speakers, vision sensors and wheels. Mobile robot has also the capability to move from one place to another place and a capability to establish a communication through dialogue. The author Rainer Stiefelhagen et al. [160] proposed an audiovisual based human robot interaction framework in which head moment, speech and gesture are used as a mode of interaction. Each mode has been handled individually like face tracking, speech recognition, gesture recognition etc. After that all the features extracted from each of the modes are fused and get a combined result.  Speech and gesture are the most prominent mode of interaction because these two are the very appealing in nature. Color based segmentation, neural network, HMM, language modelling etc. are used to handle all the modes. Ultimately a decision level as well as feature level fusion is employed to make interaction more effective and appropriate so that is well accepted by the normal people who are not reckoned to be robot specialist. Javi F. Gorostiza et al. [161] proposed a Maggie mobile robot based interaction system in which tactile, audio, visual and remote voice mode are used. A supervised learning system is developed for establishing peer to peer communication between human and machine. The author used a hybrid control architecture called AD (automatic deliberative) for this multimodal framework. Maggie has a various inbuilt facilities like voice recognition, text conversion, face detection, which are directly used in this paper and achieved a satisfactory result. In this paper Loic Kessous [162] proposed a multimodal technique based on Bayesian classifier. Information obtained from body movements, facial expressions, gestures and speech are combined using Bayesian classifier. In this technique first, classifiers are trained for each mode individually. After that, all the classified data are fused at feature level and decision level. Finally, confusion matrix is calculated for each mode of interaction. Using these confusion matrix performances of multimodal technique is compared with the unimodal and bimodal techniques. All these results show that the emotions are better recognized in terms of confusion matrix in multimodal technique in comparison to other two techniques. Brice Burger's et al. [163, 164, 165, 166] done his thesis on multimodal command based interaction between human and machine. Different body parts like head, hand and leg are used for giving instructions to machines. A machine first tracks these body parts and try to extract the meaning coming from it. Some time machine fails or some time machine identifies correctly, these recognition rates depend on the environmental conditions. For increasing the recognition rate the author mixed speech modality with these body functions. After recognizing the commands the robot performed those commands like "Give him a glass" etc. Another work is done by Christophe Mollaret's [167] in 2011 to 2015 where he developed an interactive system so that we include robots in our day to day life. First part of his work on visual tracking where he focuses on head poses and shoulder moments. For tracking these moments Particle Swarm optimization algorithm is applied. The RGB - D method is used for measuring the distance between the user and the robot because distance is the major issue in any interactive system. Further, this visual tracking system is fused with the speech mode, which is performed with the collaboration with the RAP research team. Carlos A. Cifuentes et al. [168] proposed a gait based multimodal human robot interaction framework which is helpful for the multitudes who are unable to walk or need service. Gait data have been collected using laser range finders and inertial measurement unit. Later that these data are combined using data fusion technique and bring forth an optimal dataset. This human walker system is based on the three parameters, angular velocity, linear velocity and orientation. These three parameters are computed for both human and walker and then set the interaction parameters which combines both human and walker. This system is implemented using weighted Fourier linear combiner and get a satisfactory result and also helpful to the society.

There are diverse ways through which a merger has been performed like data level fusion, feature level fusion and decision level fusion. These fusion strategies have been performed at different level of any recognition system such as data level fusion is done at the beginning of the system, feature fusion is done after the feature extraction process. At this level two different features have been mixed. And the final one is the decision level fusion, it is performed after the classification of gesture or words or any of the medium. Hartwig Holzapfel et al. [169] proposed a fusion strategy for multimodal application for Kitchen scenario. Speech is used as a main medium of interaction with respect to the gesture. Gesture is used for avoiding the ambiguities occurred with voice communication mode. A stereo vision is used for collecting the 3D point gestures and parsing is used for speech processing. Fusion is performed on the basis of time correlation. The performance of the system has been evaluated on the basis of time delay and achieved a maximum of 50 millisecond delay.

Arun Ross et al. [170] discussed a 3 types of feature level fusion strategy. The feature vector obtained using PCA and LDA of the face are fused. In next approach the LDA coefficients of RGB image of a face are fused at the feature level. Lastly the fusion is done by mixing the features of hand and face images extracted by applying PCA and LDA. In all the scenario author observed that the performance of LDA is much better than other classifier. In case of weak classifier the match score fusion strategy performs better than the feature level fusion. A feature level fusion strategy has been discussed for palm print identification [171]. Here the Gabor filter is used for calculating the phase and magnitude of the image. Further, these two images are fused using proposed fusion code and compare it with existing fusion code and found similar results with existing approaches. A novel algorithm for feature level and decision level fusion methodology have been proposed by A.H. Gunatilaka et al. [172]. A data obtained from multiple sensors is fused using priori probability or class conditional probability (Bayes rule). This fusion strategy is done at the feature level. Also, he has proposed an optimal decision level fusion strategy in which a performance of the system is measured by minimizing the Bays risk factor. Soft decision and hard decision level fusion rules have also been defined. A threshold has been defined for both EMI and IR sensor using good threshold approximation methods proposed in this paper. Finally the performance has been evaluated by generating the ROC curve. V. Chatzis et al. [173] used various types of clustering algorithm like fuzzy k-means, Median radial basis function, fuzzy vector quantization algorithm etc. These clustering methods are used for fusing the results obtained from different modes after classification. Among all Median radial basis functions performs better in comparison to other classical methods. Two methodologies have been proposed by K. Veeramachanen et al. [174] for combining two biometric classifiers at the decision level. One is optimal fusion based on the chair Varshney rule (CVR) and the likelihood ratio test and another one is a particle swarm optimization (PSO). In both the methods a threshold value for each of the biometric has been predefined and then fusion is done on behalf of these threshold values.

## 2.8 Observations from Literature Review

- Up to now no reliable testing of multimodal fusion is available.
- Very limited work has been done on Indian Sign Language gestures.
- There is no benchmark dataset available for ISL.
- Literatures that are available for Hindi speech recognition is very circumscribed and is not applicable in human robot interaction scenario.
- None of them have considered the distance issue between human and robot.
- Time complexity is the biggest challenge in real time scenario which has not been considered till now.

# Chapter 3

# Indian Sign Language: An Isolated Gesture Based Humanoid Learning using Discrete Wavelet Transform and MFCC Techniques

Gesture in the form of sign language is the most appropriate medium of communication for hearing impaired society. Usually the normal individuals do not want to learn sign language. Therefore, this community becomes isolated from others, they cannot express their self. If we program a robot in such a fashion that it can translate sign language to some words or text format, the divergence between the normal people and the deaf community can be downplayed. Every nation possesses its own sign language like American has American Sign Language, British has British sign language and so forth in the same way Indian has its own sign language which is called as an Indian sign language (ISL). ISL [175] is a best way of establishing a communication with hearing impaired society present in India. It consists of both dynamic and static hand movements. Recognizing any gesture consists of lots of challenges like segmentation, tracking, illumination, background variation etc.[209][210][211][212]. The major challenges we have solved in this chapter are:

   a) Hand segmentation of upper half of the body image.
   b) Extracting appropriate information from malformed gestures.
   c) The boundaries of each gesture may vary from one person to another.

To minimize all such challenges we have proposed a novel framework which recognizes an ISL gesture in a real time scenario. Our first segment the hand from upper half of the body image. After solving this, the next step is to find appropriate features for gesture recognition like shape, orientation, spatial temporal motion etc. We have used a combination of discrete wavelet transform (DWT) and Mel frequency Cepstral coefficient (MFCC) features extraction technique for picking out an unknown ISL gesture with the assistance of support vector machine (SVM) and K nearest neighbor (KNN) as a classifier. The combinations of these two feature extraction techniques are never been applied earlier for the ISL gesture recognition purpose and besides it is really effective against translation, scaling, orientation, background variation and light variation when gestures are executed in real time environment. After DWT, 12 MFCC coefficients of each frame are taken as a feature vector. The suggested technique is examined on the ISL dataset created in robotics and artificial intelligence laboratory, IIIT-Allahabad, India. It has likewise been tried out on Sheffield Kinect Gesture (SKIG) dataset [177]. All the experiments are executed in various background conditions with different illuminations like red, yellow etc. and we get 98 percent classification results on in-house. Comparative analysis of proposed techniques with existing techniques like MFCC, orientation histogram (OH) etc. have been done and found that the DWT with MFCC techniques provides maximum classification result than any other technique. Later in the classification process humanoid learning is performed by a HOAP-2 robot using WEBOTS robotic simulation software.

## 3.1 Proposed Framework

We have proposed a gesture recognition method where MFCC features have been extracted by processing of transforming images by wavelet descriptors. The proposed framework for ISL gesture recognition has been presented in Figure 3.1.

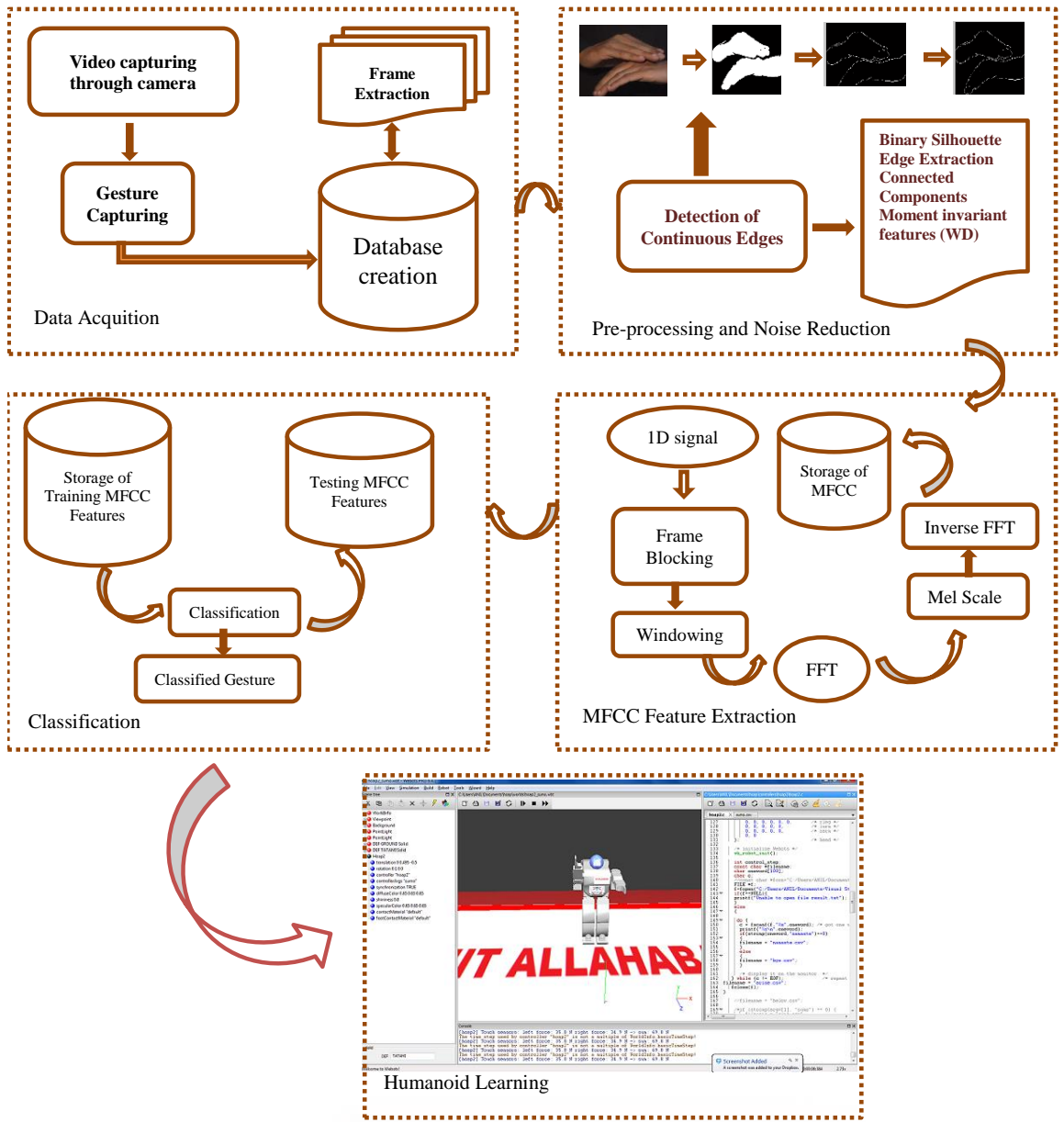

Figure 3.1 Proposed Framework

**Justification of Framework**

In Figure 3.1 we have addressed the complete framework which has been split up into 5 modules: Data Acquisition where data collection has been done through webcam and Canon EOS then each video is split up into a sequence of frames.

These figures further lead to the pre-processing phase in which hand extraction, silhouette image formation, and boundary point extraction have been done. In this module silhouette images are converted into 1D signal and then wavelet transform has been applied for finding boundary points which are moment invariant. These characteristics are further processed for finding MFCC coefficients which have been done in feature extraction module. The MFCC feature extraction technique has been widely practiced in speech processing here we explored MFCC in hand gesture recognition in combination with DWT and see the performance on image frames. This is the new type of experiment which has never been applied to ISL gesture recognition. This feature extraction technique gives a high recognition rate of speech, due to its spectral envelop quality and less time complexity due to reduction of size using DWT. Further probes are classified using SVM and KNN in the classification phase. In this module these classified gestures have been performed by a HOAP-2 robot using the WEBOTS simulation platform. For performing any gesture by HOAP-2 robot we interfaced WEBOTS software to MATLAB software by calling .wbt function. All the modules of the proposed framework are described in detailed in subsequent sections.

### 3.1.1 Database creation

Sir William Tomkins mentioned 100 signs, posture and guaranteed about them of being true Indian Signs in his book Universal Indian Sign Language [175]. We have created a database of 60 ISL static as well as dynamic gestures among the 100 signs by using simple Logitech HD 720p camera where few of them have been shown in Table 3.1 and Figure 3.2. The complete database is listed as: Counting (one, two, three, four, five, six, seven, eight, nine and ten), above, below, Above (dynamic), Across(dynamic), Aboard(dynamic), All(dynamic), All gone(dynamic), Advance(dynamic), Afraid(dynamic), Below(dynamic), Add(dynamic), After(dynamic), Arrest(static), Arrive-here(dynamic), Arrive-there(dynamic), Baby(static), Coffee(dynamic), Cold(dynamic), Color(dynamic), Come(dynamic), Depart(dynamic), Die(dynamic), Distant(dynamic), Dive(dynamic), Dog(dynamic), Effort(dynamic), End/Done(dynamic), Escape(dynamic),

Fight(dynamic), Flag(static), Give(dynamic), Grow(dynamic), Half(static), House(static), Howmany (dynamic), Keep(static), Many(dynamic), Manytimes (dynamic), Meet(dynamic), Oath(static), Opposite(static), Question(static), Quiet down(dynamic), Thick(static), Thin(static), Up(dynamic), Wait(static), Walk(dynamic), Watch(static), This database contains both dynamic gestures as well as static gestures recorded in different light conditions (yellow light, red light, white light etc.) and different background like white, black, red etc. The complete dataset is created with black full sleeves dress. We have captured half body dataset (excluded face) where portion of both the hands come. In our database we have considered both types of gestures containing single hand as well as both the hands. Also, we have done experiments on a Sheffield Kinect gesture dataset (SKIG) [29] which is shown in Figure 3.3.

Table 3.1: Database of static and dynamic gestures in different illumination condition

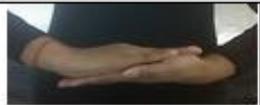

Figure 3.2 Other ISL Data Set Created in Robotics and Artificial Intelligence Laboratory

Figure 3.3 SKIG Dataset [176]

### 3.1.2 Pre-processing

Gesture videos are first divided into a sequence of frames by size (m, n) represented as $I_i$ (m, n), where i is the number of frames in each video. Hand region is extracted from the whole image frame shown in Figure 3.4.

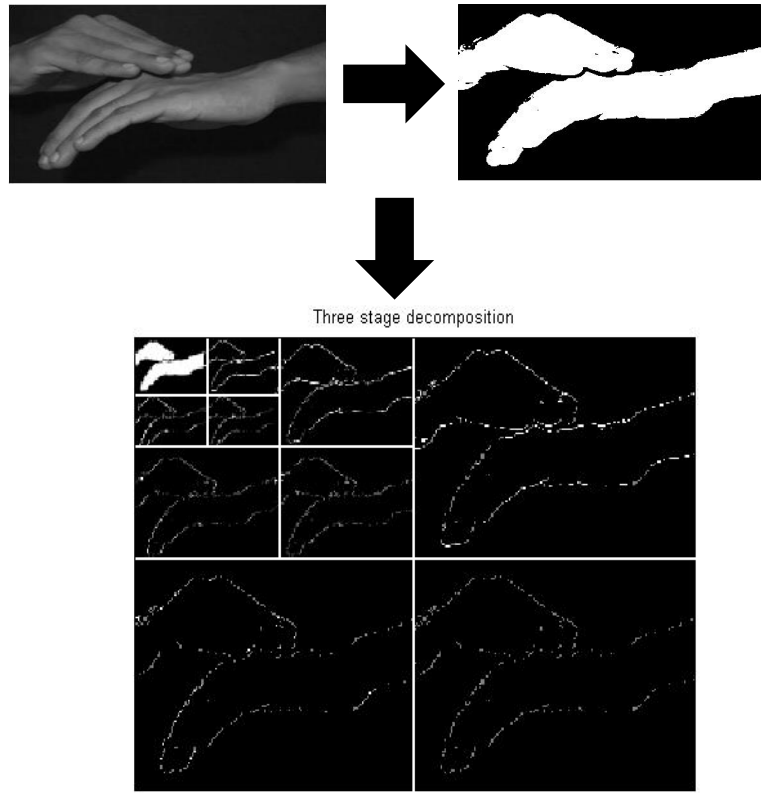

Figure 3.4 DWT decomposition up to level 3

These images are further converted into binary images using threshold T. Binarisation of an image is necessary for extracting good edges which are continuous in nature. Contour obtained are distorted at some points, therefore binarisation of an image is necessary. This binarisation of an image is performed using Otsu's thresholding method [178] where the threshold is chosen in such a manner that the intra - class variance of black and white pixel is minimized. Then after DWT is applied for reducing the dimension of an image and extracting moment invariant features which does not contain noise.

**Invariance Proof against Scale and Orientation through Discrete Wavelet Transform:** DWT [179, 180] has been applied up to the $3^{rd}$ level of decomposition of a binary image frame. The decomposition of image matrix in each level is done by row wise multiplication of images with wavelet filter followed by column wise multiplication and getting the four components of single image [LL, LH, HL and HH] where LL is the approximate coefficients which discards all high frequency coefficients and LH, HL, HH is the high frequency sub bands where LH contains horizontal components, HL contains vertical components and HH has a diagonal components of original image. The decomposition is performed as:

$$LL = D_{w_\emptyset}(j,p,q) = \frac{1}{\sqrt{PQ}} \sum_{x=0}^{P-1} \sum_{y=0}^{Q-1} f(x,y) \emptyset_{j,p,q}(x,y) \quad (3.1)$$

$$i = LH, HL, HH = D_{w_\Omega^i}(j,p,q) = \frac{1}{\sqrt{PQ}} \sum_{x=0}^{P-1} \sum_{y=0}^{Q-1} f(x,y) \Omega_{j,p,q}^i(x,y) \quad (3.2)$$

Where $\emptyset$ and $\Omega$ is a scaling and wavelet coefficients, P and Q are the dimension of image matrix, j is the level of decomposition, f(x,y) is the image matrix obtained after binarisation and p, q is the row and column of the image matrix.

Suppose Image matrix of size (4, 4)-

$$I = \begin{bmatrix} c11 & c12 & c13 & c14 \\ c21 & c22 & c23 & c24 \\ c31 & c32 & c33 & c34 \\ c41 & c42 & c43 & c44 \end{bmatrix}$$

Coefficients of Daubachies wavelet filter H is defined as:

Low pass: $\frac{1}{4\sqrt{2}}(3 - \sqrt{3},\ 3 + \sqrt{3}, 1 + \sqrt{3}, 1 - \sqrt{3})$

High pass: $\frac{1}{4\sqrt{2}}(1 - \sqrt{3},\ -1 - \sqrt{3}, 3 + \sqrt{3}, -3 + \sqrt{3})$

$$H = \begin{bmatrix} W11 & W12 & W13 & W14 \\ W21 & W22 & W23 & W24 \\ W31 & W32 & W33 & W34 \\ W41 & W42 & W43 & W44 \end{bmatrix} \quad (3.3)$$

Where W11, W12 …… are wavelet filter bank coefficients.

$$Ir = I * H$$

$$a_r^1 | d_r^1 \approx \frac{1}{4\sqrt{2}} \begin{bmatrix} r11, r12, r13, r14 \\ r21, r22, r23, r24 \\ r31, r32, r33, r34 \\ r41, r42, r43, r44 \end{bmatrix} \quad (3.4)$$

Where $1/4\sqrt{2}$ is a normalizing factor. It normalizes the approximate $a_c$ and detail $d_c$ coefficients such that

$$\| a_c \| = \| d_c \| = 1.$$

$$Ic = H' * Ir \tag{3.5}$$

$$a_c^1 | d_c^1 = \frac{1}{4\sqrt{2}} \begin{bmatrix} \begin{pmatrix} D11, D12 \\ D21, D22 \end{pmatrix} | \begin{pmatrix} D13, D14 \\ D23, D24 \end{pmatrix} \\ \overline{\phantom{xx}} \\ \begin{pmatrix} D31, \ D32 \\ D41, \ D42 \end{pmatrix} | \begin{pmatrix} D33, \ D34 \\ D43, D44 \end{pmatrix} \end{bmatrix} \tag{3.6}$$

Where Ic is an orthogonal matrix which preserves magnitude and angle of any vector is belongs to $|R^n$. Prove of these properties are given below:

---

Suppose z is a one dimensional vector of $|R^n$ and Ic is a orthogonal matrix obtained after decomposition of an image I then proof that

$$\|zIc\| = \|z\| \tag{3.7}$$
$$\|Icz\|^2 = (Icz)^T(Icz) \tag{3.8}$$
$$= z^T Ic^T Icz \tag{3.9}$$
$$= z^T Iz \tag{3.10}$$
$$= z^T z \tag{3.11}$$
$$= \|z\|^2 \tag{3.12}$$

This shows that Ic preserves magnitude of any vector z $\varepsilon$ $|R^n$ i.e. it is scale invariant.

---

Suppose z1 and z2 is two vectors of $|R^n$ and Ic is a decomposed matrix obtained after DWT decomposition of an image I then proof that the angle between z1 and z2 is always same.

Let us assume that the angle between z1 and z2 is $\phi$

$$\text{Cos } \phi = \frac{z1.z2}{\|z1\|\|z2\|} \tag{3.13}$$

If Ic is an orthogonal matrix, then angle between Icz1 and Icz2 is

$$\text{Cos } \delta = \frac{Icz1.Icz2}{\|Icz1\|\|Icz2\|} \tag{3.14}$$

$$= \frac{(Icz1)^T(Icz2)}{\|z1\|\|z2\|} \tag{3.15}$$

$$= \frac{z1^T Ic^T Icz2}{\|z1\|\|z2\|} \tag{3.16}$$

$$= \frac{z1^T z2}{\|z1\|\|z2\|} \tag{3.17}$$

$$= \frac{z1.z2}{\|z1\|\|z2\|} \quad (3.18)$$

$$= \cos \phi \quad (3.19)$$

This proves that, in DWT, the angle between two vectors are always same i.e. phase invariant.

---

These two properties of DWT show that the angle and amplitude of original image do not change when it converted into the transformed image means the image will be less distorted after transformation.

The equation 3.4 shows the final matrix obtained after 1st level decomposition which has 4 components, termed as LL, LH, HL and HH. Each component of the matrix is of size (m/2, n/2). In second level decomposition LL component is further decomposed into four parts because it contains maximum information with minimum noise similar to the original image. Whereas LH, HL and HH are high frequency coefficients having low signal to noise ratio. This process continues up to $3^{rd}$ level of decomposition and finally we get 4 coefficients FF, FV, VF and VV. After the decomposition process contour of an image is calculated by any known contour detection method. Then moment invariant features are deliberated by converting 2-D contour image (G(x, y)) into 1 dimension. For this conversion, 2-D image G(x, y) in x-y plane is converted into r-θ plane

G(r, θ) described as: x=r cos θ and y= r sin θ.

$$G_{ab} = \iint G(r,\theta) g_a(r) e^{jb\theta} r \, dr d\theta \quad (3.20)$$

Where r is the radius of the circle, θ is the orientation angle, $G_{ab}$ is the moment of hand, $g_a(r)$ is a radial basis function and a, b are constants. In case of wavelet descriptor $g_a(r)$ has been treated as a wavelet basis function and replaced with

$$\vartheta^{p,q}(r) = \frac{1}{\sqrt{p}} \vartheta\left(\frac{r-q}{p}\right) \quad (3.21)$$

p and q are the dilation and shifting parameter.

---

IIIT-A, Robotics and AI Lab    53

Now convert a 2D image into 1D form for reducing feature extraction problem and increasing performance quality. We choose cubic B-spline (Gaussian approximation) function as a mother wavelet define us:

$$\vartheta(r) = \frac{4p^{n+1}}{\sqrt{2\pi(n+1)}} \sigma_y \cos(2\pi g_0(2r-1)) \times \exp\left(-\frac{(2r-1)^2}{2\sigma_y^2(n+1)}\right) \quad (3.22)$$

Analyzing a moment in the shape of an image the values of dilation and shifting parameter p and q are chosen to be discrete expressed as:

$$p = p_0^m, \; m \; is \; an \; integer \; and \; q = nq_0p_0^m, \; n \; is \; an \; integer$$

$p_0 >1$ or $p_0<1$ and $q_0 >0$. These constraints have been considered so that $\vartheta(\frac{r-q}{p})$ covers the complete shape of gesture. Here we considered circle for representing shape of an image, whereas ($r \leq 1$). The values we choose is (p0 and q0 = 0.5). Then the wavelet basis function $\vartheta^{p,q}(r)$ has been modified as:

$$\vartheta_{m,n}(r) = 2^{\frac{m}{2}} \vartheta(2^m r - 0.5n) \quad (3.23)$$

$\vartheta_{m,n}(r)$ Defines for any orientation along the radial axis r. It is used for finding the local as well as global features of hand by varying the values of m, n. After that we define moment invariant wavelet feature vector as:

$$\|G_{m,n,b}^{wavelet}\| = \left\|\int f_b(r) \cdot \vartheta_{m,n}(r) r \, dr\right\| \quad (3.24)$$

Comparing equation 3.20 and 3.24 we get ga(r) =$\vartheta_{m,n}(r)$ and $f_b(r) = \int G(r,\theta) e^{jb\theta} d\theta$ shows the b$^{th}$ frequency feature of image $G(r,\theta)$ in r- $\theta$ plane where $0 \leq \theta \leq 2\pi$. $\|G_{m,n,b}^{wavelet}\|$ is the wavelet transform of fb(r)r. It analyses the signal in both time domain as well as frequency domain and extracts features which are locally descriptive in nature. Features shown in equation 3.24 are moment invariant for each gesture with feature vector $\|G_{m,n,b}^{wavelet}\|$. Where m=0, 1, 2, 3 and n= 0, 1…2m+1.

$\|G_{m,n,b}^{wavelet}\|$ is the generalization of moment $f_b(r)$ at mth scale level and nth sift position.

---

In WD $\|G_{m,n,b}^{wavelet}\|$ represents the moment invariant property of image I, if this image is rotated by an angle α then moment invariant property $\|G_{m,n,b}^{wavelet\ rotated}\|$ defined as:

$$G_{m,n,b}^{wavelet\ rotated} = G_{m,n,b}^{wavelet}\ e^{jb\alpha}$$

Since

$$\|G_{m,n,b}^{wavelet\ rotated}\| = \sqrt{(\|G_{m,n,b}^{wavelet\ rotated}\|)(\|G_{m,n,b}^{wavelet\ rotated}\|)^*} =$$

$$\sqrt{(\|G_{m,n,b}^{wavelet}\ e^{jb\alpha}\|)(\|G_{m,n,b}^{wavelet}\ e^{-jb\alpha}\|)} = \|G_{m,n,b}^{wavelet}\| \qquad (3.25)$$

shows the moment invariant properties of DWT.

### 3.1.3 MFCC feature extraction:

MFCC coefficients of each image has been calculated. Each image behaves like a signal. On this signal, the MFCC feature extraction technique is applied for calculating the spectral envelope of each frame. The spectral envelop property of MFCC feature [181] provides robustness towards background noise as here in the case of getting false edges of hands. Spectral envelope is a property of an image which gives the knowledge about image intensity over frequency by providing the smooth curve or regular curve having no discontinuity in the frequency domain. It is a curve in the frequency, amplitude plane, which tightly connects all the points of the magnitude spectrum, linking the peaks. These peaks represent the highest intensity value of the image, carries maximum information. It derived from a Fourier magnitude spectrum.

With the help of the spectral envelope graph which is shown in Figure 3.9 we see the nature of the signal, how tightly the curve will cover the signal. The steps of MFCC features and spectral envelop calculation are:

(a) In noisy background condition, there is a large variation in consecutive pixel values. Thus, by assuming it statistically stationary in a short period of time, the column vector block broken up into small sections called frames which is having N samples/frame. These frames are separated by M samples (N>M) therefore overlapping is done with N-M samples.

(b) For removing the side ripples (spectral distortion) and maintaining continuity in the signal windowing is performed using hamming window function. It provides lower side lobe and narrower main lobe in the image frame. Which is expressed as:

$$\text{hm}(n) = 0.54 - 0.46 \cos\left(\frac{2\pi n}{N-1}\right), \quad 0 \leq n \leq N-1 \quad (3.26)$$

The resultant signal y(n) is defined as:

$$y(n) = x(n) * \text{hm}(n). \quad (3.27)$$

Where hm (n) is the window signal, x (n) is the wavelet coefficients.

(c) After that Fast Fourier Transform (FFT) is applied to resultant signal y(n) which converts time domain image frame into the frequency domain image frame represented as $S_k$.

$$S_k = \sum_{n=0}^{N-1} y(n) * e^{-\frac{j2\pi kn}{N}} \quad (3.28)$$

Where k=0, 1, 2……N-1.

(d) Periodogram based power spectral density ($P_i(k)$) is calculated by taking absolute value of complex Fourier transform and square the result as.

$$P_i(k) = \frac{1}{N}|S_i(k)|^2 \quad (3.29)$$

Where positive frequencies $0 \leqslant f < fs/2$ correspond to values $0 \leqslant n \leqslant N/2 - 1$, while negative frequencies $-fs/2 < f < 0$ correspond to values $N/2 + 1 \leqslant n \leqslant N - 1$.

(e) The irrelevant frequencies are mixed up with very closely spaced relevant frequencies and this effect is more prominent with increased frequency. Thus to get a clear idea about exact energy amplitude at various frequencies, power spectral density (PSD) coefficients are binned and correlated with each filter from a Mel filter bank. Mel is a nonlinear scale represented as-

$$\text{Mel Frequency} = 2595 * \log\left(1 + \frac{fs}{700}\right) \quad (3.30)$$

Where fs is the frequency range used for generating filter bank is designed in next step.

(f) This Mel scale has been used by filter banks as their Centre frequency follows this Mel frequency and thus the filters near the lower frequencies are having the narrow bandwidth and as the frequency increases the filters width increases.

$$H(k, b) = \begin{cases} 0, & \text{if } f(k) < fc(b-1) \\ f(k) - \dfrac{fc(b-1)}{fc(b) - fc(b-1)}, & \\ & \text{if } fc(b-1) \leq f(k) < fc(b) \\ f(k) - fc(b+1))/fc(b) - fc(b-1), & \\ & \text{if } fc(b) \leq f(k) < fc(b+1) \\ 0 & \\ & \text{if } f(k) < fc(b-1) \end{cases} \quad (3.31)$$

Where fc (b) is the central frequency of the filter, b is the number of filters used in filter bank and the f (k) is the Mel frequency.

(g) Find the log energy output of each of the Mel frequencies.

$$S(b) = \sum_{k=0}^{N-1} H(k, b) * P \quad (3.32)$$

where $b = 1, 2, \ldots m$ and m is the number of filter and P is PSD matrix

(h) Coefficients $me_1$, $me_2$…..are generated by applying Discrete Cosine Transform (DCT) on theses Mel frequencies.

$$me = DCT\,(s(b)) \quad (3.33)$$

Where me is the number of Cepstral coefficients. These coefficients are saved as a feature vector shown in Figure 3.5, 3.6.

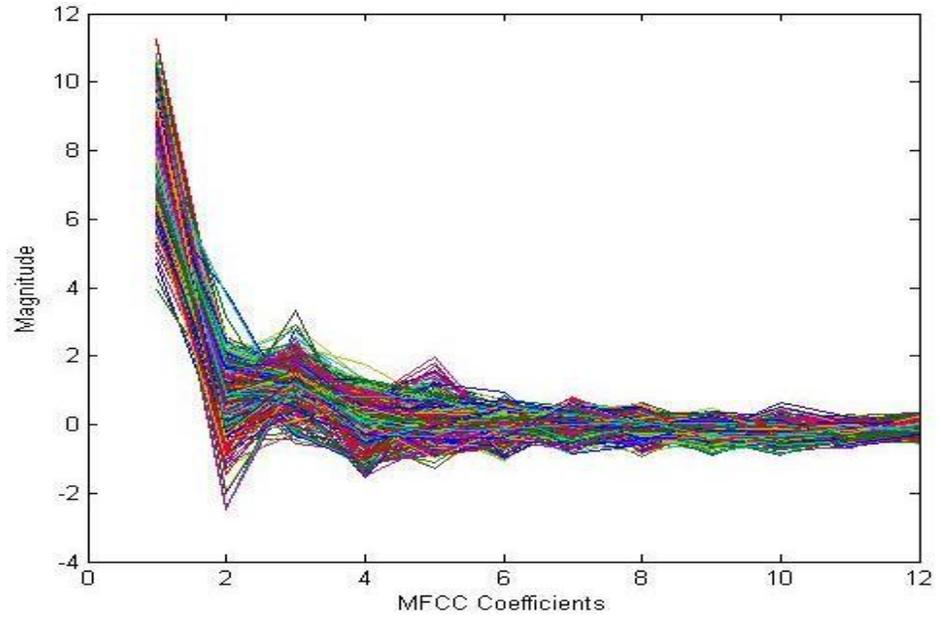

Figure 3.5 MFCC plot for gesture above

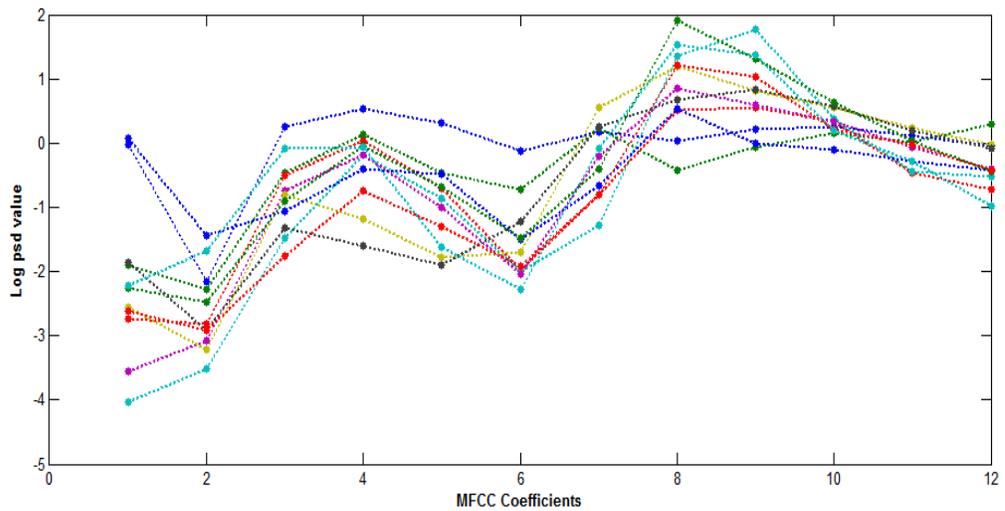

Figure 3.6 MFCC with DWT plot for gesture above

Graph shown in Figure 3.5 and 3.6 represents the noise level of MFCC and the proposed DWT with MFCC technique. Here X axis represent the number of MFCC coefficients and the Y axis represents the log magnitude value corresponding to that coefficients. After comparing both the graphs we have seen that the proposed technique eliminates maximum number of noise in comparison to single MFCC technique. The graph of proposed DWT with MFCC technique depicts the clear representation of all the 11 frames. Both the graphs have been plotted for each blocked frame of one image corresponding to their mentioned gesture. 12 MFCC coefficients have been taken for each image frame. Here the single image of one gesture is blocked into 11 frames and the graph shows MFCC plot for those 11 frames. The spectral envelope in MFCC features extraction technique is expressed as:

The spectral envelope of any signal has been calculated by multiplying the windowing function with the Cepstral coefficients obtained in the last step of the MFCC feature extraction technique.

1- Number of bins present in any spectral envelope of the signal has been taken by dividing the range of frequency $f_t$ up to Nyquist frequency $f_s/2$. Where $f_s$ is the sampling rate. Here we divide the frequency into equal parts up to $f_s/2$.

$$f_t = t\frac{f_s/2}{n}, \quad t=1\ldots\ldots n \qquad (3.34)$$

2- Calculate the angular frequency

$$w_t = f_t \frac{2\pi}{f_s} \quad \text{where } w_t \text{ is the angular frequency.} \qquad (3.35)$$

3- Finally, we calculate the spectral envelop $\vartheta_t$ of frequency $f_t$ as:

$$\vartheta t = \exp(\sum_{i=1}^{n} me_i * \cos iw_t) \qquad (3.36)$$

Where $me_i$ is the cepstral coefficient.

The spectral envelope of above gesture is shown in Figure 3.7 and 3.8. From this graph we see that number of bins of frequency in which the spectral envelope lies is 10 and the solid line shows the spectral envelope of the signal which is curved generated by combining the peaks of the signal.

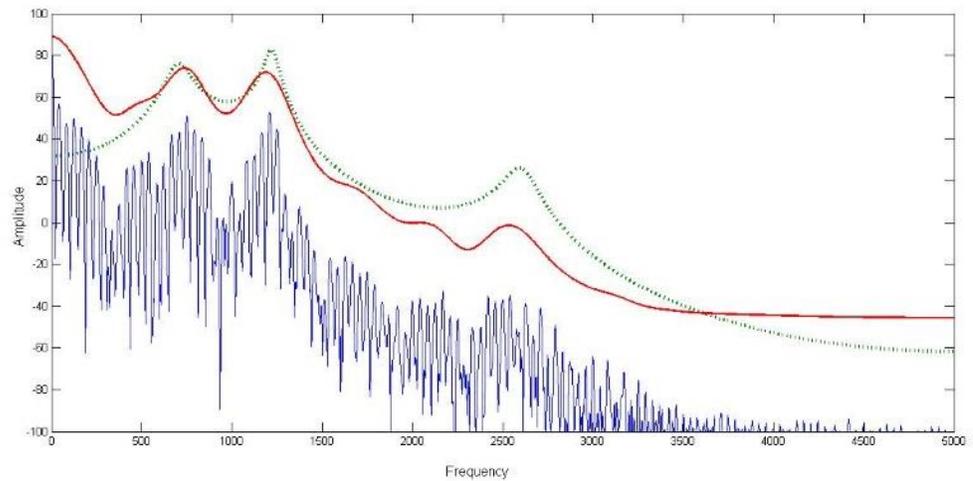

Figure 3.7 spectral envelope of MFCC features

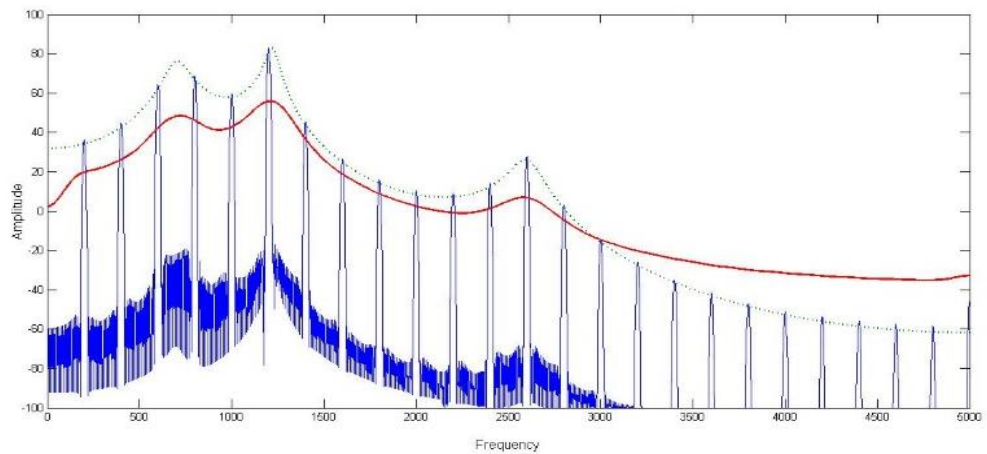

Figure 3.8 spectral envelope of DWT with MFCC features

From Figure 3.7 and 3.8 we see the nature of the curve of spectral envelope and found that in case of simple MFCC the curve is not tightly bound. When the signal becomes invisible it becomes a straight line, but in case of DWT with MFCC the spectral envelope curve is more tightly bound to contain peaks of the signal and we properly discriminate the changes in the signal.

### 3.1.4 Classification:

MFCC coefficients of various gestures are classified using K nearest neighbor (KNN) [182] and support vector machine (SVM) [183].

**3.1.4.1 K Nearest Neighbors (KNN):** It is a simple but elegant classifier used for classification of an unknown gesture. It is based on nearest neighbor method presents nearest to the unknown gesture shown in Figure 3.9. In KNN we initially assume the value of k. The distance between testing and training vectors of gestures using Euclidean distance has been measured. Then select k closest vectors whose Euclidean distance are minimized. On behalf of majority voting the class label has been assigned. Here proposed algorithm is tested on k=1, 3, 5 and 7.

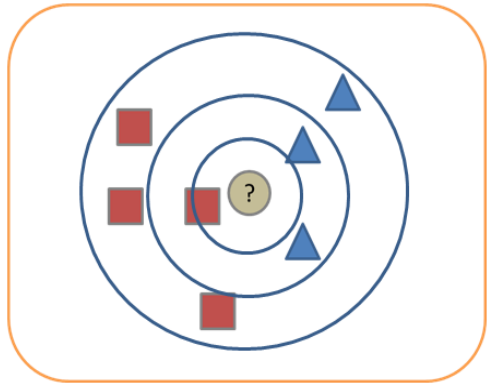

Figure 3.9 KNN classification example for k=1, 3, 5 or 7

In Figure 3.9 we have seen that for k=1, unlabeled data assigned to square class, for k=3, unlabeled data assigned to triangle class and k=7, unlabeled data assigned to square class.

**3.1.4.2 Support Vector Machine (SVM):** It is a classifier, handles nonlinearity present in features by transforming it into higher dimension. Here we use the linear kernel function for discriminating different classes because linear kernel avoids over fitting problem.

$$y_i\ (w.tr_i - k) \geq 0 \qquad (3.37)$$

$tr_i$ is the training class where i represents the class number, . is the dot product, w is the normal vector $\|w\| = 1$, $y_i$ is the class label for two class problem the value of

$y_i$ is 1 and -1. k is the constant. Constant $\frac{k}{||w||}$ sets the width of the hyper plane between different classes. If k=2 then there will be two classes. For multiclass problem SVM uses multiple binary classifiers. Let we have a t classes, then generate t binary classifiers $f_1, f_2\ldots f_t$. Each classifier is trained with one class from the rest of the class. Combine all the binary classifiers to get a classification for multiclass SVM expressed as:

$$argmax_{i=1\ldots t} F^i(x) \text{ where } F^i(x) = \sum_{n=1}^{m} y_n \, \alpha^i_n K(te, tr_n) + k^n \quad (3.38)$$

$\alpha^i_n$ is the lagrangian multiplier, K is the linear kernel function defined in equation 3.38.

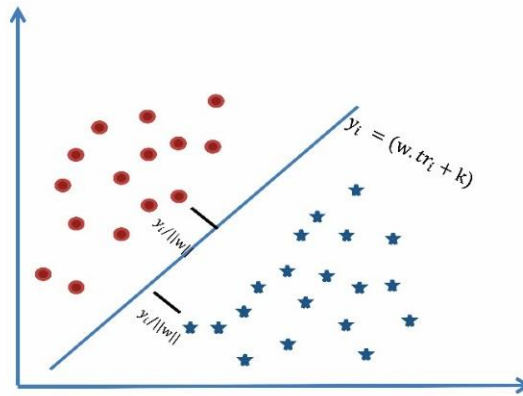

Figure 3.10 SVM classification

For classifying an unknown test data te within t classes then $\frac{t.(t-1)}{2}$ binary classifiers have been applied and count the number of times the test data te belongs to a particular class. Those classes having maximum number of class labels, the test data te belong to that class. The binary SVM classification is shown in Figure 3.10. Here we use "multisvm" function of MATLAB for classifying a test vector of an unknown gesture w.r.t. known gesture (one verses all strategy).

### 3.1.5 Humanoid learning:

After the classification process humanoid learning is performed by HOAP-2 robot in WEBOTS robotic simulation platform. It has various types of robots like social robots, industrial robots etc. This software platform has been used for establishing an interaction between human and robot. Gestures are used for establishing interaction which a human performs and then all these gestures are performed by the humanoid robot HOAP- 2. HOAP-2 is a humanoid open architecture platform [184] having 25 degrees of freedom (DOF) (6 DOF on foot*2, 4 DOF on arm*2, 1 DOF on waist, 1DOF on hand*2, 2 DOF on neck). A humanoid HOAP-2 robot joint description has been shown in Figure 3.11. We have used this robot for experimenting with multimodal interactions. Gesture requires sequential movements of joints for which we need the joint angle data of each joint of the human hand. When the user performs the gesture, the hand of a person moves horizontally as well as vertically, or ups or down. At each instant of time the joint angle value changes with respect to initial coordinate frame. These joint angle values have been captured using atan2 function. Smoothing of these joint angle values have been done for the smooth motion of HOAP-2 joints. Finally, these values are passed to the .csv file which are uploaded in the robot controller program. Here we have linked WEBOTS software to the MATLAB where gestures are classified. The range of each joint of HOAP-2 is shown in Table 3.2.

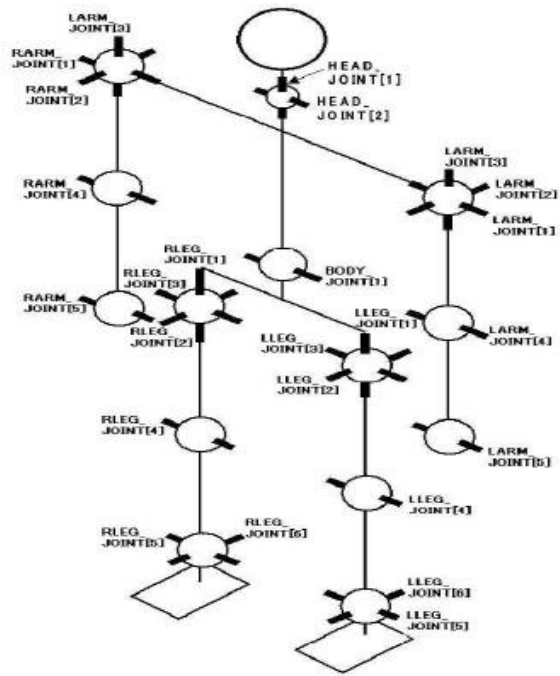

Figure 3.11 Joint names of HOAP-2 robot [184]

Table 3.2 Allowable range of joint angle for Humanoid Robot HOAP-2

| Joint Name | Min Angle (degree) | Min counter (decimal) | Max Angle (degree) | Max counter (decimal) | CSV columns |
|---|---|---|---|---|---|
| RLEG_Joint[1] | -91 | -19019 | 31 | 6479 | C |
| RLEG_Joint[2] | -31 | -6479 | 21 | 4389 | D |
| RLEG_Joint[3] | -82 | 17138 | 71 | -14839 | E |
| RLEG_Joint[4] | -1 | -209 | 130 | 27170 | F |
| RLEG_Joint[5] | -61 | -12749 | 61 | 12749 | G |
| RLEG_Joint[6] | -25 | 5225 | 25 | -5225 | H |
| RARM_Joint[1] | -91 | -19019 | 151 | 31559 | I |
| RARM_Joint[2] | -96 | -20064 | 1 | 209 | J |
| RARM_Joint[3] | -91 | 19019 | 91 | -19019 | K |
| RARM_Joint[4] | -115 | 24035 | 1 | -209 | L |
| LLEG_Joint[1] | -31 | -6479 | 91 | 19019 | M |
| LLEG_Joint[2] | -21 | -4389 | 31 | 6479 | N |
| RLEG_Joint[3] | -82 | -17138 | 71 | 14839 | O |
| LLEG_Joint[4] | -1 | 209 | 130 | -27170 | P |
| LLEG_Joint[5] | -61 | 12749 | 61 | -12749 | Q |
| LLEG_Joint[6] | -25 | 5225 | 25 | -5225 | R |
| LARM_Joint[1] | -91 | 19019 | 151 | -31559 | S |
| LARM_Joint[2] | -1 | -209 | 96 | 20064 | T |
| LARM_Joint[3] | -91 | 19019 | 91 | -19019 | U |
| LARM_Joint[4] | -115 | -24035 | 1 | 209 | V |
| Body_Joint[1] | -1 | 209 | 90 | -18810 | W |
| RARM_Joint[5] | -60 | - | 60 | - | X |
| LARM_Joint[5] | -60 | - | 60 | - | Y |
| Head_Joint[1] | -60 | - | 6- | - | Z |
| Head_Joint[2] | -15 | - | 60 | - | AA |

From all the joints shown in Table 3.2 we used only 10 joints (RARM Joint 1, 2, 3, 4, 5 and LARM Joint 1,2,3,4,5) and corresponding to their column values present in .csv file for performing any gestures, all other joint values kept as constant. Position of joint of humanoid robot in a particular position is defined using joint angle values expressed as:

$$Po = A \times \varphi \qquad (3.39)$$

Where Po is the position value, $\varphi$ is the joint angle in degree and A is the change of coefficients in pulses/deg. In HOAP-2 robot architecture the movement of joint angle values are performed in pulses because each joint having its own motor. Performing 1 degree of motion, the motor of each joint requires 209 pulses. The movement of HOAP-2 has been performed in both clockwise and anticlockwise direction. So the comma separated value .CSV controller needs to know which joint moves in which direction with changing the value in +ve direction as well as

–ve direction. +ve direction represents the motor moves between 0 to 90 degree and –ve direction represents the motor moves in between 0 to -90 degree. Which has been calculated by multiplying number of degrees of motion to the pulse value.

## 3.2   Experimental Results and Analysis

Data set of 60 ISL gestures like big, bring, below etc. (37 static and 23 dynamic) are created in three light conditions and different background conditions. 5 videos of each gesture are taken as a training gesture and 3 videos are taken as a testing set where each video is having 150 frames. For training we take 5 persons dataset and for testing 7 person's dataset where 6 are males and 6 are females (all are the students of Indian Institute of Information Technology, Allahabad). The MATLAB Image Acquisition Toolbox has been used for capturing ISL video data. The toolbox is configured as: colour space:   RGB (320*240), Triggering mode: immediate, Number of triggers: 10, Frame rate: 30 frames per second.

After data acquisition and preprocessing the features of each isolated gestures have been extracted using proposed DWT with MFCC based feature extraction technique.

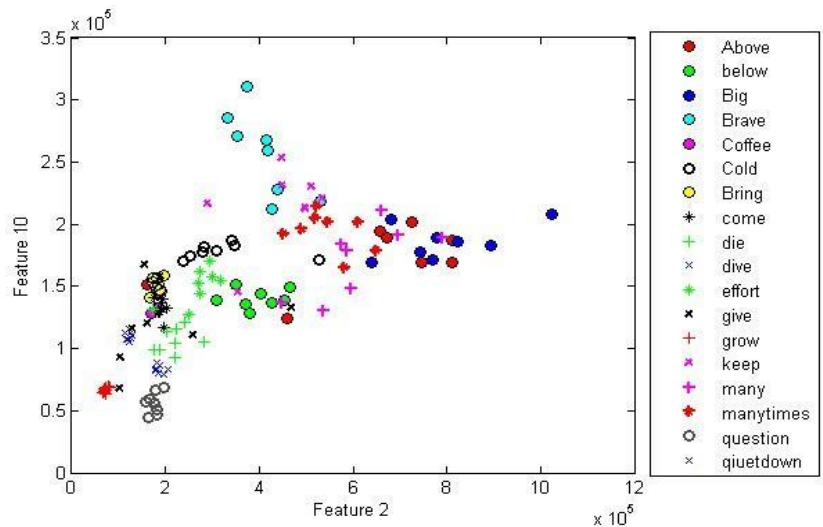

Figure 3.12 Feature space plot between feature 2 vs feature 10 MFCC features

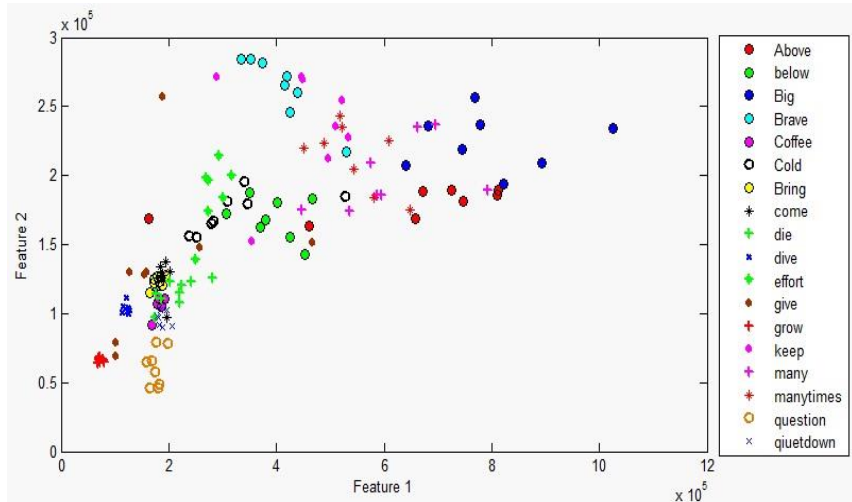

Figure 3.13 Feature space plot between feature 2 vs feature 10 DWT with MFCC features

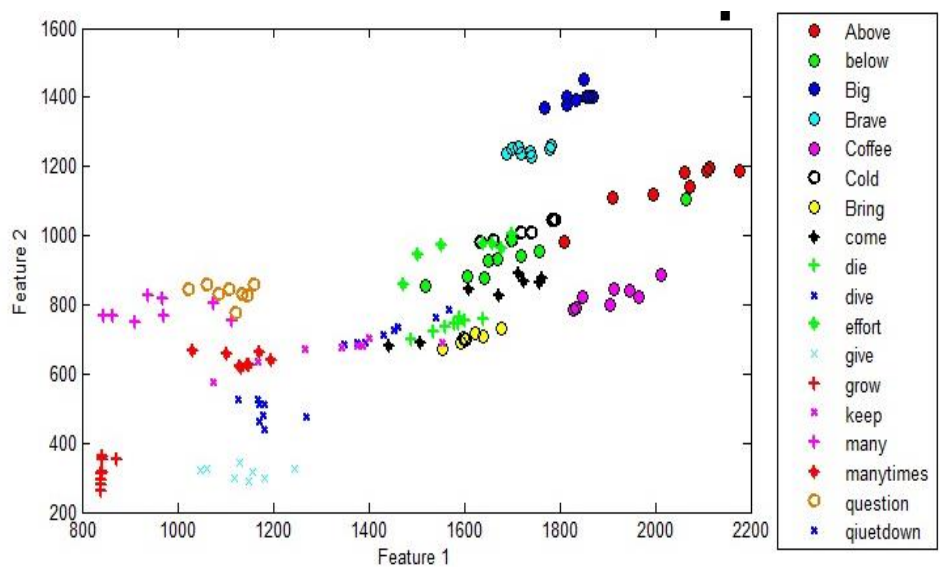

Figure 3.14 Feature space plot between feature 1 and feature 2 MFCC Features

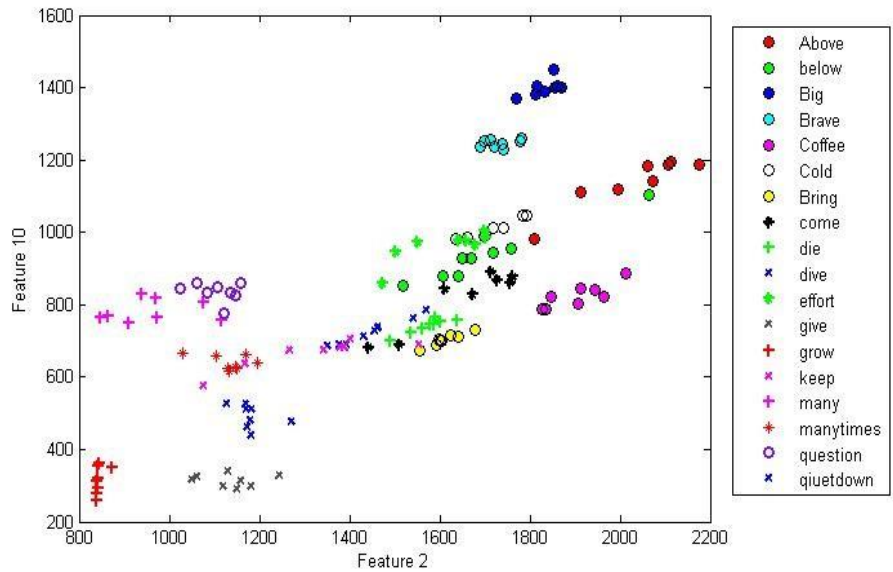

Figure 3.15 Feature space plot between feature 1 and feature 2 DWT with MFCC features

These features have been demonstrated through a feature space plot shown in Figures 3.12, 3.13, 3.14 and 3.15. We have compared two feature extraction techniques one is MFCC based and second one is DWT with MFCC based. From Figures 3.12, 3.13, 3.14 and 3.15 we have found that the proposed technique has more discriminating capability to separate all 18 classes' than simple MFCC based technique. After comparing Figures 3.14 and 3.15 we observe that the overlapping between different gesture classes is minimum in proposed approach as compared to MFCC based approach, due to the multiresolution property of DWT. To prove the discriminating capability of our proposed approach (DWT with MFCC) we have also performed two types of statistical analysis one is calculation of mean of variance of each class and second one is the between class distance. These two statistical analysis have been shown in Figures 3.17, 3.18, 3.19 and 3.20.

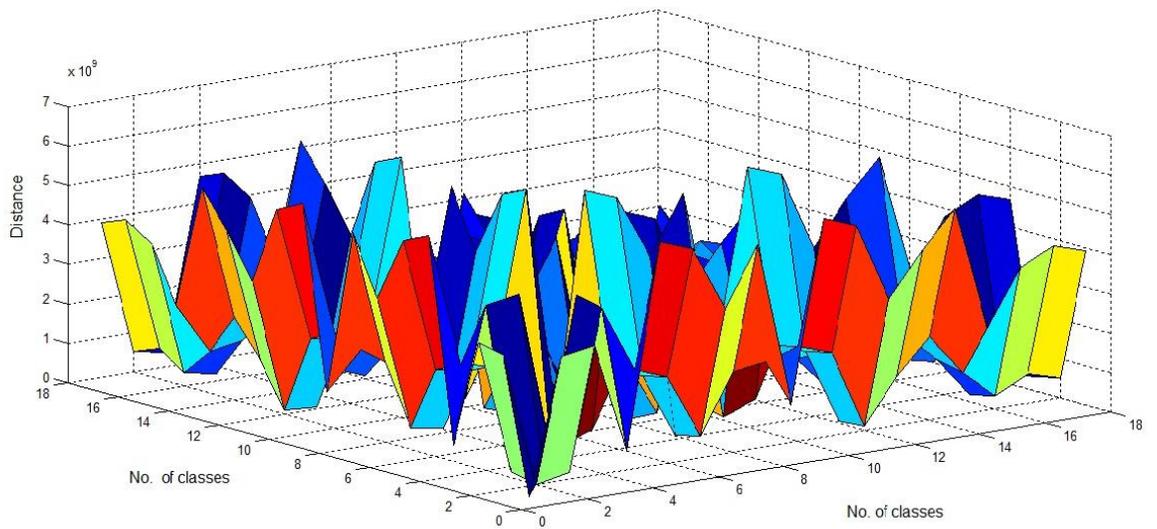

Figure 3.16 Distance plot for MFCC features

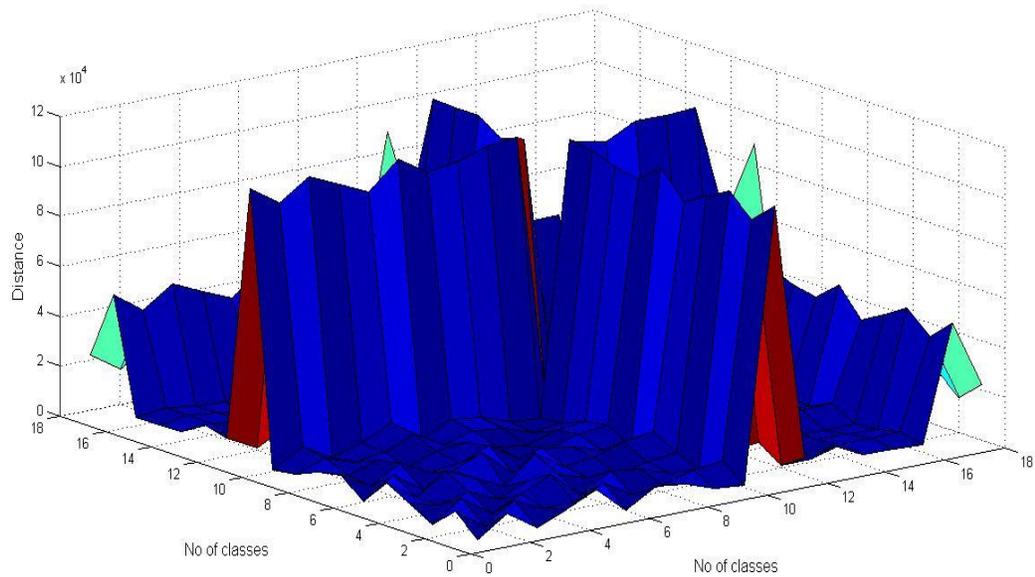

Figure 3.17 Distance plot for DWT with MFCC features

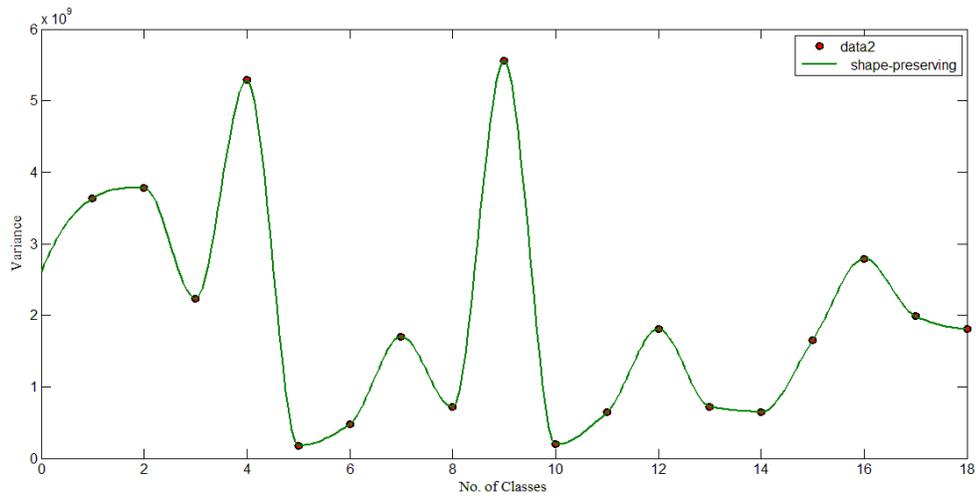

Figure 3.18 Variance plot for MFCC feature

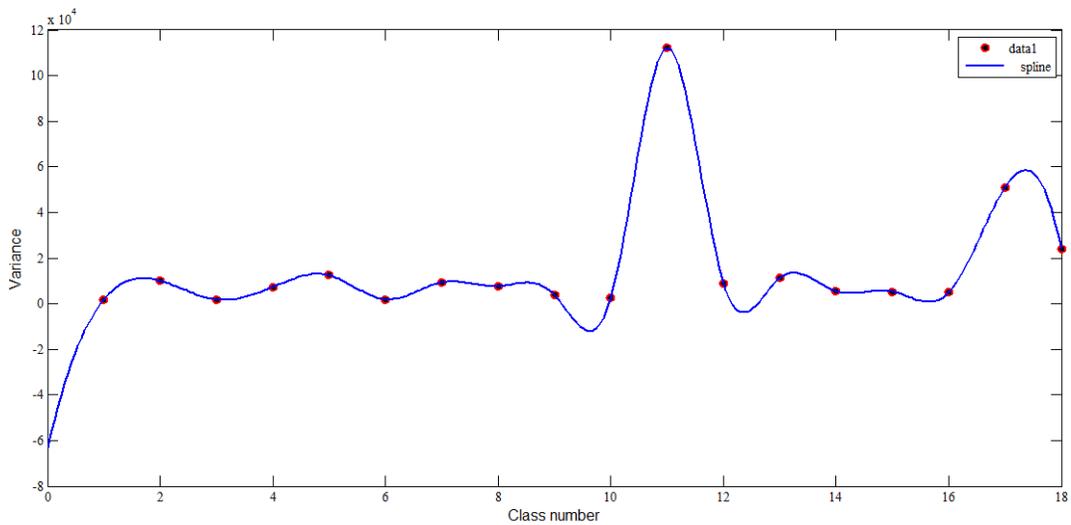

Figure 3.19 Variance plot for DWT with MFCC features

The graph shown in Figures 3.16 and 3.17 represents the between class distance plot of various gestures. After analysing these two figures we observed that the proposed approach provides greater distance between two different gesture classes as compare to MFCC based approach. Again in DWT with MFCC based approach, within class distance matric is zero whereas in MFCC based approach, within class distance matric is nonzero. This shows the discriminating nature of proposed approach. Similarly, Figure 3.18 and 3.19 signifies the variance plot of within class matric of proposed as well as MFCC based approach. After comparing Figure 3.18 and 3.19 we see that the mean of the variance of each class lies below a threshold value (2db) in case of DWT with MFCC based approach whereas in MFCC based approach, it varies very frequently for each gesture. All these statistical analysis proves that the DWT with MFCC based approach has better discriminating capability than MFCC based approach.

Percentage of classification rate and confusion matrix has been calculated for performance analysis of our proposed method where KNN and SVM are used as classifier.

Classification rate =
$$\text{No. of frames misclassified /total no. of frames)*100} \qquad (3.40)$$

In KNN, experiments have been performed at various values of k (k=1, 3, 5, 7 etc.) where Euclidean distance is used for finding the nearest neighbour of a particular class. From these experiments we obtained similar results for k=3, 5 and 7 may lead to the classification of ISL gesture. Similarly in SVM, we used the linear kernel function for classifying an unknown gesture. Here Linear kernel function has been used for avoiding the over fitting problem.

Table 3.3 Frame based Classification results for dynamic gestures using SVM and KNN.

| Class | For KNN (K=3) | | SVM(Linear Kernel) | |
|---|---|---|---|---|
| | Frame Misclassified | Accuracy (%) | Frame Misclassified | Accuracy (%) |
| Above | 60 | 86 | 12 | 97 |
| Add | 56 | 88 | 2 | 99.55 |
| Big | 102 | 77 | 10 | 97.77 |
| Below | 60 | 86 | 5 | 98.88 |
| Bring | 50 | 89 | 3 | 99 |
| Coffee | 115 | 77 | 14 | 97 |
| Colour | 45 | 90 | 12 | 97 |
| Code | 49 | 89 | 1 | 99.98 |
| Come | 50 | 89 | 1 | 99.98 |
| Die | 68 | 85 | 1 | 99.98 |
| Effort | 75 | 83 | 1 | 99.98 |
| Grow | 54 | 88 | 15 | 96.66 |
| Drive | 27 | 94 | 2 | 99.78 |
| Many | 130 | 71.45 | 33 | 92.66 |
| many times | 50 | 89 | - | 100 |
| give | 50 | 89 | 6 | 98.72 |
| Keep | 55 | 87 | 12 | 97.2 |
| quite down | 45 | 90 | 1 | 99.98 |
| Question | 69 | 84.34 | 15 | 96.66 |

From Table 3.3 we have seen that the SVM gives better classification accuracy than KNN because SVM handles non linearity presents in the dataset by using kernel function. But KNN is a simple nearest neighbourhood based classifier which is not appropriate for classifying non- linear data and also it is very sensitive against noise.

Performance of our proposed algorithm is also tested by calculating confusion matrix as shown in Table 3.4. Here 3 videos per gesture for training and 5 videos for testing have been taken for classification. Each video contains 150 frames. Confusion matrix is created by calculating True positive, true negative, false positive and false negative for each gesture classes.

True positive=50, True negative=1018, False positive=8, False negative=7, Total positive =58,

Total negative=1025, Total population=1083.

Accuracy = (true positive + true negative)/total population       (3.41)

Accuracy= (50+1018)/1083=1068/1083=98.61%

Table 3.4: Confusion matrix for MFCC with DWT based classification for dynamic gestures (SVM as classifier)

| Gesture | Above | Brave | Big | Below | Bring | Coffee | Color | Cold | Come | die | Effort | grow | Dive | Many | Many times | Give | keep | Quiet down | Question |
|---|---|---|---|---|---|---|---|---|---|---|---|---|---|---|---|---|---|---|---|
| Above | 67 | 0 | 0 | 0 | 0 | 0 | 0 | 33 | 0 | 0 | 0 | 0 | 0 | 0 | 0 | 0 | 0 | 0 | 0 |
| Brave | 0 | 100 | 0 | 0 | 0 | 0 | 0 | 0 | 0 | 0 | 0 | 0 | 0 | 0 | 0 | 0 | 0 | 0 | 0 |
| Big | 0 | 33 | 67 | 0 | 0 | 0 | 0 | 0 | 0 | 0 | 0 | 0 | 0 | 0 | 0 | 0 | 0 | 0 | 0 |
| Below | 0 | 0 | 0 | 67 | 0 | 0 | 0 | 0 | 0 | 33 | 0 | 0 | 0 | 0 | 0 | 0 | 0 | 0 | 0 |
| Bring | 0 | 0 | 0 | 0 | 67 | 0 | 33 | 0 | 0 | 0 | 0 | 0 | 0 | 0 | 0 | 0 | 0 | 0 | 0 |
| Coffee | 0 | 0 | 0 | 0 | 0 | 100 | 0 | 0 | 0 | 0 | 0 | 0 | 0 | 0 | 0 | 0 | 0 | 0 | 0 |
| Color | 0 | 0 | 0 | 0 | 0 | 67 | 33 | 0 | 0 | 0 | 0 | 0 | 0 | 0 | 0 | 0 | 0 | 0 | 0 |
| Cold | 0 | 0 | 0 | 0 | 0 | 0 | 0 | 100 | 0 | 0 | 0 | 0 | 0 | 0 | 0 | 0 | 0 | 0 | 0 |
| Come | 0 | 0 | 0 | 0 | 0 | 0 | 0 | 0 | 100 | 0 | 0 | 0 | 0 | 0 | 0 | 0 | 0 | 0 | 0 |
| Die | 0 | 0 | 0 | 0 | 0 | 0 | 0 | 0 | 0 | 100 | 0 | 0 | 0 | 0 | 0 | 0 | 0 | 0 | 0 |
| Effort | 0 | 0 | 0 | 0 | 0 | 0 | 0 | 0 | 0 | 0 | 67 | 0 | 33 | 0 | 0 | 0 | 0 | 0 | 0 |
| Grow | 0 | 0 | 0 | 0 | 0 | 0 | 0 | 0 | 0 | 0 | 0 | 100 | 0 | 0 | 0 | 0 | 0 | 0 | 0 |
| Dive | 0 | 0 | 0 | 0 | 0 | 0 | 0 | 0 | 0 | 0 | 0 | 0 | 100 | 0 | 0 | 0 | 0 | 0 | 0 |
| Many | 0 | 0 | 0 | 0 | 0 | 0 | 0 | 0 | 0 | 0 | 0 | 0 | 0 | 67 | 33 | 0 | 0 | 0 | 0 |
| Many times | 0 | 0 | 0 | 0 | 0 | 0 | 0 | 0 | 0 | 0 | 0 | 0 | 0 | 0 | 100 | 0 | 0 | 0 | 0 |
| Give | 0 | 0 | 0 | 0 | 0 | 0 | 0 | 0 | 0 | 0 | 0 | 0 | 0 | 0 | 0 | 100 | 0 | 0 | 0 |
| Keep | 0 | 0 | 0 | 0 | 0 | 0 | 0 | 0 | 0 | 0 | 0 | 0 | 0 | 0 | 0 | 0 | 100 | 0 | 0 |
| Quiet down | 0 | 0 | 0 | 0 | 0 | 0 | 0 | 0 | 0 | 0 | 0 | 0 | 0 | 0 | 0 | 0 | 0 | 100 | 0 |
| Question | 0 | 0 | 0 | 0 | 0 | 0 | 0 | 0 | 0 | 0 | 0 | 0 | 0 | 0 | 0 | 0 | 0 | 0 | 100 |

From Table 3.3 and 3.4 we see that our proposed method gives satisfactory results for all the 19 dynamic gestures. We have tested our experiments on static gestures also. For static gestures, training is performed with one video and all other videos in different light condition are used for testing. 150 frames are there in each video. Accuracy based on confusion matrix shown in Table 3.5.

True positive=7, False negative=1, False positive=1, True negative=55, Total positive =8, Total negative=144, Total population=64; Accuracy=62/64=96.875%.

Table 3.5 Confusion matrix for KNN k=3 based classification for static gestures in different illumination condition.

| Static gesture | Cross | Hide | Hold | Stable | Straight | Together | Warm | With |
|---|---|---|---|---|---|---|---|---|
| Cross | **100** | **0** | **0** | **0** | **0** | **0** | **0** | **0** |
| Hide | 0 | 100 | 0 | 0 | 0 | 0 | 0 | 0 |
| Hold | 0 | 0 | 100 | 0 | 0 | 0 | 0 | 0 |
| Stable | 0 | 0 | 0 | 100 | 0 | 0 | 0 | 0 |
| Straight | 0 | 0 | 0 | 0 | 100 | 0 | 0 | 0 |
| Together | 0 | 0 | 0 | 0 | 0 | 100 | 0 | 0 |
| Warm | 0 | 0 | 0 | 0 | 0 | 0 | 100 | 0 |
| With | 0 | 0 | 0 | 0 | 0 | 0 | 0 | 100 |

Table 3.5 shows that DWT and MFCC based method gives 100 percent classification result for all the static gestures because of no changes between the frames and its orientation. Hence all the environmental noises and movement variation problems have been easily eliminated using DWT and MFCC method. We have also observed that all the static gestures are constant w.r.t. time whereas dynamic gestures vary with respect to time.

We have also tested our algorithm on Sheffield Kinect Gesture data set (SKIG) created by L. Liu at university of Sheffield [176] [177]. These datasets are dynamic in nature, containing 10 types of actions like circle, triangle, up-down etc. Here data set is captured with 3 backgrounds (white wooden board, white plain paper, white paper with symbols), 2 light conditions (dark light, poor light) and 2 poses (clockwise and anti-clock wise). Therefore, each gesture has $3*2*2 = 12$ sample. We have tested our proposed methodology on total 360 SKIG gesture dataset which is shown in Figure 3.3. Table 3.6 shows the classification results on a SKIG dataset with different light conditions and different backgrounds. From the results obtained on SKIG dataset we observe that the performance of proposed method goes down as the light conditions varies drastically (white light to red light). We also observe that when background colour is similar to the skin colour, the performance of proposed method degrades.

Table 3.6 Classification Results for SKIG Kinect Gestures Dataset

| Class | Wooden board | | | | White plain paper | | | | Paper with characters | | | |
|---|---|---|---|---|---|---|---|---|---|---|---|---|
| | Strong light | | Poor light | | Strong light | | Poor light | | Strong light | | Poor light | |
| | Frame Misclassified | Acc. % | Frame Misclassified | Acc. % | Frame Misclassified | Acc. % | Frame Misclassified | Acc. % | Frame Misclassified | Acc. % | Frame Misclassified | Acc. % |
| Circle | 5 | 92 | 6 | 90 | 9 | 82 | 7 | 86 | 5 | 92 | 4 | 93 |
| Triangle | 22 | 56 | 4 | 92 | 10 | 80 | 6 | 90 | 7 | 90 | 5 | 92 |
| Up-down | 6 | 90 | 3 | 94 | 8 | 84 | 8 | 84 | 5 | 92 | 8 | 84 |
| Right-left | 13 | 73 | 4 | 92 | 9 | 82 | 10 | 80 | 4 | 93 | 6 | 90 |
| Wave | 5 | 92 | 5 | 92 | 11 | 80 | 8 | 84 | 3 | 94 | 4 | 93 |
| Z | 4 | 94 | 3 | 94 | 8 | 84 | 20 | 60 | 3 | 94 | 7 | 86 |

## 3.3 Performance Comparison

Comparative analysis of various feature extraction methods like (MFCC, Orientation histogram (OH) and DWT with MFCC (proposed method) have been performed on the basis of classification rate which has been shown in Figure 3.20 which clearly shows that DWT with MFCC out performed others. Here we have performed experiments on various parameters of various methods explain in Table 3.7.

Table 3.7: Performance parameters

| Method | Training Dataset (150 Frames) | Testing Dataset (150 Frames) | Parameters | KNN | SVM |
|---|---|---|---|---|---|
| OH [26] | 5video | 3 video | 9,18,27,36 | K=1,3,5,7(Euclidean distance) | Linear Kernel |
| MFCC [25] | 5video | 3 video | 10,12,13,14,15,17 | K=1,3,5,7(Euclidean) | Linear Kernel |
| WD+MFCC (proposed method) | 5 video | 3 video | HAAR, Daubachies wavelet/ 10,12,13,14,15,17 | K=1.3.5.7euclidean | Linear Kernel |

In OH experiments have been performed on 9, 18, 27, 36 bins (no. of features), in MFCC we take 10,12,13,14, 15 and 17 cepstral coefficients for analyzing the recognition accuracy and finally we tested our proposed algorithm on HAAR wavelet, Daubachies wavelet with 10,12,13,14,15 and 17 MFCC coefficients. Here number of decomposition levels in DWT is 3.

The graph shown in Figure 3.20 represents that DWT with MFCC method provides high classification rate than other methods because we believe that the multi-resolution and movement invariant properties of DWT and spectral envelop property of MFCC helps to reduce ambiguity.

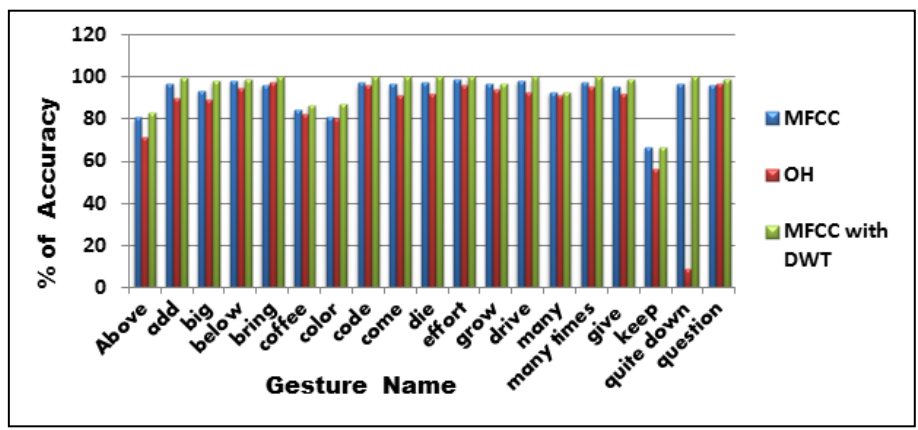

Figure 3.20 Comparative results of MFCC, OH, DWT with MFCC method

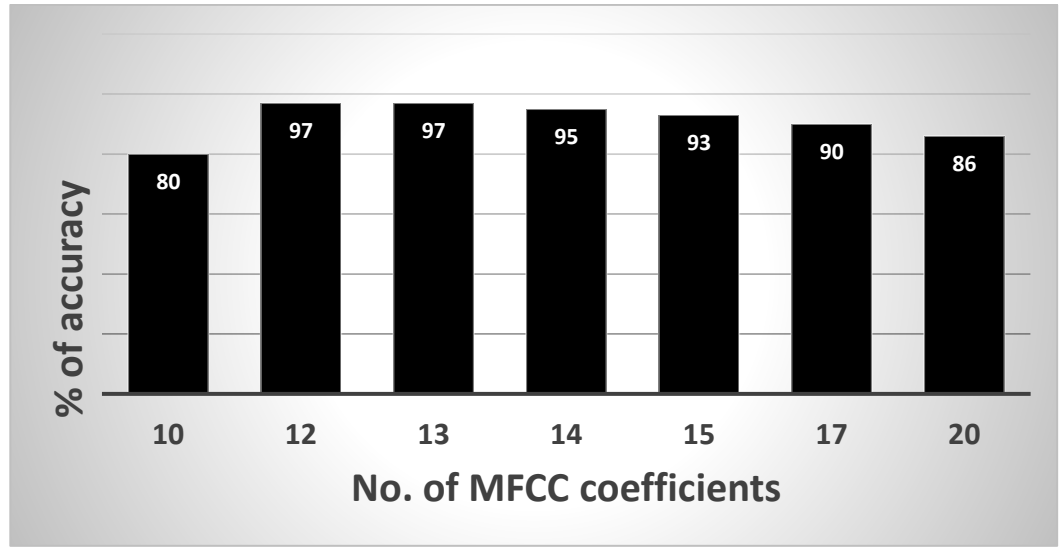

Figure 3.21 Tradeoff between accuracy and no. of MFCC Coefficients

We have also compared the classification accuracy of the proposed gesture recognition algorithm at various numbers of MFCC coefficients like 10, 12, 13…..20 which is shown in Figure 3.21. From this graph we have found that the recognition accuracy is maximum at 12 and 13 MFCC coefficients in comparison to other MFCC coefficients (10, 15, 20) when it's applied with DWT. This is because the more number of MFCC coefficients represents very frequent change in the energy which degrades the performance of the system and similarly less number of MFCC coefficients unable to present the energy change. We have also compared the results of the proposed method with the existing methods OH and MFCC. The results are shown in Figure 3.22.

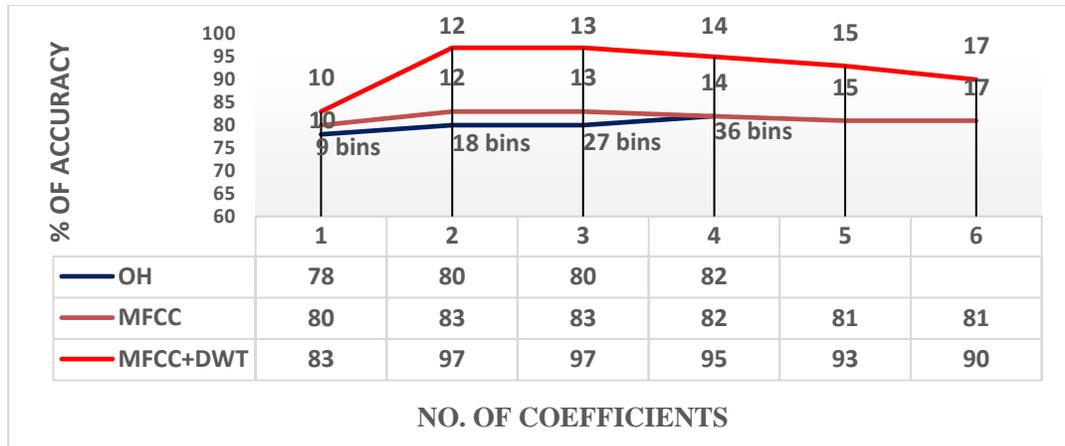

Figure 3.22 Comparative analysis of % of accuracy at different no. of MFCC coefficients

For analysing the graph shown in Figure 3.22 we have compared the proposed technique with the other two existing techniques and found that as the complexity (background variation, illumination condition etc.) of dataset increases, the recognition accuracy decreases in both the existing techniques. Because these two existing methods are unable to locate the exact change in the hand shape. This problem has been minimized in our proposed method which provides approximately 97% average accuracy with 12 DWT with MFCC coefficients.

We have also compared the time complexity of proposed method with other existing methods which is shown in Table 3.8. All the experiments are performed on core i5 processor with 4GB RAM and MATLAB 2013 software.

Table 3.8 Avg. processing time of 19 dynamic gestures

| Total Avg. time of OH | Total Avg. time of MFCC | Total Avg. time of DWT with MFCC |
|---|---|---|
| 18.08 sec | 22.69 sec | 18.65 sec |

From Table 3.8 we observed that the average processing time (CPU time) of DWT with MFCC method is approximately similar with the other two existing methods (OH and MFCC) whereas it provides higher amount of accuracy than others.

Table 3.9 Average Classification accuracy for static gestures

| Features | % of Classification Rate |
|---|---|
|  | Static gestures |
| OH | 97 |
| MFCC | 96 |
| DWT with MFCC | 100 |

Comparative analysis of all the three features has also been done on static gestures which is shown in Table 3.9. This analysis shows that all the three methods give same recognition accuracy for static gestures because there is no variation between intermediate frames with respect to time. We have also performed the comparative analysis of all three features of SKIG dataset.

Table 3.10 Comparative Results of SKIG dataset for different feature extraction technique

| Features | % of Classification Rate |
|---|---|
|  | SKIG data set |
| OH | 84 |
| MFCC | 85 |
| DWT with MFCC | 93 |

From Table 3.10 we have seen that the performance of our proposed method of SKIG dataset is good as compare to other feature extraction technique like OH and MFCC. This is because the proposed method minimized the ambiguity problem which is major concern in any gesture recognition system.

### 3.4 Validation in Real Robot

After classification of an unknown gesture, it has been validated in real robot. This section introduces an underlying concept of learning gestures for humanoid

HOAP-2 robot, in order to perform several tasks eventually. The real time robotics simulation software (WEBOTS) has been used for generating various types of HOAP-2 gestures perfectly.

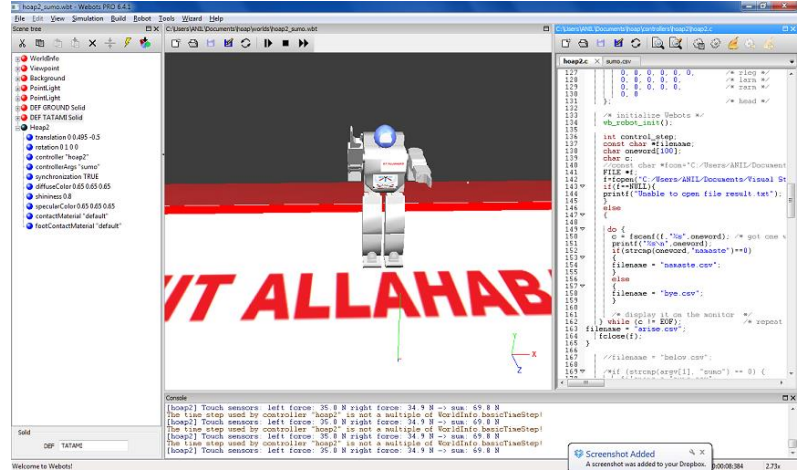

Figure 3.23 WEBOTS simulation of 'ARISE' gesture.

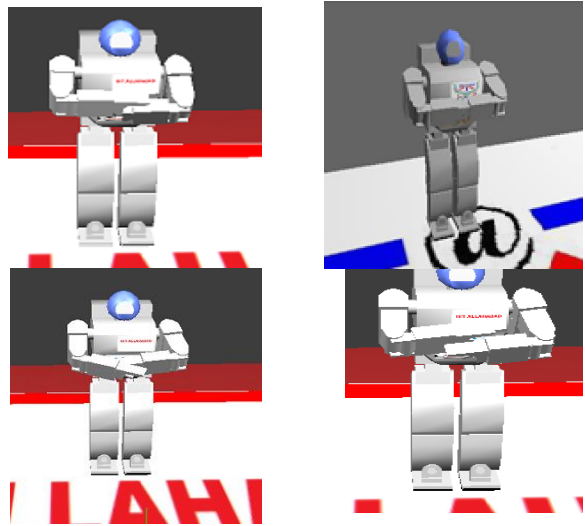

Figure 3.24 HOAP-2 performing gesture 'Below', 'Add' 'Across' and ' Above'.

Various types of HOAP-2 gestures ('Add', 'Across', 'Above' and Below) and there simulation environment have been shown in Figure 3.23 and 3.24. The working details we have already discussed in humanoid learning section.

**3.5 Conclusion**

In this chapter, an attempt has been made to analyse the characteristics of ISL gestures with the effect on classification process and used them for imitation learning by humanoid HOAP-2 robot on WEBOTS simulation platform. In this work, we proposed a novel DWT with MFCC based ISL gesture recognition method in which both the hands have been used for performing any gestures. DWT analyses the image or signal with respect to time as well as frequency. It also provides a moment invariant properties about any gesture. These properties make it invariant against scale, orientation, moment, phase etc. It reduces the 1/4 feature space of the data set, in the first level of decomposition which is a solution for the reduction of time complexity as well as space complexity. After reduction and noise elimination of images, moment invariant features is extracted by converting 2d contour image into 1d plane. From these moments invariant features spectral envelope of MFCC features has been calculated using the MFCC feature extraction technique. This technique has been applied for extracting the spectral envelope of the image signal. This technique is helpful against different backgrounds and different illumination conditions, than OH technique and also it is very effective for discriminating one gesture to another gesture. This is because the DWT minimizes the ambiguity issues by using its multiresolution property and spectral envelop property of MFCC solves the problem of hand shape extraction. MFCC has been generally used for speech/voice recognition as it has the quality to discriminate the vocal disturbance form background noises due to its Cepstral features. We tried to use MFCC technique for gesture image analysis and it can be said that it is an effective technique for voice recognition as well as for gesture recognition. When we use it with DWT it gives 98% recognition accuracy with dark background and various lighting conditions. Gestures are also classified using different types of classifiers like KNN, SVM. Analysis has also been done by calculating False Positive (FP), False Negative (FN), True Positive (TP) and True Negative (TN). By using these values a confusion matrix has been created. We also observed from these confusion matrix that the proposed technique gives better accuracy than other methods. Proposed technique has also been tested on a SKIG data set which was published by University of Sheffield. In this data set 2160 video sequences are there in different backgrounds and different illuminations. DWT with MFCC based algorithm has also provided 93 percent accuracy towards this data set, but other techniques like OH and MFCC provides only 84 and 85 percent accuracy which is much less than our proposed method.

After classifying an unknown gesture, it has been validated on HOAP-2 humanoid robot using WEBOTS simulation software. The controller program for HOAP-2 has been associated with .CSV files to generate various gesture patterns accordingly. The software has been developed to support imitation learning by humanoid robot for each classified ISL gesture on WEBOTS environment.

The three constraints, we have considered in this proposed approach are: background, cloth and exclusion of face region. In this chapter we have collected the data with single colour background with black full sleeves dress. Also, we have excluded the face region from all the dataset. We try to reduce all these constraints in the next few chapters.

# Chapter 4

# Continuous Sign Language Gesture Recognition using Possibility Theory Based Hidden Markov Model

In chapter-3 we have discussed about the isolated gestures and proposed an Isolated ISL gesture recognition framework using DWT with MFCC technique. In our day to day life we normally exchange our thoughts, conversations in the form of sentences like "how are you", "I agree" (continuous gestures) etc. which are a combination of an isolated words (isolated gestures). This chapter has been dedicated towards addressing the continuous gesture recognition problem. The following challenges have been solved:

1. Continuous gestures have been recognized in **real time** environment correctly.

2. It is then to be communicated unambiguously to a humanoid robot.

3. Subsequently the system's response has been evaluated to show the effectiveness of our development.

While appropriate trajectory planning and motion control have reached to a level when real time communications are feasible, the main challenge still is in solving problem 1, the real time recognition problem. In this chapter we have developed a new methodology for real time gesture based communication using Possibility Theory Based Hidden Markov Model (PTBHMM). The main objective is to reduce the computational complexities of conventional HMM, while explaining its beautiful theoretical framework. We have here addressed all the three basic problems of HMM and formulated corresponding three basic PTBHMM problems [218][219][220]. We have compared there time complexities in the subsequent sections of this chapter.

## 4.1 Background of Possibility Theory:

Uncertainty is the measurement of ambiguity where it is difficult to find the perfect solution of any problem. It occurs due to linguistic error, loss of information, variation in data and data selection for a particular event. For solving the issue of uncertainty, various theories have been proposed in the past few decades like fuzzy logic, probability, possibility, rough set, Dempster Shafer theory, etc. Among all we have applied possibility theory for solving all the three fundamental problems of HMM because it deals with both uncertainty as well as imprecision. Here uncertainty defines the ambiguity and imprecision defines the inexactness (inaccurate). The concept of possibility was introduced by Zadeh in 1978 [186]. After that Didier Dubois et al. [187] [188] discuss their views on the possibility theory by taking the example of prices of lottery. They explained how the possibility theory can be applied and what are the benefits of using possibility theory. They also explain the quantitative and qualitative aspects of the possibility theory as described below:

### 4.1.1 Quantitative possibility:

It represents a degree of possibility of an event which means how much a particular event can occur. For example, if a women is tall therefore, it measures how much a women is tall?

### 4.1.2 Qualitative possibility:

Let us consider a finite set S = ($s_1$, $s_{2...}$ $s_n$) of the universe of discourse U. Then possibility distribution Π can be expressed as:

$$\Pi(U) = Sup_{u \in A}(\pi(u)) \qquad (4.1)$$

Where A is a finite set of U and u is an element of A lies between [0, 1].

and necessity can be defined as:

$$N(A) = \inf_{u \in A}(1 - \pi(u)) \qquad (4.2)$$

Table 4.1 Possibility distribution of different materials of two factories

|  | Crude oil | Natural Gas | Coal | Nuclear |
|---|---|---|---|---|
| $\pi(cell)$ | 0.9 | 0.7 | 0.8 | 0.3 |
| $\pi(chevron)$ | 0.4 | 0.5 | 0.9 | 0.2 |

Where $\Pi(u)$ shows how much an event can exist and N(A) reflects that an event A can imply with a degree $\pi$. It satisfies two properties union and intersection, which are defined in equation 4.1 and 4.2. All the properties defined in equation 4.5, 4.6 and 4.7 have been explained with the help of an example given below. Let there be two factories in which different types of material like (crude oil, coal, natural gas and nuclear) are stored. The possibility distribution of each material in each factory is shown in Table 4.1. Define the universe of discourse $U_d$ = (crude oil, coal, natural gas and nuclear) and a subset A = (crude oil, coal) then possibility degree $\Pi(A)$ can be defined as:

$$\begin{aligned} \Pi(A) &= \sup(d \in A)\, \pi(d) \\ &= \sup(0.9, 0.7) \\ &= 0.9 \end{aligned} \qquad (4.3)$$

In a similar way necessity N(A) can be defined as:

$$\begin{aligned} N(A) &= \inf(1 - 0.8, 1 - 0.3) \\ &= \inf(0.2, 0.7) \\ &= 0.2 \end{aligned} \qquad (4.4)$$

## 4.2 Properties of Possibility Theory in Comparison with Probability Theory

Fundamentally the probability and possibility theories are similar in Bayesian concept, but it totally differs when the concept of fuzzy theory evolved which is based on possibility theory.

There are various properties like:

1. Additivity:

$$\left.\begin{array}{r}\sum_{i=1}^{n} P(A_i) = P\left(\cup_i^N A_i\right) \quad \max i \in N \text{ (in probability)} \\ \prod(A_i) = \prod(\cup_i^N A_i) \quad \text{(in possibility)}\end{array}\right\} \quad (4.5)$$

2. Monotonicity:

$$\left.\begin{array}{r}P(A) \geq 0 \text{ (in probability) given } A, B \in v \\ If\ A \subseteq B, then\ \Pi(A) \leq \Pi(B)\ (in\ possibility)\end{array}\right\} \quad (4.6)$$

3. Boundary Conditions:

$$\left.\begin{array}{r}P(\cup) = 1 \quad \text{(in probability)} \\ \Pi(\varphi) = 0 \text{ and } \Pi(\cup) = 1 \text{ (in possibility)}\end{array}\right\} \quad (4.7)$$

We have given two possible set of events A and B. In possibility theory, both possibility and necessity can be defined through belief and plausibility functions which is defined as:

Bel (A ∩ B) = min (bel (A),bel(B))

It can also be represented as:

η(A ∩ B) = Min(η(A),η(B))  and

Po(A ∪ B) = Max(Po(A),Po(B))               (4.8)

It can also be represented as:

$\Pi(A \cup B) = Max(\Pi(A), \Pi(B))$ and complement of A can also be defined as:

η(A) = 1 – Π A$^c$

## 4.3 Proposed Algorithm

We have proposed a novel possibility theory based hidden Markov model (PTBHMM) where all the three basic problems of HMM have been redefined using possibility theory and have been solved subsequently with rigorous complexity analysis. Subsequently, it has been tested on continuous gesture recognition. Here we used an axiomatic approach of possibility which is very efficient against real word problems. Let us consider the example of N basket having M number of distinct balls. Find out the possibility of picking a ball from each of the baskets. These balls are picked up from a basket in a sequential manner like $O_1, O_2, \ldots\ldots O_M$. These sequences are known as observation sequence. The notations used in this chapter are summarized in Table 4.2.

Table 4.2 Description of notations

| N | Number of states |
|---|---|
| M | Total number of distinct observation symbol |
| T | Length of observation sequence |
| $O=O_1, O_2,\ldots\ldots O_M$ | Total Number of observation symbols |
| $\Psi= \Psi(i)=Po(i_1=i)$ | Initial possibility at t=1 |
| $\theta = (s_{ij}) = Po(i_{t+1} = j/i_t = i)$ | State transition possibility from time t+1 to t |
| $\pi = (\pi_i(k)) = Po(O_k$ at $t/i_t = i)$ | The possibility of coming an observation symbol $O_k$ (Emission probability) at time t at which the state is $i_t$) |
| $s_t$ | Represents states at t time stamp |
| $\zeta = (\theta,\pi,\psi)$ | Concrete representation of PTBHMM |
| $\Pi$ or Po | Symbol for possibility function |

There are three problems of PTBHMM which are explained as:

## 4.4 Redefining the Problems of HMM in Terms of Possibility Theory

1. **Evaluation Problem:**

   We have given a parameter ($\theta, \pi, \psi$) where $\theta$ is the state transition possibility matrix, $\pi$ is the observation possibility matrix and $\psi$ is the initial possibility of each state. The possibility of occurrence of observations $O_1, O_2, O_3\ldots\ldots$ $O_M$ given model $\zeta = (\theta, \pi, \psi)$ i.e. $\Pi (O/\zeta)$.

## 2. Decoding Problem:

In this problem we have selected the state sequence ($s_1$, $s_2$, $s_3$........) in such a way that $\Pi(O, S/\zeta)$ is maximized. Where O= $O_1$, $O_2$, $O_3$....... and S = ($s_1$, $s_2$, $s_3$........). This problem is solved using possibility based decoding algorithm.

## 3. Learning Problem:

The third problem is the learning problem. In which state transition possibility matrix $\theta$, observation possibility matrix $\pi$ and initial possibilities $\psi$ have been updated. This process continues till all the parameters of the PTBHMM model $\zeta = (\theta, \pi, \psi)$ have been optimized. In this chapter all the three problems of HMM has been solved using possibility algorithm.

## 4.5 Solution of PTBHMM

All the three problems of PTBHMM have been solved by remodeling the existing algorithms where probability theory has been replaced with possibility theory.

### 4.5.1 Possibility Based Forward Algorithm:

Problem 1 has been solved by calculating the possibility $\Pi(O/\zeta)$. $\Pi(O/\zeta)$ has been calculated with the help of $Po(O,S/\zeta)$ possibility function where S = ($s_1,s_2,s_3........s_N$) (N no of states) and $\zeta$ is the model.

$$\Phi_t(i) = \Pi_i(O_1,O_2,O_3....O_t, s_t = s/\zeta) \qquad (4.9)$$

Equation 4.9 has been proved using induction method.

- For t=1,

$$\left. \begin{array}{rl} \varphi_1(i) &= \Pi_i(O_1, S_1 = s/\zeta) \\ &= \Pi_i[Po(S_1 = s/\zeta)Po(O_1/\zeta)] \\ &= \max[\max(\psi_i), \max(\pi_i(O_1))] \ 1 \leq i \leq N \end{array} \right\} \qquad (4.10)$$

- Equation 4.9 is true for t+1 time stamp, where t=1, 2......T-1

$$\begin{aligned}
\varphi_{t+1}(j) &= [\Pi_i(O_{t+1}, S_{t+1} = s)/\zeta] \\
&= [\Pi_i[Po(S_{t+1} = s/\zeta)Po(O_{t+1}/\zeta)] \\
&= \max[\max(\varphi_t(i)), \max(S_{ij}), \max(\pi_j(O_{t+1}))] \; 1 \leq j \leq N
\end{aligned} \quad (4.11)$$

- To prove that equation 4.9 is true for t=1, 2,..... T where s=1....... N

$$Po(O/\zeta) = \sum_{i=1}^{N} \Phi_T(i) = \max(\Phi_T(i)) \; i=1...... N \quad (4.12)$$

Equation 4.12 has been proven with the help of Bayes rule of possibility theory which is similar to Bayes' theorem of probability. Let us consider the observed sequence of length 2 ($O_1$, $O_2$) which means observation $O_1$ will come first from any of the state $s_1$, $s_2$...... $S_n$. After observation $O_1$ state $s_2 = j$ will come then observation $O_2$) appears which is represented as:

$$\varphi_2(j) = PO(O_1, O_2, s_2 = j)$$

Equation 4.9 has been proved by generalizing equation 4.12.

$$\begin{aligned}
\varphi_2(j) &= \Pi_i[Po(O_2, S_2 = j, O_1 \, from \, state \, i) \, Po(O_1 from \, state \, i) \\
&= \Pi_i[Po(O_2/S_2 = j, O_1 from \, state \, i) \, Po(S_2 = j/O_1 from \, state \, i) \\
&\quad Po(O_1/S_1 = i)Po(S_1 = i)] \\
&= \Pi_i[Po(O_2/S_2 = j)Po(S_2 = j/S_1 = i)Po(O_1/S_1 = i)Po(S_1 = i)] \\
&= \max[\max(\pi_j(O_2)), \max(S_{ij}), \max(\pi_i(O_1)), \max(\psi_j)] \\
&= \max[\max(\psi_j), \max(\pi_i(O_1)), \max(S_{ij}), \max(\pi_j(O_2))] \\
&= \max[\max(\Phi_1(i), \max(S_{ij})), \max(\pi_j(O_2))]
\end{aligned} \quad (4.13)$$

From above equation we found that

$$\begin{aligned}
Po(O) &= \Pi_i[Po(O/S_T)Po(S_T = s)] \\
&= \Pi_i[Po(O, S_T = s)] \\
&= \Pi_i(\Phi_T(i))
\end{aligned} \quad (4.14)$$

This proves that the conditional possibility is also true of the N=T observation sequence.

**Complexity of the algorithm:**

In HMM the complexity of forward/backward algorithm is of order of $N^2T$. In our proposed approach, complexity has been calculated as:

- Each step having N number of loops, therefore complexity will be N+1. For t=1, it has only one loop, therefore having complexity 1 which is a constant number. Therefore the complexity for N loops is N.

- In our proposed algorithm max function is applied for finding the possibility of outcome events. This max function has been applied for all the N loops and this has been continued for j=1..... N and t=1... T-1. Therefore the final complexity has been represented as:

    (N+N)(T-1)= 2 NT

- In the final step the complexity will be $(N + N..... + n * N)(T) = n * N * T$ where n is a constant. Therefore, the final complexity will be of the order of the NT which is N times less than the complexity of forward algorithm.

### 4.5.2 Possibility Based Backward Algorithm:

The process of analysis is similar to the earlier process. It is helpful for finding the optimal value of HMM model $\zeta$ at each observation sequence from backward direction. Here we find the parameter $\gamma_t(i) = Po(O_{t+1}, O_{t+2},......O_T/i_t = i, \zeta)$.

(a) For T=1

$$\left.\begin{aligned}\gamma_t(i) = \Pi_i(O_t, S_t = s/\zeta) &= \max[\Pi_i((S_t = s)/\zeta), \Pi_i(O_t/\zeta)] \\ &= \max[\max(\psi_i), \max(\pi_i(O_T)) \\ &= max[1,1] = 1\end{aligned}\right\} \quad (4.15)$$

(b) For t = T − 1, T − 2.......1, 1 ≤ i ≤ N

$\gamma_t(i) = \max[\max(\varphi_t(i)), \max(S_{ij}), \max(\pi_j(O_{t+1}))]$, 1 ≤ j ≤ N     (4.16)

(c) The proof is very much similar to the earlier problem.

$$\Pi(O/\zeta) = \prod_{i=1}^{N}(\psi_i(O_1)\gamma_i(i)) \quad (4.17)$$

*IIIT-A, Robotics and AI Lab*                                                                                                           90

Its complexity has been calculated and found to be similar to the earlier solution (NT).

### 4.5.3 Possibility Based Viterbi Algorithm:

The second problem, the so called decoding problem in which we maximize Po(O,S/ζ) where O = $O_1, O_2, O_3 .... O_T$, S = ($s_1, s_2, s_3........s_T$) and ζ is the possibility based HMM model. Then finding out the optimal path from observation sequence $O_1, O_2....... O_T$ means extracting the maximum possibility of states coming to the particular observation symbol. It is similar to dynamic programming approach.

$$[max_{s_t}]_{t=1}^{T}(\Pi(O, s_1, s_2, ...... s_T/\zeta)) \qquad (4.18)$$

### 4.5.4 Possibility Based Baum-Welch Algorithm:

Learning problem is the most important problem of HMM which has been solved in this section using possibility theory. The solution is based on optimizing the value of Po(O/ζ). The parameters (θ, π, ψ) of HMM model have been updated using proposed possibility theory. The values of (θ, π, ψ) have been updated till the difference between two consecutive values of (θ, π, ψ) becomes zero. The Initial value of (θ, π, ψ) has been defined as:

$$\psi_s = \frac{\text{number of times in state s at t=1}}{\text{Total number of times at t=1}} \qquad (4.19)$$

$$\theta = S_{ij} = \frac{\text{Number of possible transition from i to j}}{\text{no. of transitions from i}} \qquad (4.20)$$

$$\pi = \pi_i(O_t) = \text{The emission probability matrix} \qquad (4.21)$$

For optimizing the value of θ, π and ψ, few parameters are necessary like $\xi_t(i)$, $\phi_t(i)$ etc. which have been calculated as:

$$\xi_t(i) = \Pi(i_t = i/O, \zeta) \qquad (4.22)$$

$\xi_t(i)$ represents the possibility measurement where the observation sequence $O = O_1, O_2, O_3.... O_T$. $O_T$ and model ζ has been given to state i at time t. We expand equation 4.22 with the help of Bay's Law as:

$$\xi_t(i) = \frac{\Pi(i_t = i/O, \zeta)}{\Pi(O, \zeta)} = \frac{\max(\gamma_t(i)\, \phi_t(i))}{\Pi(O, \zeta)} \quad (4.23)$$

Where $\gamma_t(i)$ and $\varphi_t(i)$ are the solution of possibility based forward algorithm. $\varphi_t(i)$ defines the possibility of observations $O_1, O_2, O_3 \ldots O_t$ where $O_t$ is at state i and the time stamp is t. $\gamma_t(i)$ measures the possibility of observations $O_{t+1}, O_{t+2}, \ldots O_T$ at time t where state is i. Next we also define parameter $£_t(i, j)$ which supports for finding the optimal state transition possibility matrix.

$$£_t(i,j) = \Pi(i_t = i, i_{t+1} = j/O, \zeta) = \frac{\Pi(i_t = i, i_{t+1} = j/O, \zeta)}{\Pi(O, \zeta)} \quad (4.24)$$

Where $O = O_1, O_2, O_3 \ldots O_T$

$$\Pi(i_t = i, i_{t+1} = j/O, \zeta) = \Pi(i_t = i,$$
$$O_1, O_2, O_3 \ldots O_t, O_{t+1}, O_{t+2}, O_{t+3} \ldots O_T, i_{t+1} = j/\zeta)$$
$$= \max[\Pi(i_t = i, O_1, O_2, O_3 \ldots O_t/$$
$$\zeta)\, \Pi(O_{t+1}, O_{t+2}, O_{t+3} \ldots O_T, i_{t+1} = j/\zeta)]$$
$$= \max[S_{ij}, \pi(O_{t+1}), \gamma_{t+1}(i)]$$

$$£_t(i,j) = \frac{\max[S_{ij}, \pi(O_{t+1}), \gamma_{t+1}(i)]}{\Pi(O, \zeta)} \quad (4.25)$$

$\prod_{t=1}^{T-1}(\phi_t(i))$ shows the number of possible evolutions from state i.
$\prod_{t=1}^{T-1} £_t(i,j)$ represents the possibility of transitions from state i to state j.

Now we define the possibility of finding an optimal value of model parameters $\psi$, $\theta$ and $\pi$.

$$\psi_i = \phi_t(i), \quad 1 \leq i \leq N \quad (4.26)$$

$$\theta_{ij} = \prod_{t=1}^{T-1} £_t(i,j) / \prod_{t=1}^{T-1}(\phi_t(i)) \quad (4.27)$$

$$\pi_j(k) = \prod_{t=1, O_t = k}^{T}(\phi_t(i)) / \prod_{t=1}^{T-1}(\phi_t(i)) \quad (4.28)$$

All these values are updated till difference between two consecutive values of the transition matrix and emission matrix has been 0 or less than a threshold value.

## 4.6 Implementation of PTBHMM on Continuous ISL Gesture Recognition

This section focuses on continuous ISL gesture recognition using proposed PTBHMM model. Any gesture recognition system has five modules data acquisition, preprocessing, key frame extraction, feature extraction and classification. Here PTBHMM has been used for classifying an unknown (probe) gesture. After data collection background modelling and segmentation has been done for hand subtraction and extraction of the start and end point of informative gestures. Gesture segmentation is done using gradient based key frame extraction method. This helps us to break each sentence into a sequence of words (isolated gestures) and also obliging for extracting frames of meaningful gestures. Finally feature extraction and classification have been done. The performance of PTBHMM and HMM are analyzed and compared on 10 sentences database of continuous ISL.

### 4.6.1 Dataset creation:

Dataset consisting of a collection of signs where single hand or both the hands have been used for performing continuous ISL gestures. Ten sentences database have been created. Each sentence consists of two, three and four types of gestures. Every sentence is a combination of static and dynamic gestures. The dataset has been created using an external camera with the configuration of Canon EOS with 18-55mm lens, 30 frames per second and resolution is 18 mega pixels. Here we used a single camera for creation of gesture dataset. Black background is used for database, creation of ISL gestures. Here we concentrate on the upper body part only. Movement of the upper body part is acceptable. Position of camera is very important (camera calibration), for clarity of the dataset and for removing many backgrounds related problems like background noise, body motion, etc. Here we have taken 10 Indian sign language sentences of 5 different people, where each sentence has been recorded 10 times, 6 for training and 4 for testing. Every video is divided into a sequence of frames by size 640X480. Let us consider a continuous gesture "it is closed today". It is made of 3 gestures ("it", "today", "closed") shown in Figure 4.1. Similarly, we take 10 sentences of ISL database as shown in Figure 4.6.

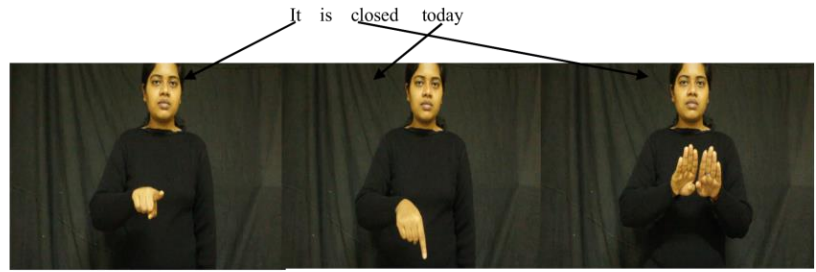

Figure 4.1 Gestures of sentence, it is closed today

### 4.6.2 Preprocessing:

In this step silhouette images of every hand gestures are created. Here we extract foreground image from complete image means it removes the background of an image and get the skeleton of upper body parts. Then hand region has been subtracted from these foreground images by eliminating largest connected region which is face. Finally, we get the hand portion from upper body. We first convert each video into a sequence of RGB frames. Each frame has dimension 640*480. Skin color segmentation is applied for extraction of skin region, which is divided into a number of chunks. For finding skin region, each frame is converted into HSV (Hue, saturation, value) plane where only H and S value having a threshold ($H > 0.55$ or $S <= 0.20$ or $S > 0.95$) is used for finding non skin region of an image. Then this region marked as zero for extracting skin region. Median filter is applied to preserving outer boundary (edges) of segmented region. It mainly removes salt and pepper noise and impulsive noise for edge preservation. Images obtained after median filtering are converted into binary form. At the end of preprocessing subtract largest connected region, which is the face. Eliminate facial region of the upper half of the body and will get hands gestures. Each step of pre-processing is shown in Figure 4.2.

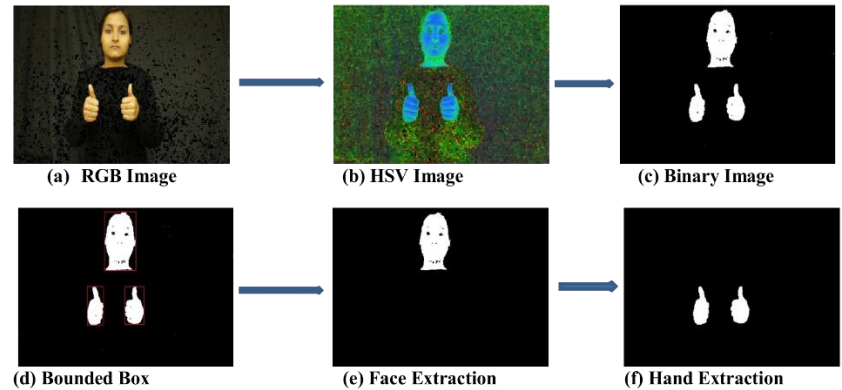

Figure 4.2: Pre-processing steps of each RGB frame

### 4.6.3 Key frame extraction:

Each video consists of a sequence of frames and each frame can be any gesture frame or non-gesture frame. So we have to extract those frames that belong to any gesture and remove non-gesture frames because these frames creates an extra affliction of processing. Here we used a gradient based key frame extraction method for extracting key frames of each video sequence. We take frames, do segmentation and calculate the gradient of each frame. Gradient helps us in finding the overlapping frames between gestures. A key frame extraction graph of "I agree" and "Why are you sad" gestures are shown in Figure 4.3 and 4.4.

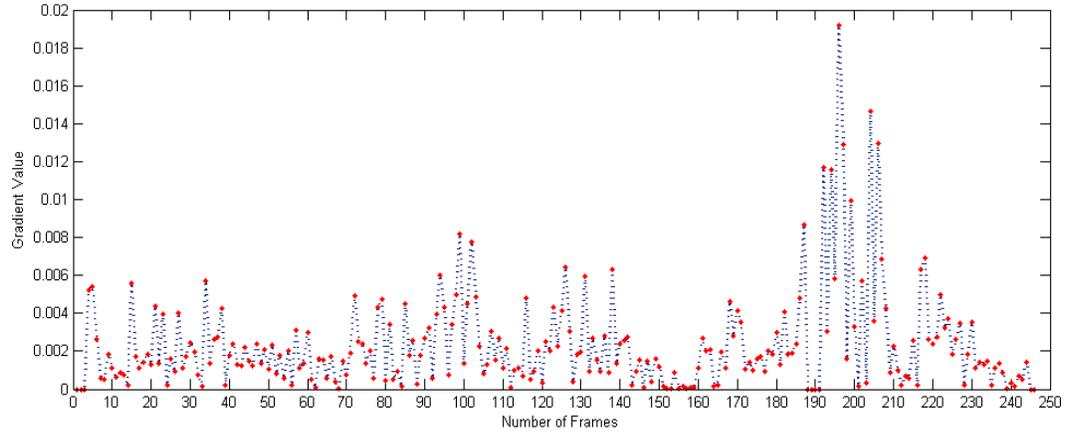

Figure 4.3: Key frame extraction of "I agree" gesture

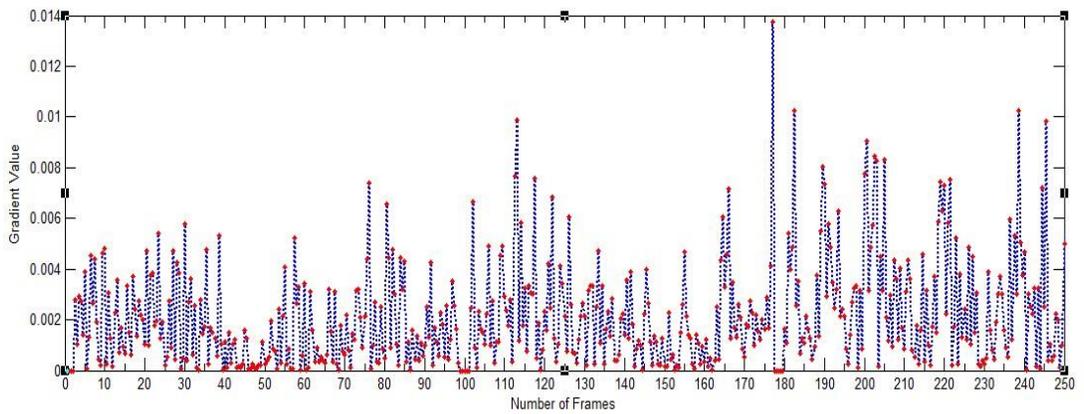

Figure 4.4: Key frame extraction of "Why are you sad?" gesture

The graph shown in Figure 4.3 shows that gesture1 starts in between the frame $0^{th}$ to frame $4^{th}$ and end in between frame 188 to frame 192. From $0^{th}$ to frame $4^{th}$, there is a constant gradient value and similarly from 188 to frame 192 we also obtained constant gradient value which shows the end of one gesture or the start of another gesture. In the second graph as shown in Figure 4.4, $1^{st}$ gesture starts at $0^{th}$ to frame $4^{th}$, $2^{nd}$ starts in between 99 to 102 and ends at 176 and last gesture starts at 178 and ends at 248. In all these positions the value of gradients is 0. Gradient value 0 shows maximum overlapping between two consecutive frames. Suppose if gesture1 ends at frame 188 then this is also the start of next gesture. In this way we can calculate the total number of frames in each gesture. In this approach we have considered the 15 middle frames of each isolated gestures which carries maximum information.

### 4.6.4 Feature extraction:

We applied orientation histogram as a feature extraction technique for extracting most appropriate features of each gesture. It offers advantages to scene illumination changes and even light condition changes. The edges of the sequences of images would still be the same. All the continuous ISL gestures have been captured in normal lighting conditions where pixel intensities can be suggested to change the scene lighting. Another advantage which is employed on orientation histogram is translation invariant property. It demonstrates that the same frames in different position of gestures would produce the same feature vectors. It is achieved to calculate the histogram of the local orientations for all the frames of the moving gestures. Translation of the frame in the gesture does not change the local orientation histogram. The steps of the orientation histogram algorithm are:

1. Sub-sample the 640 ∗ 480 images into 60 ∗ 40, which reduces the space complexity and makes the processing time fast with no loss of information.

2. Finding the edges of an image using 3-tab derivative filter x= [0 -1 1] y= [0 1 -1]. It helps us in finding the image gradient in the x-direction and y-direction.

3. The gradient in x-direction as well as in y-direction.

$$dx = \frac{\partial g(x,y)}{\partial x} = \frac{g(x+1,\ y) - g(x-1,\ y)}{2} \quad (4.29)$$

$$dy = \frac{\partial g(x,y)}{\partial y} = \frac{g(x,\ y+1) - g(x,\ y-1)}{2} \quad (4.30)$$

Where g (x, y) represents the intensity function at (x, y) pixel position.

4. Find out the gradient direction using arctan2 function which is expressed as:

$$X(x, y) = atan2\frac{\partial g/\partial x}{\partial g/\partial y} = atan2\frac{\partial y}{\partial x} \qquad (4.31)$$

The value of X lies between $[-\pi/2\ \pi/2]$.

$$\text{Magnitude} = mg(x, y) = \sqrt{dx2 + dy2} \qquad (4.32)$$

5. Convert these values into a column vector so that the radian values have been converted into degrees. Here 180 degrees is divided into 18 and 36 bins, in 18 bins each bin is about 10 degrees and in 36 bins each bin is about 5 degrees. The polar plot for 18 bins and 36 bins of "How" gesture is shown in Figure 4.5. This polar plot shows that the angle of variation in the hand at the time of performing the gesture.

6. After the feature extraction process, principle component analysis (PCA) [24] is applied for dimensionality reduction. Large amount of data set is very difficult to handle. Therefore, PCA is applied to reduce the dimension of the data. The Eigen value generated from PCA gives the projection direction of the confused data set.

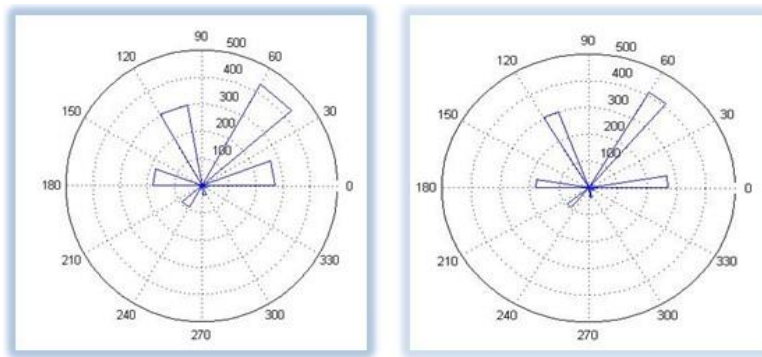

Figure 4.5: Polar plot of orientation histogram of "How" gesture (for 18 bins and for 36 bins)

### 4.6.5 Classification:

After the feature extraction process HMM and PTBHMM have been applied for classifying an unknown gestures. These two methods have been applied on codebook generated using Linde Buzo Gray (LBG) vector quantization algorithm [198]. Here codebook is generated for each gesture. The steps of LBG algorithm are:

- For every gesture we have considered 15 frames, among 15 frames we select 3 frames as initial centroid.

- For each frame we have calculated the Euclidean distance between feature vector and centroid points.
$$D_m = \sum(j = 1)^m \sum K_i \in X_j \|K_i - C_j\| \qquad (4.33)$$

- Again calculate a new centroid value(C).

$$C_j = \frac{1}{(|X_j|)} \sum K_i \in X_j\, K_i \qquad (4.34)$$

- This process continues till centroid value stop changing.

$$C_{j-1} - C_j < \in (a\ small\ number) \qquad (4.35)$$

- Our codebook consist of 1(for centroid 1), 2(for centroid 2), 3(for centroid 3).

- Features having minimum distance belong to the class of a particular centroid and then replace feature vector with class numbers where it belongs.

- Codebook having length 15 and having values are numbers of centroid to which frame belongs.

    Where $X = [K_1, K_2, K_3,..........K_n]$;

After codebook generation HMM and PTBHMM have been applied for classification and recognition of gesture data set. Here we have taken 4 states and 4 observation symbols for each HMM and PTBHMM model and then probes are tested using possibility based decoding algorithm for PTBHMM and for HMM Viterbi algorithm is used. The training and testing steps of PTBHMM are explained as:

### A. Training Steps:

- We have taken q number of states say $s_1, s_2, s_3,....s_q$.

- Then the state transition possibility matrix $\theta_{ij}$=[0.8,1,1,0.4;0.5,0.6,0.1,1;1 0.3,1; 0.5, 0.5,1,1]; and emission possibility matrix $\pi_{jk}$ = [0.1,0.2,0.3,0.4;0.2,0.1,0.5 0.2;0.1,0.1, 0.4,0.4;0.2,0.1,0.3,0.4] are taken.

- Emission matrix is created with l observation symbols where Number of observations would be equal to the number of clusters present in the codebook (LBG Algorithm).

- Randomly selected Initial possibilities of all q states $p(s_1), p(s_2), p(s_3),......,p(s_q)$ = 1/q.

- Found the state sequence for a particular observation sequence obtained using codebook $\Pi(s/O)$.

- Update $\theta_{ij}$, $\pi_{jk}$ and initial possibility $\psi_i$ using solution three explained in 4.

- When the difference between two consecutive values of $\theta_{ij}$, $\pi_{jk}$ are 0 or less than some threshold value then learning stops. This is a PTBHMM model for first gesture HT1 ($\theta, \pi, \psi$).

- Similarly PTBHMM model for each gesture has been generated.

### B. Testing Steps:

- An observation sequence $v_1, v_2... v_n$ has been generated for testing gestures using vector quantization codebook generation algorithm.

- PTBHMM model ($\theta, \pi, \psi$) of each training gesture has been used for calculating the state sequence at a particular observation sequence of testing gesture using possibility theory based forward algorithm.

- Finding the most probable state sequence at a particular observation sequence using equation 4.36.

$$\text{Predicted gesture} = argmax_i f(v_1, v_2,...v_T; PTBHMM) \quad (4.36)$$

- Maximum likelihood value of testing gesture corresponding to each training class has been calculated.

## 4.7 Experimental Results and Analysis:

Experiments are performed on 10 types of sentences as shown in Figure 4.6. Here each sentence have 2, 3 or 4 gestures. Each continuous gesture is made up of static as well as dynamic gestures. Each sentence has been recorded 10 times, 6 times for training and 4 times for testing. In every sentence out of n number of frames, 20 frames of each isolated gesture has been considered (n will vary from sentence to sentence) for training and 10 frames of each gesture for testing. After key frame extraction only those frames are taken which are present in the middle because it contains maximum information. After that an unknown sentence has been classified using PTBHMM. All the steps of testing phase are similar as training phase. After classification of probe gestures a text formation has been performed. The results obtained from PTBHMM are then compared with the results obtained from HMM. All the experiments are performed on 18 bins as well as on 36 bins which means 180 degrees is clustered into 18 bins and 36 bins where each bin having 10 degree and 5 degree resolution. The classification results are shown in Table 4.3.

We have clearly seen that the PTBHMM gives 92% classification accuracy which is similar to the classification rate given by HMM. We also observe from Table 4.4 that PTBHMM takes very less time (processing time) compared to HMM. Theoretically as well as experimentally we found that the processing time of PTBHMM is N times less than the processing time of classical HMM. In possibility theory comparison functions (maximum and minimum) have been used which has less computation. Whereas in classical HMM multiplicative and additive (summation of all) functions have been used which has higher computations. Therefore, HMM has higher time complexity than PTBHMM. From both experimentally as well, theoretically we have seen that PTBHMM has less time complexity than HMM.

| Sentences | Gesture 1 | Gesture 2 | Gesture 3 | Gesture 4 |
|---|---|---|---|---|
| How are you? | How | You | | |
| I am agree. | I | Agree | | |
| Are you coming? | You | Coming | | |
| I am studying. | I | Studying | | |
| It is closed today. | It | Today | Closed | |
| Are you hearing? | You | Hearing | | |
| You do not have a car? | You | Car | Not | |
| Your home is big or small? | Your | Home | Big | Small |
| Why are you sad? | You | Sad | Why | |
| Is this your own room? | Your | This | Room | Own |

Figure 4.6: 10 sentences gesture dataset

All the tests are executed on 32 bits (windows 7) machine having an Intel Core I-5 processor with 2.50 GHZ CPU speed. A machine having 8GB of internal memory. Therefore PTBHMM is suitable for real time applications where we need less processing time.

Table 4.3 Classification results using HMM and possibility theory

| Gesture name | 18 bins(%) | | 36 bins(%) | |
|---|---|---|---|---|
| | HMM | **PTBHMM** | HMM | **PTBHMM** |
| How are you? | 85 | 85 | 91 | 92 |
| I am agree | 91 | 90 | 93 | 93 |
| Are you coming | 90 | 90 | 94 | 93 |
| I am studying | 94 | 94 | 95 | 95 |
| It is closed today | 93 | 91 | 94 | 95 |
| Are you hearing | 92 | 93 | 95 | 94 |
| You do not have a car | 89 | 87 | 91 | 91 |
| Your home is big or small | 92 | 93 | 94 | 94 |
| Why are you sad | 91 | 91 | 94 | 94 |
| Is this your room | 90 | 92 | 93 | 93 |

Table 4.4 Processing time of HMM and PTBHMM

| Methods | Avg. processing time(sec) |
|---|---|
| For HMM | 28.92 |
| For PTBHMM | 16.987 |

## 4.8 Conclusion

The dynamic gesture recognition has been performed using PTBHMM. We have demonstrated through rigorous experiments that PTBHMM out performs conventional HMM in terms of handling uncertainties in all the three problems of HMM. We have used the possibility theory which has a better uncertainty handling capability while maintaining the beautiful properties of HMM in terms of Markov chain and Bayes decision making. Data uncertainty has been modeled using possibility theory as a kernel in PTBHMM in all the three classical problems of HMM. The axiomatic approach of possibility theory is successful in reducing the time complexity from $O(N^2T)$ of HMM to $O(NT)$ for PTBHMM. These findings may be extremely useful in handling real time "time series" data and more rigorous experiments may further establish this innovative idea.

The proposed PTBHMM model gives similar accuracy as the conventional HMM. Here we also have a general image processing constraint on background colour means in all the experiments the background must be in single colour. This limitation has been overcome in chapter 5.

# Chapter 5

# Development of a NAO based Human-Robot Interactions Framework with Continuous Indian Sign Language

We have discussed a Continuous Indian Sign Language gesture recognition using possibility theory based Hidden Markov Model in the preceding chapter. Where orientation histogram is used as a feature extraction technique, but this feature alone is not sufficient for recognizing continuous gestures in real time environment. In any gesture recognition system an extraction of suitable feature is exceedingly significant. After studying various literatures we found that single feature like speed works well in one scenario, but it runs out in another scenario, similarly the same thing occur with other features too. Therefore, we have proposed a framework in which various characteristics (speed, moment and orientation) of the hand are extracted and blended. Combining these three features provides maximum information about any gesture. In this chapter we have developed a novel framework in which NAO humanoid robot recognizes continuous ISL gestures in real time environment and then translates into a text / speech format. These texts are then matched with the knowledge database of NAO robot using similarity measure. This matching increases the classification accuracy. Here the database has been created using NAO vision sensors which are accessed through NAO MATLAB API. This database contains several sentences of ISL, commands and so on which is helpful for normal persons to understand commands through robot. These signs are helpful in the communication established between human and robot. In any gesture recognition system, preprocessing and coarticulation detection (start and end point of each gesture) are the major issues which are also being handled in this chapter[215][216][217].

After recognizing a sentence or commands, NAO [209][221]converts it into a speech format or answer to deaf persons into gesture format. Thus, the divergence between the normal people and the deaf community becomes minimized and then communication ratio between Deaf and Dumb community and normal persons can be improved. Here continuous gestures are tested in real time environment with black full sleeve dress using an NAO humanoid robot. It is a humanoid robot having 25 degrees of freedom developed by a French company Aldebaran Robotics. NAO has been proved to be a good test bed for developing various technologies like social robotics, autism, sketch drawing etc. Here we have used NAO for communication establishment between human and robot. Various problems arise, when we recognize any gesture in real time environment. Among all we have considered three problems and tried to resolve it in this chapter.

A) Gesture segmentation and hand extraction is one of the biggest challenge because of spatiotemporal and coarticulation variations in successive gestures. Here we also considered gestures containing single hand as well as both the hands.

B) Let $s_1, s_2, s_3, s_4 \ldots s_n$ be a sequence of video frames having n number of gestures. Find out a start and end point of meaningful gestures from a sequence of informative and uninformative frames.

C) Deformities occurred in hand moment trajectory during hand segmentation.

All the three problems have been solved by proposing a novel ISL gesture recognition framework where hybrid features (speed, orientation and moment) are extracted using frame difference and wavelet descriptor. Ultimately this approach is being implemented on the NAO humanoid robot. First NAO recognizes a particular gesture and form it into text format. These texts are then matched with the knowledge database of NAO robot which we have stored in NAO's memory. This framework is helpful for establishing the communication between normal and hearing impaired people through NAO humanoid robot.

## 5.1 Proposed Framework

We proposed a novel NAO based human robot communication system through continuous ISL gesture as shown in Figure 5.1.

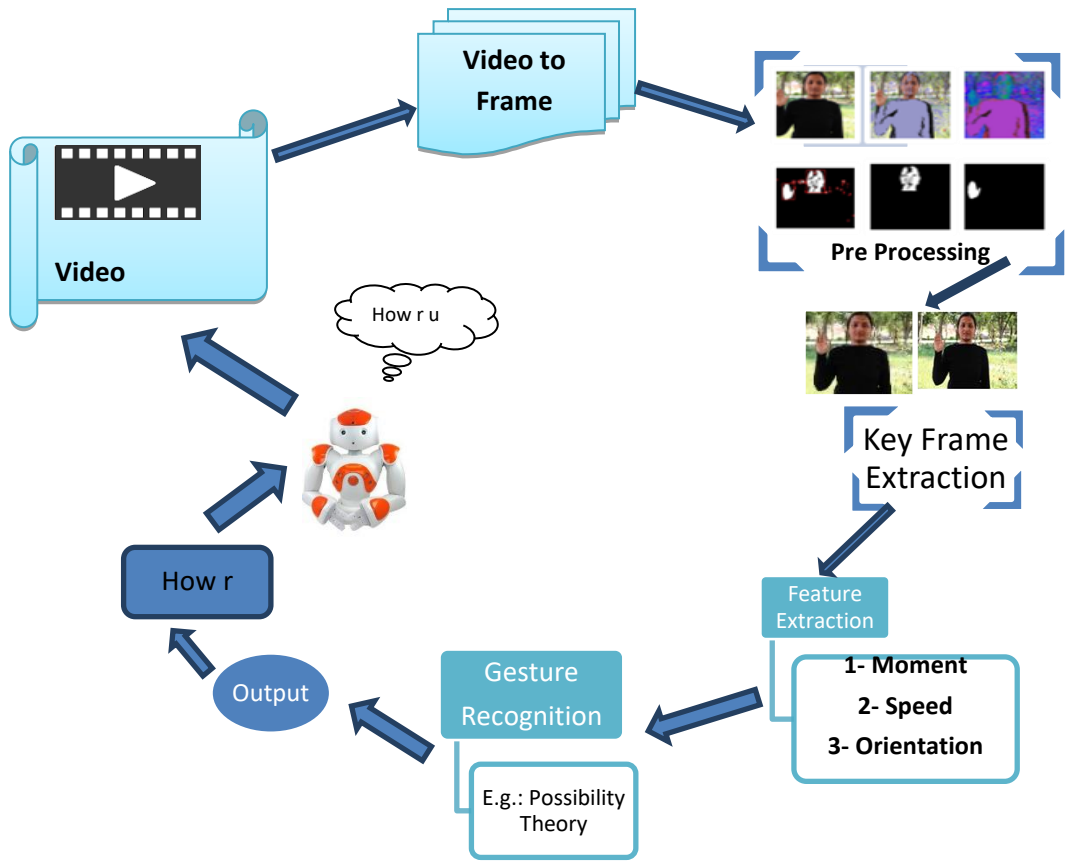

Figure 5.1 Proposed framework of gesture recognition model

**Justification of Framework**

The framework shown in Figure 5.1 has six modules:

i) Data acquisition ii) Pre-processing iii) key frame extraction iv) feature extraction v) classification vi) Text generation/speech formation. We have collected a database of 20 sentences containing 36 isolated gestures where single hand or both the hands have been used. Each sentence or commands have a start symbol (gesture) and an end symbol, when this symbol appears in front of robot, robot starts recording and recording stops when end symbol comes. This symbol has been used for preventing an unwanted recording. The video of each gesture has been recorded through the NAO's vision system. This vision system has been used for creating a video sequence of continuous gestures through MATLAB API of NAO robot. Here each sentence consists a mixture of two, three and four types of static as well as dynamic gestures like "I agree", "Your home is big or small" etc. All the ISL signs are used from book of Universal Indian Sign Language which is published by Sir William Tomkins [175].

### 5.1.1 Preprocessing

Pre-processing of video gestures has been done after data acquisition unit. In this portion background modelling and segmentation has been performed for hand subtraction and extraction of the start and end point of informative gestures. Before this we have been eliminating starting and ending gesture which is common in the knowledge database. This elimination is done by subtracting subsequent frames till $f_{t+1}-f_t=0$. This process continuous till $f_{t+1}-f_t \neq 0$. Similar steps have been applied for removing end symbol also. After removal of starting and ending symbol background modelling is performed.

In this modelling, foreground which contains portions of the hand are extracted from a background. The steps involved in preprocessing are:

1- Background subtraction has been done by differentiating the background to the foreground. After getting foreground and background images, HSV color segmentation technique is applied to both of the images and then the XOR operation is performed between both HSV images. The ranges of $0 \leq H \leq 50$, $0.25 \leq S \leq 0.57$ and $V \geq 40$. We consider only H and S value for extracting the silhouette images of hand.
2- Median filter is applied on binary images generating after skin color segmentation. It removes noise present during preprocessing. Median filter preserves the boundary (edges) of the images which is most important in the case of gesture.

*IIIT-A, Robotics and AI Lab*     107

3- Determine the area of the skin color region, which are face and hand. Then eliminate the biggest area which is face and get the final hand silhouette images.

The complete operation of hand segmentation is presented in Figure 5.4.

**5.1.2 Gesture Splitting**

Splitting of gesture has been done using gradient based overlapping frame extraction method. This step is the heart of the continuous gesture recognition system. In the gesture splitting process, extraction of those frames which consists of maximum changes, is the most critical issue. This issue has been solved by calculating the gradient in x direction and y direction between successive frames. Here are the steps:

i) Subsample each image from 640*480 to 60*40. This size will take less time while processing the images by keeping all the information in tact.

ii) 3 tab two derivative filters are used for finding the derivative in x direction (dx) and in y direction (dy) where filters have been defined as: x=[0 -1 1] y=[0 1 1]

iii) Gradient direction has been calculated using atan2 function which is given along a gradient (dy/dx) i.e.

$$\text{Gradient direction } (\theta) = \text{atan2 (gradient (dy/dx))} \quad (5.1)$$

iv) Steps i to iii have also been applied for succeeding frame i+1. Then find out the difference of mean of two consecutive frames, which shows the change between two subsequent frames. This process continues for all the frames.

In the whole process pause of few milliseconds occurred during the end of one gesture and start of another gesture at a time when the ($\theta$) value becomes null. This helps us to break each sentence into a sequence of words (isolated gestures) and also obliging for extracting frames of meaningful gestures.

Here we consider middle 15 frames of each isolated gesture which contains informative gesture. Three dimensional feature vector (Orientation, speed and moment invariant) is used for recognizing an unknown gesture. Then these features are classified using possibility theory. Possibility theory also handles the similarity issue means when two gestures like "it" and "you" have been represented by the same symbol.

### 5.1.3 Feature Extraction

After preprocessing and key frame extraction, 3-dimensional features are calculated.

i) Speed: It determines how fast a gesture has been performed. It can be calculated as:

$$Sp_t = \sqrt{(x_t - x_{t+1})^2 + (y_t - y_{t+1})^2} \qquad (5.2)$$

Where $x_t$, $x_{t+1}$ and $y_t$, $y_{t+1}$ is a change in x direction and y direction at time t and t+1. It varies from person to person or time to time, therefore we add a weight factor which lies between 0.1 to 0.9 and see at what weight the recognition rate is maximized.

ii) Orientation: Here orientation of the hand is determined using discrete wavelet transform (DWT) [179] [180]. DWT is also helpful for reducing frame size, which reduces space complexity as well as processing time. Let us have an image matrix I of size m×n which is multiplied by a wavelet filter W of size m×n.

$$I_w = I \times W \qquad (5.3)$$

$I_w$ is further multiplied by W' for generating approximate and detailed coefficient.

$$a^1|d^1 = I_w \times W' \qquad (5.4)$$

Equation 5.4 shows the first level of decomposition having 4 components LL (low frequency component), LH, HL (diagonal component) and HH (high frequency). LL is further decomposed upto 2nd level and we get FF (low frequency component), FG (diagonal component), GF (diagonal component) and GG (high frequency) components. FG and GF contains orientation components which is determined as:

$$\theta = \operatorname{atan}\left(\frac{FG}{GF}\right) \tag{5.5}$$

Here θ determines the change in orientation angle.

iii) Moment invariant features: After segmentation process, wavelet descriptor [199] [200] has been applied for detecting the variation in hand shape where motion chain code (MCC) is used for closed contour extraction of hand. This mapping is helpful for determining the appropriate gesture through the anatomy of the hand. Wavelet descriptors are really prominent for identifying moment invariant features due to its invariance properties against scaling, rotation, moment and translation. For extracting moment invariant features we first convert the 2-D image into 1-D signal because wavelet works much more effectively in one dimension in comparison to two dimensions. For this conversion, 2-D image G(x, y) in x-y plane is converted into r- θ plane G(r, θ) described as: x=r cos θ and y=r sin θ.

$$G_{ab} = \iint G(r,\theta) g_a(r) e^{jb\theta} r \, dr d\theta \tag{5.6}$$

Where r is the radius of the circle, θ is the orientation angle, $G_{ab}$ is the moment of hand, $g_a(r)$ is a radial basis function and a, b are constants. In case of wavelet descriptor $g_a(r)$ has been treated as a wavelet basis function and replaced with

$$\vartheta^{p,q}(r) = \frac{1}{\sqrt{p}} \vartheta\left(\frac{r-q}{p}\right)$$

p and q are the dilation and shifting parameter.

Now convert a 2D image into 1D form for reducing feature extraction problem. We choose cubic B-spline (Gaussian approximation) as a mother wavelet define as:

$$\vartheta(r) = \frac{4p^{n+1}}{\sqrt{2\pi(n+1)}} \sigma_y \cos(2\pi g_0(2r-1)) \times \exp(-\frac{(2r-1)^2}{2\sigma_y^2(n+1)}) \quad (5.7)$$

Analyzing a moment variation from a shape of an image, the values of dilation and shifting parameter p and q are chosen to be discrete expressed as:

$$p = p_0{}^m, \; m \; is \; an \; integer$$

$$q = nq_0 p_0{}^m, \; n \; is \; an \; integer$$

$p_0 >1$ or $p_0<1$ and $q_0 >0$. These constraints have been considered so that $\vartheta(\frac{r-q}{p})$ covers the complete shape of gesture. Here we considered the circle for representing shape of an image where ($r \leq 1$) and $p_0$ and $q_0 = 0.5$. Then the wavelet basis function $\vartheta^{p,q}(r)$ has been again represented as:

$$\vartheta_{m,n}(r) = 2^{\frac{m}{2}} \vartheta(2^m r - 0.5n)$$

Here $\vartheta_{m,n}(r)$ defines for any orientation along the radial axis r. It is applied for finding the local as well as global features of hand by varying the values of m and n.

After that we define moment invariant wavelet feature vector as:

$$\|G_{m,n,b}^{wavelet}\| = \|\int f_b(r) \cdot \vartheta_{m,n}(r) r \, dr\| \quad (5.8)$$

Comparing equation (5.6) and (5.8) we get $g_a(r)=\vartheta_{m,n}(r)$ and $f_b(r) = \int G(r,\theta)e^{jb\theta}d\theta$.

$\|G_{m,n,b}^{wavelet}\|$ is the wavelet transform of $f_b(r)$. It analyses the signal in both time domain as well as frequency domain and extracts features which are locally descriptive in nature. Features shown in equation (5.8) are moment invariant for each gesture with feature vector $\|G_{m,n,b}^{wavelet}\|$. Where m=0, 1, 2, 3 and n= 0, 1…$2^{m+1}$.

### 5.1.4 Union of Three Features

All three characteristics are combined using 3 dimensional feature mapping. First we combine orientation and moment feature using x-y feature mapping. Then speed S is further combined with x-y and make it x-y-s feature map. For this feature mapping normalization is necessary. It makes all three features in the same range between 0.0 and 1.0.

Normalization of moment feature: Moment features of hand changes according to the size and shape of the gesture. It is being normalized as:

$$G_{max} = max_{t=1}^{n} (G_t) \tag{5.9}$$

$$g_t = \frac{G_t}{G_{max}} \tag{5.10}$$

$g_t$ is the normalized moment, $G_{max}$ is the maximum moment from the center and $G_t$ is the moment of hand at various time interval.

Normalization of orientation: orientation θ is normalized by using chain code. Chain code has been represented either using 4-connecitivity or 8-connectivity. Where point start at 0 and as go on to the right this number increases accordingly. Here we used 8-connectivity chain code.

$$D_{x1} = X_t - ce_x \qquad D_{y1} = Y_t - ce_y$$
$$D_{x2} = X_t - X_{t+1} \qquad D_{y2} = Y_t - Y_{t+1}$$
$$\theta_1 = a\,tan2\,(D_{y1}, D_{x1}) \qquad \theta_2 = a\,tan2\,(D_{y2}, D_{x2})$$
$$\theta_{1t} = (Num\,of\,chain - (\,\theta_1/(pi/4) + 0.5 + (\,D_{y1} \tag{5.11}$$
$$< 0) * Num\,of\,chain))\%Num\,of chain$$

$$\theta_{2t} = (Num\,of\,chain - (\,\theta_2/(pi/4) + 0.5 + (\,D_{y2}$$
$$< 0) * Num\,of\,chain))\%Num\,of chain$$

Where $\theta_{1t}$ and $\theta_{2t}$ are the normalized orientation parameter.

Normalization of speed: speed is normalized by calculating the maximum speed at which the gesture is performed.

$$S_{max} = max_{t=1}^{n}(S_t) \quad (5.12)$$

$$s_t = \frac{S_t}{S_{max}} \quad (5.13)$$

Here $S_{max}$ is the maximum speed, $S_t$ is the speed at time t and $s_t$ is the normalized speed at time t. This value is also lies between 0.0 and 1.0.

Then these features are combined in x-y-s Cartesian coordinate system and get the hybrid feature.

### 5.1.5 Classification

These features are classified using possibility theory [189, 190]. In possibility theory, we measure both possibilities as well as a necessity instead of one. These two functions (possibility and necessity) handle all the three problems, uncertainty, imprecision and accuracy instead of probability which only deals uncertainty. Here possibility can be measured in terms of probability defined as:

$$\prod(C_i/s) = \frac{P(s/C_i)}{max_{class\ t}(P(s/C_t))} \quad (5.14)$$

Where $P(s/C_i)$ is the probability density function and it has been calculated using multivariate Gaussian distribution.

$$P(s/C_i) = \text{AGRMAX}\left(\frac{1}{(\sqrt{2\pi})^m |\Sigma_{C_i}|^{1/2}} exp\left\{-\frac{1}{2}(s-\mu_{C_i})' \Sigma_{C_i}^{-1}(s-\mu_{C_i})\right\}\right) \quad (5.15)$$

Where P is the probability distribution over class $C_i$. s is the feature vector of test data. $\mu_{C_i}$ is the mean of each class $\Sigma$= Co-variance, $|\Sigma|$ =determinant of Co-variance, m is the dimension of given data. Here 3 dimensional feature vector is used. From equation (5.15) we calculate the possibility of each of the unknown (probe) gesture ge lies with in a class $C_i$. Here t represents the class number equal to 20.

### 5.1.6 Text to speech conversion:

After recognizing a particular gesture by an NAO robot a text formation has been performed. Matching is performed through a knowledge base of the NAO robot using shortest distance techniques shown in Figure 5.2.

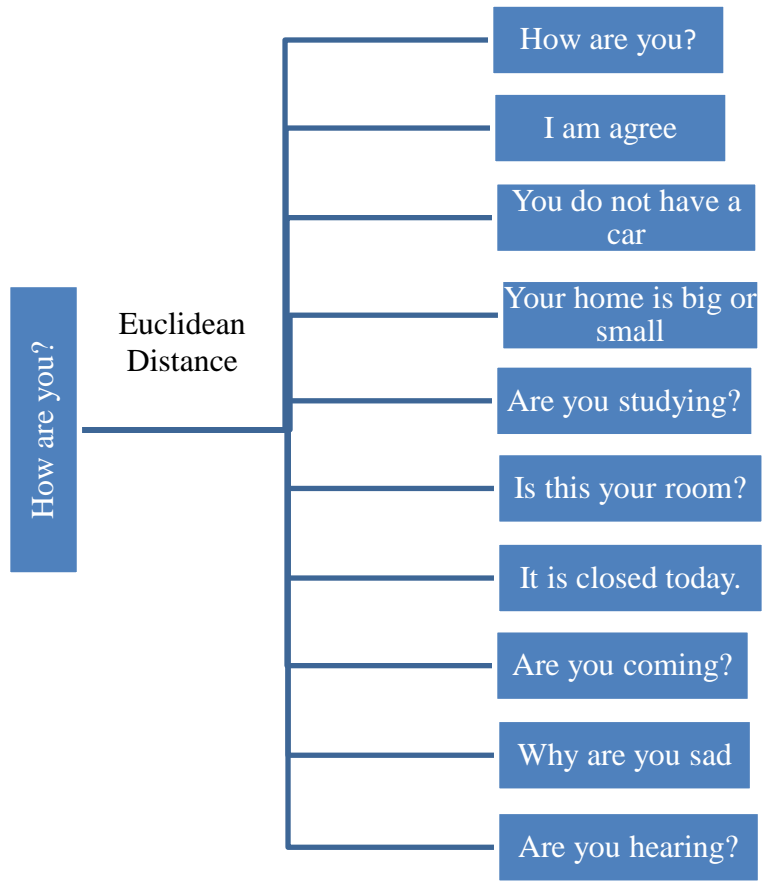

Figure 5.2 Matching with knowledge database of NAO

In this figure we have shown the matching process only for ten gestures. A similar process will be done for all the twenty sentences. These text commands are further converted into speech/gesture format using NAO text to speech or gesture conversion module. All these modules are available in NAO API's which we used directly through MATLAB. For speech conversion NAO has voice synthesizer and 2 speakers.

### 5.2 Experimental Results and Analysis

We have created a data set of twenty sentences of continuous Indian sign language, gestures using the NAO's vision system which consists of two cameras fixed in their head [201]. Each camera having resolution 640×480 with 30 frames per second. Videos of different continuous gestures are captured in different backgrounds like black, red, multiple object etc. with black full sleeves dress. Here we concentrate only on the upper half of the body. The angle of view of NAO robot has been shown in Figure 5.3.

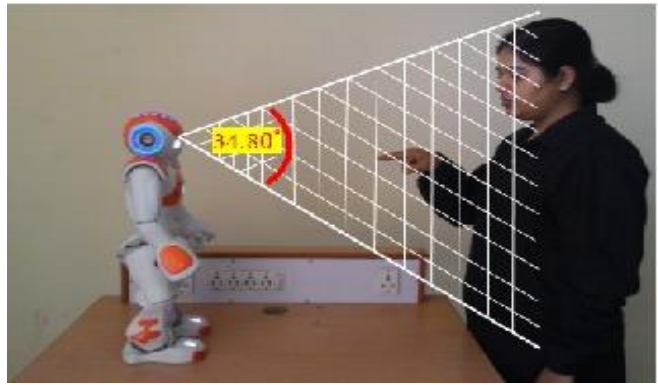

Figure 5.3 Angle of view of NAO robot

The proposed framework experiments on 10 subjects having 10 samples of each gesture. Total number of samples we have analyzed are 2000 where 50 percent will be used for training and 50 percent will be used for testing. Experiments are also done in real time environment. The database of continuous ISL gestures is listed in Table 5.1.

Table 5.1 List of Gestures

| Sr. No. | Sentences | Sr. No. | Sentences |
|---|---|---|---|
| 1 | How are you? | 11 | Is the child studying |
| 2 | I am agree | 12 | What are those? |
| 3 | I am studying | 13 | We miss you very much |
| 4 | This is your own room? | 14 | What do you do? |

| 5 | It is closed today. | 15 | I have finished my work |
| 6 | Are you coming? | 16 | Just a minute |
| 7 | Why are you sad? | 17 | The child is very weak |
| 8 | You do not have a car? | 18 | May god bless you |
| 9 | I am very lucky today. | 19 | Come in please |
| 10 | Are you hearing? | 20 | Please switch off the light |

Table 5.1 shows the list of 20 continuous gestures having thirty six individual gestures. After data acquisition, preprocessing has been performed using background modelling technique. Results obtained after background modelling are shown in Figure 5.4.

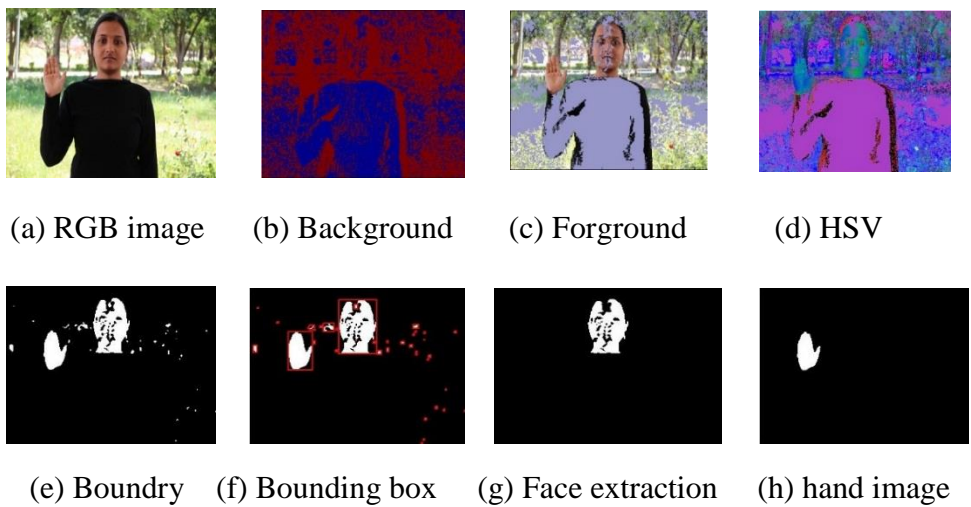

(a) RGB image    (b) Background    (c) Forground    (d) HSV

(e) Boundry    (f) Bounding box    (g) Face extraction    (h) hand image

Figure 5.4: Pre-processing steps of each RGB frame

Figure 5.4 shows, how background modelling has been performed for segmenting the hand from the upper half of the body. This method is very effective in different background conditions. After hand segmentation gesture splitting is done where continuous gestures are separated into isolated gestures using the gradient method which is explained in section 5.1.2. Results of splitting of gesture are shown in Figure 5.5.

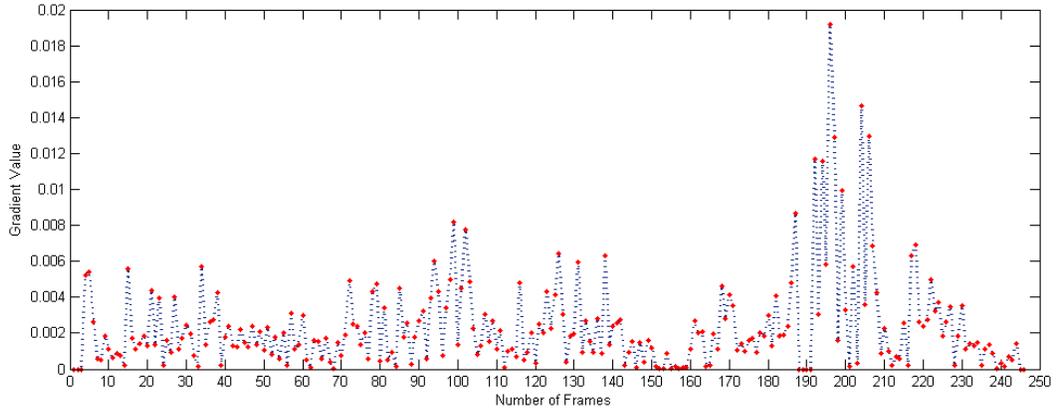

Figure 5.5: Overlapping frame of "I agree" gesture

Figure 5.5 shows splitting of I agree gesture i.e. how continuous gesture (I agree) has been split out into isolated gestures. Here graph shown in Figure 5.5 represents overlapping frames where gradient value is zero shows the end of one gesture and the start of another. In Figure 5.5 we have seen that gesture1 (gesture I) starts in between the frame $0^{th}$ to frame $4^{th}$ and end at in between the frame $254^{th}$ to frame $258^{th}$. From $0^{th}$ to frame $4^{th}$, there is a constant gradient value and similarly from $254^{th}$ to frame $258^{th}$ we obtained constant gradient value which shows the end of one gesture or the start of another gesture. Suppose if gesture1 ends at frame 256th than this is the commencement of next gesture. By this way we can calculate the total number of frames in each gesture. We consider 15 middle frames of each isolated gesture. Various features of hand gestures are extracted and combined to get hybrid feature. These hybrid features are helpful in recognizing gestures in different environmental conditions.

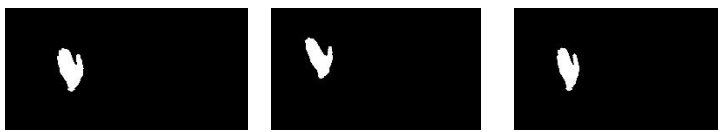

Figure 5.6: Different orientation of "coming" gesture

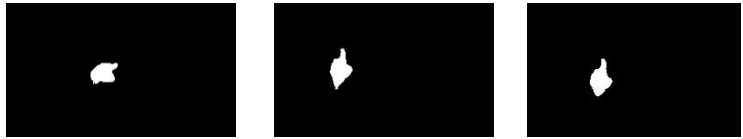

Figure 5.7: Different orientation of "you" gesture

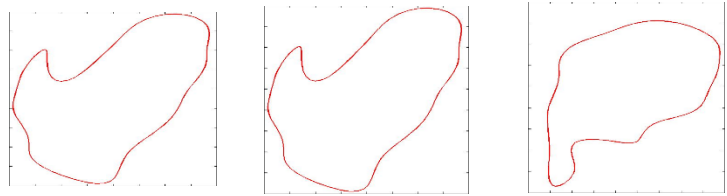

Figure 5.8: Contour of "coming" and today gesture

Several moments, orientation and its contour of "coming", "today" and "you" gesture are shown in Figure 5.6, 5.7 and 5.8. Hybrid features are brought forth by aggregating all the three characteristics (speed, orientation and instant). Further, it is classified using possibility theory. The comparative results between possibility theory, HMM and PTBHMM are shown in Table 5.2.

Table 5.2 Comparison between different classifiers on the basis of percentage of classification

| Sentences | Possibility Theory (%) | Hidden Markov Model (%) | Possibility Theory Based HMM (%) |
| --- | --- | --- | --- |
| How are you? | 88 | 85 | 85 |
| I am agree. | 91 | 91 | 91 |
| Are you coming? | 92 | 90 | 90 |
| I am studying. | 93 | 93 | 94 |
| It is closed today. | 89 | 90 | 91 |
| Are you hearing? | 89 | 92 | 90 |
| You do not have a car? | 81 | 85 | 86 |
| Your home is big or small? | 81 | 88 | 84 |

| Why are you sad? | 92 | 91 | 90 |
| Is this your own room? | 88 | 90 | 92 |

After analyzing the results we observe that possibility theory, HMM and PTBHMM gives the similar amount of accuracy. But possibility theory and PTBHMM handles the ambiguity problem present in the dataset effectively. The average classification accuracy, we achieved is 92 percent, which is higher the accuracy achieved in paper [130] which is 89%.

Also, we have compared the processing time of all the three classifier we have used in this chapter and found that PTBHMM takes much less time than possibility theory and classical HMM which is very helpful in real time applications.

Table 5.3 Average processing time of different classifier

| Method | Average processing time (Sec) |
|---|---|
| HMM | 28.92 |
| Possibility Theory | 20.65 |
| PTBHMM | 16.982 |

We also observe that the misclassification rate and average processing time of possibility theory is higher than the PTBHMM but lesser then the classical HMM. Classified gesture sentences are further matched with the knowledge database of NAO robot. It increases the accuracy about 96%, which is 4-5 percent higher in nature. Finally, these texts are converted either in speech format or in gesture format, so that communication becomes easier. The complete system is shown in Figure 5.9.

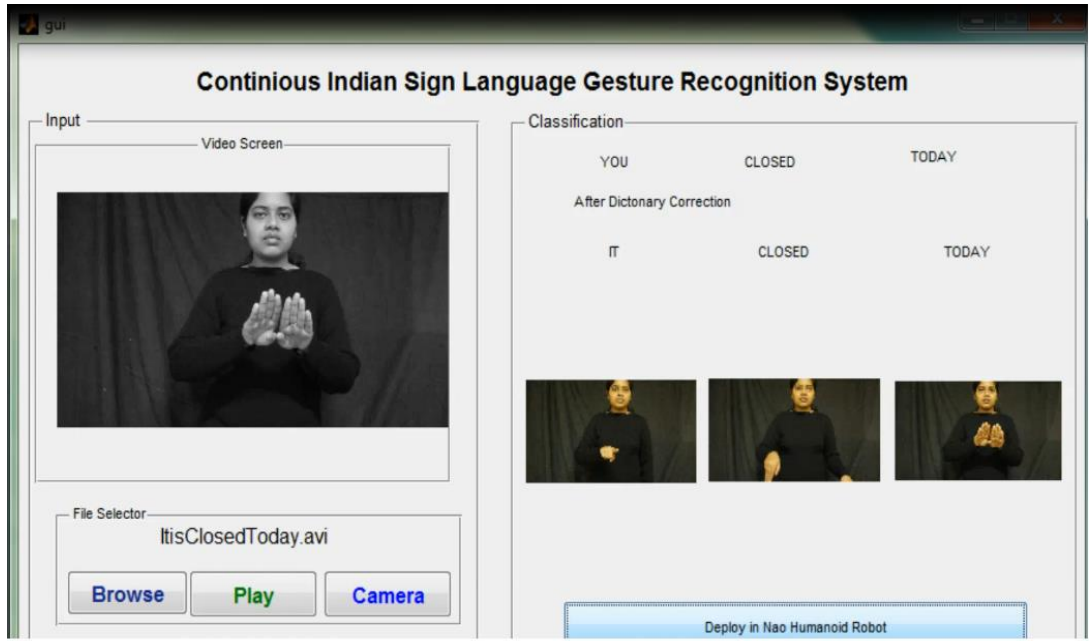
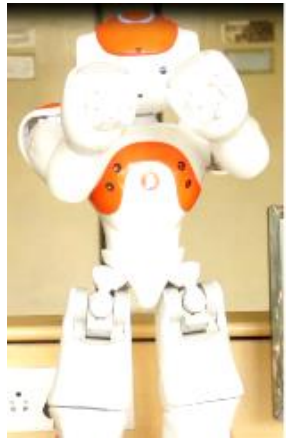
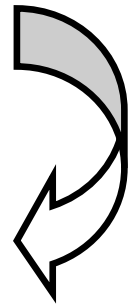

Figure 5.9 MATLAB framework of complete gesture recognition system

## 5.3 Important Findings of This Chapter

In our everyday conversations we used sentences and instructions. These sentences are made up of words using grammar. In a similar way hearing impaired persons also used sentences for communication, but in a different form called gestures. For understanding these continuous gestures a novel NAO humanoid robot based framework has been proposed which is helpful for hearing impaired society. The robot first understands the continuous gestures and then interpret into the text format. The knowledge database of the NAO is further used for matching which improves the recognition rate. Matched continuous gestures are converted into speech or gesture format. In gesture recognition, gesture segmentation and gesture spotting are the two major issues which have been solved in this proposed work where background modelling is applied to hand segmentation from upper half body image and then the gradient method is introduced for finding informative gestures. During this process some of the information like orientation, movement etc. of hand gesture are changed or missed. Our proposed framework based on hybrid features extracted using wavelet descriptor improves feature quality. Holding invariance property against translation, rotation, scaling and movement, our technique works well in real time environment with possibility theory and PTBHMM which is much more effective than probability theory because of uncertainty and imprecision handling property. All the experiments are executed on our own dataset created using NAO vision sensors. In this dataset some of gestures are performed either using single hand or both the hands or are also inactive as well as dynamic in nature. After performing all the experiments we found that our proposed algorithm gives 92% initial average accuracy of the continuous ISL dataset has been described in chapter 6 which is quite appreciable. A further extension of our work can lead to effective bidirectional human robot interactions in which robot will respond according to human's command or perform the task given by deaf and dumb people.

# Chapter 6

# Multimodal Human Robot Interaction

In the last three chapters we have discussed about the vision based human robot interaction. There are other modes speech and multimodal also used for establishing the communication between human and robot. Among two speech is the most common medium of interaction. Any human being can use it in her/his day to day life. Speech based interaction [142, 146] has been performed using natural human voice and also has many difficulties like noise, behavior of human, ascent of spoken word etc. Identification of an individual speech considers the variation present in a human voice. Because every person has its unique way of speaking like some persons speaks very quickly, some of them speaks slowly etc. Therefore, the objective of this chapter is to develop a robust speech recognition system which is very prominent in recognizing each word correctly. Where each word is articulated by different speakers.

For recognizing Hindi speech, we have proposed two approaches one is HTK toolkit based Hindi speech recognition system and another one is DWT with HFCC based speech recognition system. Both the methods have been tested on Hindi words like ek, do, sawan, bhado etc. The collection of 100 words has been taken into account. All these words are recorded by different speakers of Robotics and Artificial Intelligence Laboratory, Indian Institute of Information Technology, Allahabad, India using audacity software. This software is well recognized by the various researchers working in the field of speech. Both the methods have been tested on Hindi as well as on an English speech dataset and then comparative analysis has been performed. After comparing the results, we found that DWT with HFCC methods works well in comparison to HTK and also handling of DWT with HFCC methods is easier than HTK toolkit. Any single mode in not sufficient for the establishment of good interaction between human and robot because each mode has its own limitations which can be minimized through multimodal mode of interactions. It is a mode of interaction in which two or more mediums are combined and get the final decision. This method increases the overall accuracy of the system significantly. After obtaining the final decision the gesture will be performed by the HOAP-2 humanoid robot.

**6.1 HTK Toolkit Based Speech Recognition Technique**

HTK toolkit [202] has been developed by the Machine Intelligence Laboratory of Cambridge University Engineering Department (CUED). It is an HMM based toolkit primarily developed for English speech recognition. It simply consists of some tools and libraries written in C source code. Using these tools one can implement Hidden Markov Model in any of the applications like speech recognition, speech synthesis, DNA sequencing and character recognition.
The general framework of ASR using the HTK toolkit has been shown in Figure 6.1.

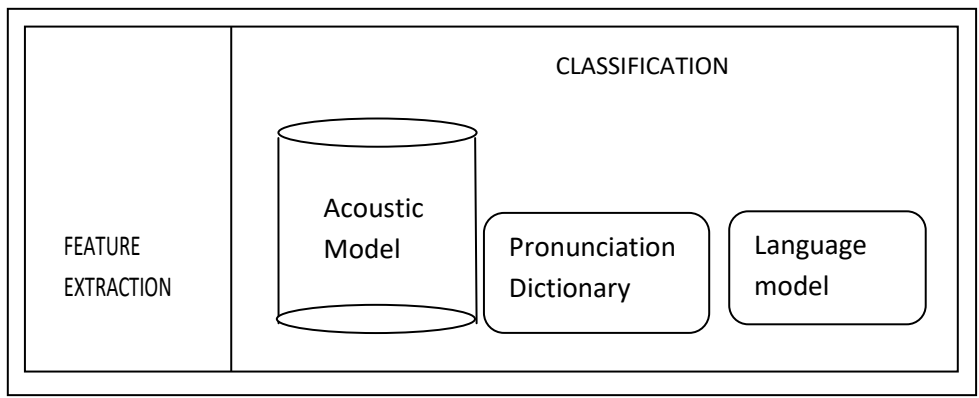

Figure 6.1 Automatic Speech Recognition (ASR) Overview

This toolkit consists of various modules like feature extraction, language generation etc. It is based on bigram language model in which every sequence is a combination of two adjacent sequences. These adjacent sequences are either letters, phonemes or words. Past literature shows that HTK toolkit is very effective in recognizing English speech, therefore we think that we can apply the HTK toolkit for Hindi Speech recognition because Hindi words are more complex than English words. Hindi words are very difficult in writing as well as in speaking and also there is no standard phonemes are there which is acceptable by the researcher like English. In Hindi speech everyone declares its own phonemes. Therefore, it is very difficult to process such speech signals.

### 6.1.1 Proposed Methodology:

Complete framework which recognizes Hindi speech using HTK toolkit is represented in Figure 6.2.

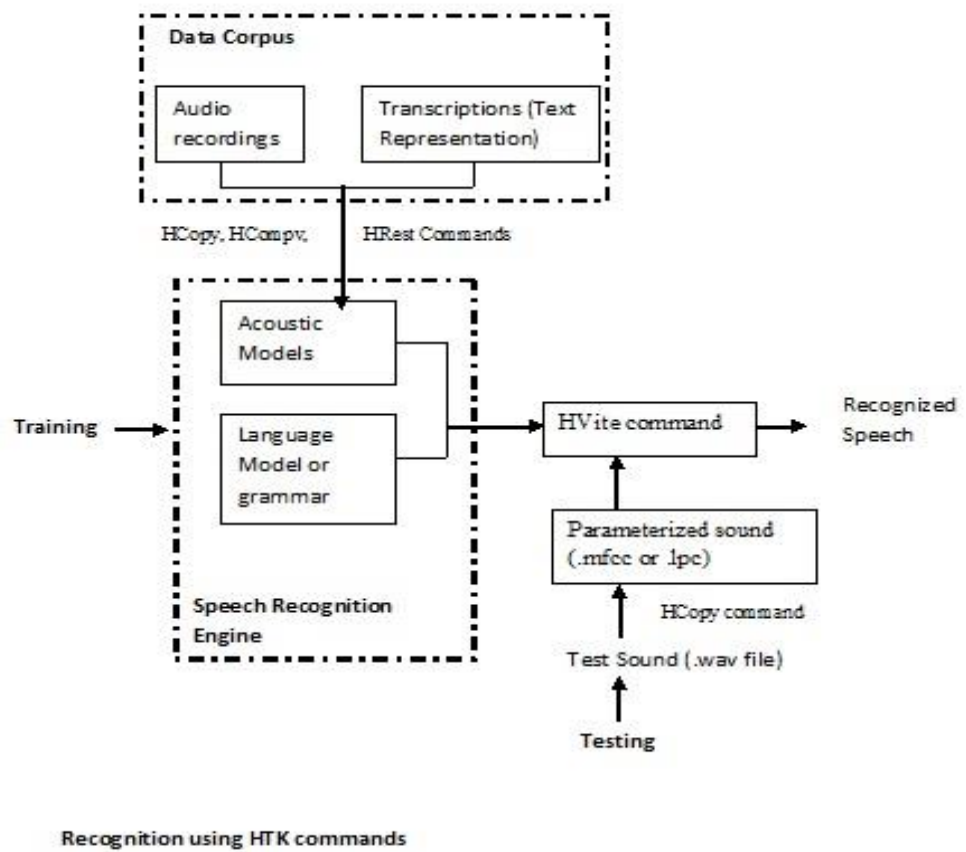

Figure 6.2 Methodology for speech recognition using HTK commands

There are three main modules of this HTK toolkit [207] one is database generation and transcriptions, second one is training where feature extraction and language model or grammar has been created. Final module is the testing module where we give an unknown speech, from this unknown speech, important features have been extracted and classify using some command. The detailed description of each of module has been discussed in the latter on.

### 6.1.2 Data Corpus Preparation:

A train and test data set with some voice samples has been created. For sound recording and sound processing purposes a well-known application, Audacity [203] has been used. Multiple sounds can be exported after being recorded all at a time. The sound files recorded are exported to the wav folder residing in the data folder, while the labeled files of the sound files are kept in a separate directory named label, which also resides in the same directory named data. First, the sound samples have been recorded and then after the segments of the sounds have been labeled using Audacity. A train database with sound samples of 35 words with 5 samples per word has been created. The voice samples have been taken from 2male and 3 female speakers. Thus, the train database consists a total of (35 x 5 x 5 =) 875 sound samples. Similarly, there is a test database with voice samples from all of the 5 speakers with one test sample per word per speaker. Thus, the test dataset consists of (35 x 1 x 5 =) 175 sound samples, having one sample sound per speaker of all of the 35 words, in the case of speaker dependent environment, while in the case of speaker independent systems, the test dataset consists of (35 x 1 x 1=) 35 sound samples of only one speaker who is not present in the train dataset.

### 6.1.3 Feature Extraction:

We have a targetlist.txt file which consists of a list, each entry of which is the source file in .wav format and the target file in either .mfcc format or .lpc format, whichever is required.

We also have a file analysis.conf, which is a configuration file that consists of the parameters required for the feature extraction phase.

These two files targetlist.txt and analysis.conf are taken as input to the HCopy command to extract the features. The feature files are written in the target file, prescribed in the file targetlist.txt.

### 6.1.4 Creating the Acoustic Models:

After that we create an acoustic model for all of the 35 words using HInit command. This command has been used to initialize the parameters of the HMM of each word in the database.

HCompv is used to initialize the HMM parameters of the words to global means and variances.

Then after, HRest is used for re-estimation of the HMM parameters again for all of the 35 words.

Re-estimation is done three times for each of the word. The final hmm models for the words are kept in the directory hmm3.

### 6.1.5 Creating the Language Model

For creating a language model we define some rules in the gram.txt file to be followed by the words in the system. HParse function has been used to create a word lattice by taking gram.txt as input, while output is written to net. slf.

HSGen has been used to produce different sets of words according to the rules described under gram.txt using the word lattice existing in net.slf, which has been used as input for the command.

After the language models and the acoustic models of the words have been created, and the pronunciation dictionary is there, HVite function has been used for the recognition of the words. The inputs to the command HVite are hmmsdef.mmf, net.slf, dict.txt, hmmlist.txt files, among which hmmsdef.mmf is the master macro file, that consists of the HMM models of the words and the rest of the files have the same meaning as previously described. We give the test files in the testfile.txt and the HVite command gives the output recognized as a transcription file in rec.mlf.

### 6.1.6 Experimental Results and Analysis:

The system has been tested in two different environments- "Speaker Dependent" and "Speaker Independent". For both types of environments system performances have been computed using both of the Feature Extraction Techniques. HMM using the HTK Toolkit has been implemented as a classifier. Results for the system have been computed using both of the techniques for the training database with different number of speakers. Again the accuracy is computed as

$$Accuracy = \frac{(number\ of\ correctly\ identified\ test\ samples)}{total\ number\ of\ test\ samples\ taken} \times 100$$

The performances of the system have been evaluated on the dataset consists of sound samples from different speakers having the same number of sound samples per speaker.

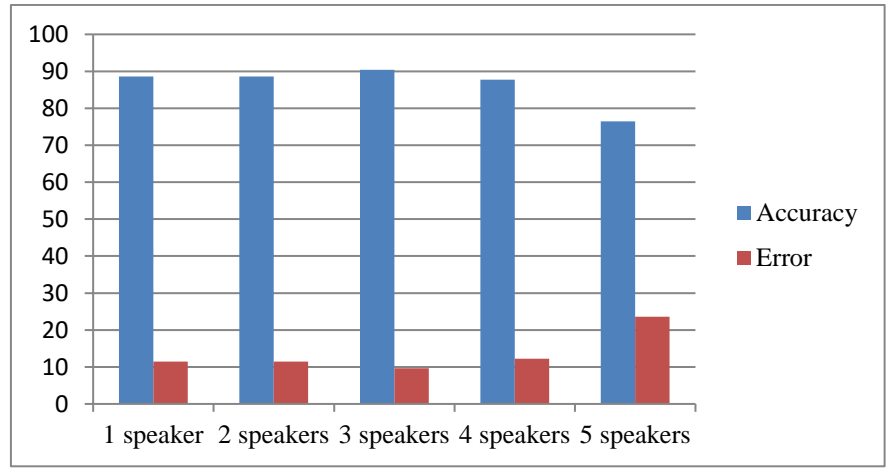

Figure 6.3 Diagram showing system performance in Speaker Dependent Environment using MFCC

The bar chart shown in Figure 6.3 demonstrates that initially the accuracy of the system is in between 88% to 90% with 1 to 5 speakers, which has been decreased to 76.44% when the number of speakers is increased to 5 speakers.

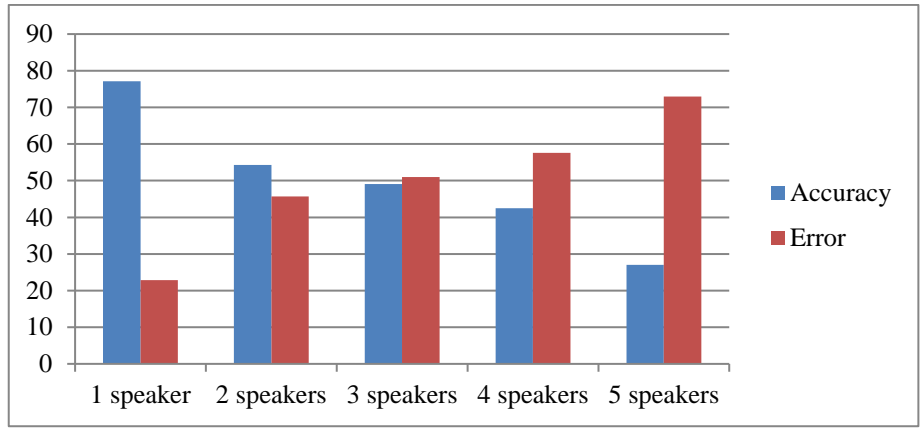

Figure 6.4 Diagram showing system performance in Speaker Dependent Environment using LPC

Diagram presented in Figure 6.4 is the depiction of the system performance with sound samples of varying number of speakers, in a speaker dependent environment using LPC as feature extraction technique. Again, it is clear from the demonstration that initially when there are the sound samples of only one speaker in the database, the accuracy is maximized with the value, 77.14%, which eventually decreases as the number of speakers in the train database increases. From the graph it can be seen that, it is 54.28% for the 2 speaker dataset, 49.038% for the 3 speaker dataset, 42.45% for the 4 speaker dataset and finally minimum with 27.02% for the 5 speaker dataset.

The performance of Speech Recognition System using the MFCC feature extraction technique in Speaker Independent Environment is shown in Figure 6.5 which shows that the performance of the system degrades in case of speaker independent environment. This HTK toolkit [202] doesn't sustain the recognition accuracy, when the training speakers are different from the testing speakers.

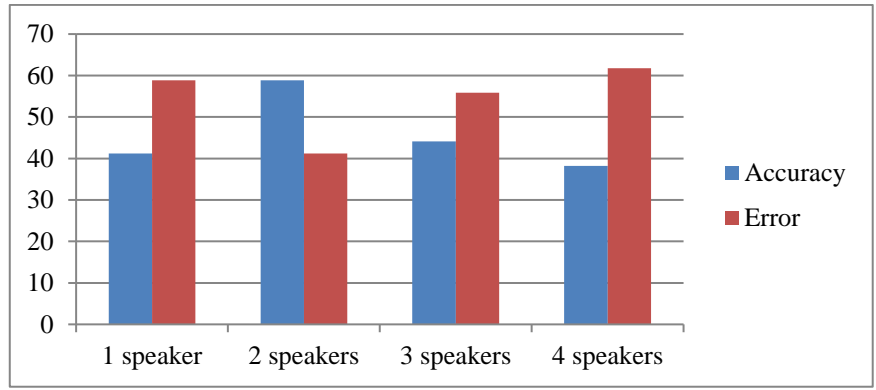

Figure 6.5 System performance in Speaker Independent Environment using MFCC

In case of LPC features in speaker independent system, it is seen that the performance goes down gradually with increased number of speakers in the database. The system performance deteriorate in this case which is shown in the Figure 6.6.

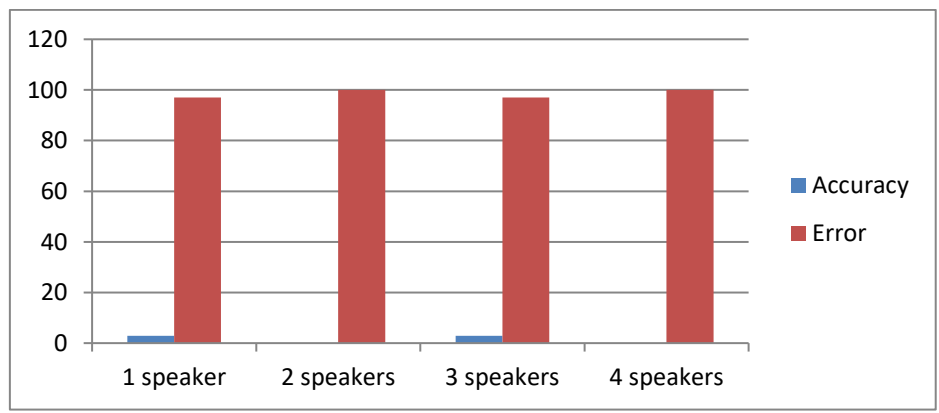

Figure 6.6 System performance in Speaker Independent Environment using LPC

It is very clear from the graph shown in Figure 6.7 that the system performs very well in a speaker dependent environment using MFCC as feature extraction technique. It gives the worst performance in a speaker independent environment using the LPC feature extraction technique. The system initially gives good performance in a speaker dependent environment using LPC as compared to speaker independent environment using MFCC technique, which afterwards leads to somewhat similar performances. In LPC, each sample of the speech signal is represented as a linear combination of the preceding samples of the signal whereas MFCC follows the human peripheral auditory system which handles the nonlinearity of the auditory system. Therefore, performance of MFCC is better than LPC.

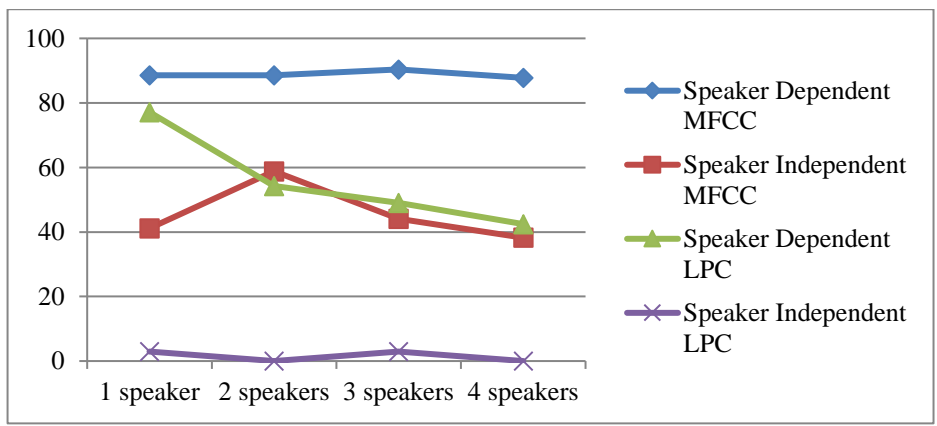

Figure 6.7 Performance comparison between Speaker Dependent (MFCC, LPC) and Speaker Independent (MFCC, LPC) system

### 6.1.7 Conclusion

A novel HTK toolkit based Hindi speech recognition system has been proposed where the cepstral coefficients of the speech signal has been extracted using MFCC and LPC feature extraction techniques. We have created our own dataset having 875 voice samples of 5 different speakers in the training as well as the testing database consists of 175 voice samples. Experiments have been performed in two different conditions one is Speaker Dependent and another one is Speaker Independent environments. In this way there are following 4 cases which has been considered: Speaker Dependent environment with MFCC, Speaker Independent environment with MFCC, Speaker Dependent environment with LPC, Speaker Independent environment with LPC.

After rigorous analysis of all the four cases with in house dataset we found that in each case, the accuracy gets decreased with the increased number of speakers in the training database. Comparing all the cases with each other we conclude that MFCC works better than LPC in all the cases and the system performs very poorly in a speaker independent environment using LPC as feature extraction technique. MFCC represents frequency spectrum in perceptual domain which is closely related with the way that human can hear whereas LPC represents frequency spectrum using linear coding which is not suitable when signal varies very frequently. The major limitation of this work is that it only supports MFCC and LPC feature extraction technique. Also, we cannot modify all these feature extraction techniques. Its interface requires a Linux Type environment which is not user friendly because the HTK toolkit originally developed for Linux, only compatible versions are available for windows. This compatible version has so many issues which creates problem throughout the implementation. To minimize the problems related with HTK toolkit, we have come up with a novel technique which has been explained in section 6.2.

**6.2 DWT with HFCC based Speech Recognition Technique**

We have proposed a novel speech recognition technique where DWT [208] [213][214] is used for reducing the noise present in the speech signal before HFCC features are extracted. Schematically the proposed method is shown in Figure 6.8.

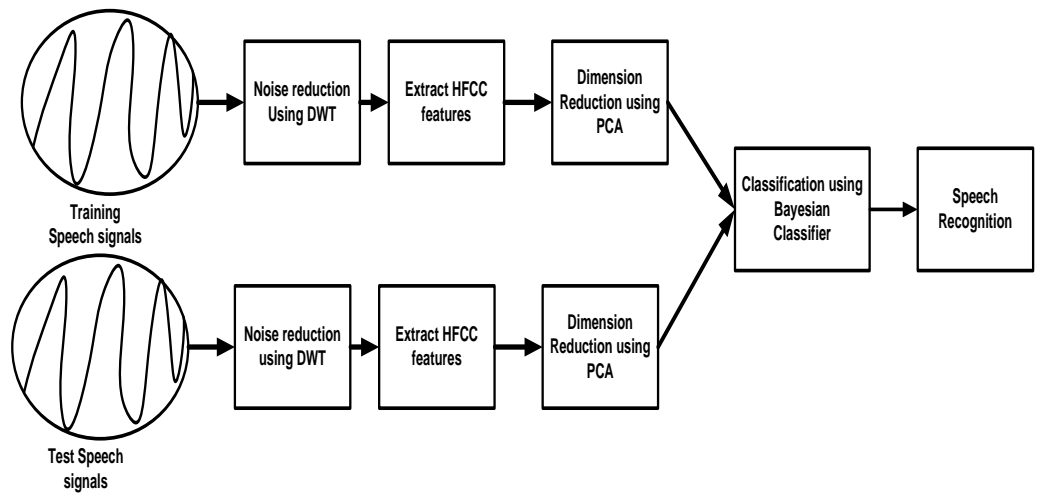

Figure 6.8: Proposed Speech Recognition Technique using Bayesian Classifier

### 6.2.1 Database:

We have collected the database of isolated Hindi words. Twelve words of Hindi months are recorded. Each word is recorded six times by six people, two are males and four are females. Each word is recorded using the audacity software at 16000 Hz frequency of mono signals. Four words are used for training and the two words are used for testing purposes. Plot of the original speech signal of SAWAN, there decomposition process and the signal after decomposition is shown in Figure 6.9, 10 and 6.11.

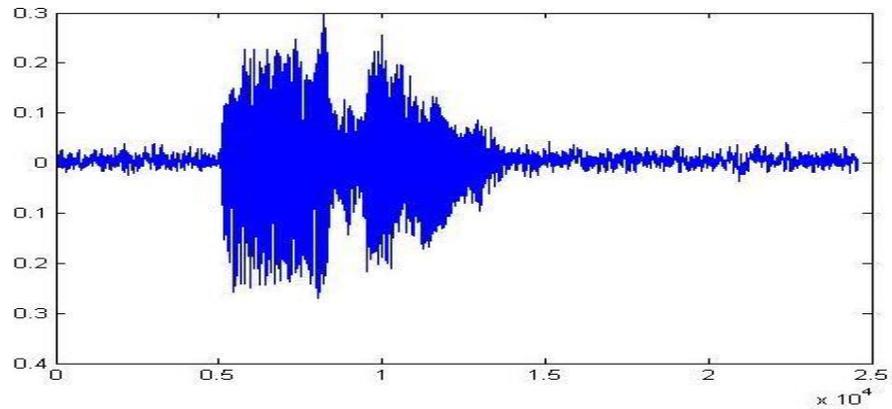

Figure 6.9: Original speech signal (SAWAN)

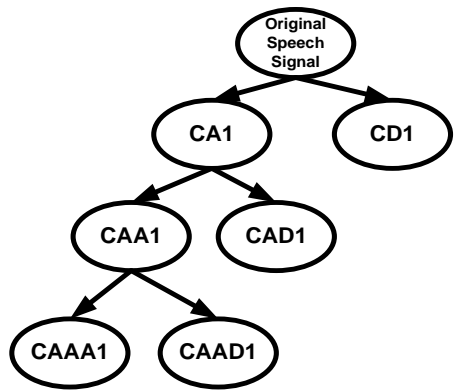

Figure 6.10: Third level decomposition of an audio signal using DWT

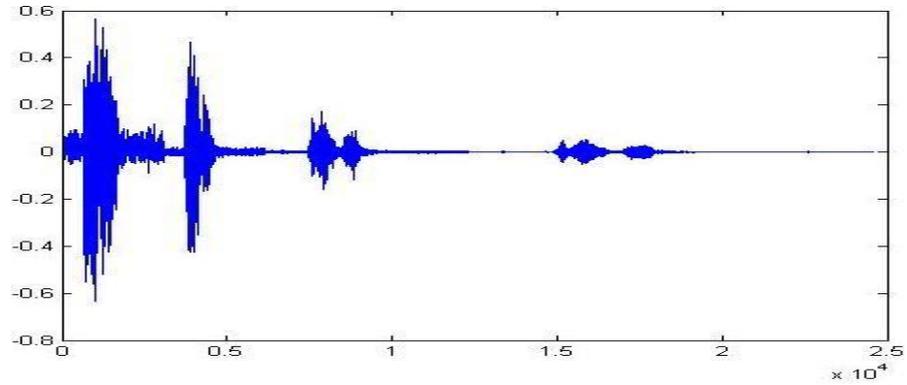

Figure 6.11: Speech signal after DWT decomposition (SAWAN)

### 6.2.2 Noise Reduction:

Discrete wavelet transform [143, 152] is applied for reducing the noise present in the original audio signal. In DWT the audio signal is decomposed up to L level of approximate and detail coefficients where L is any integer number. The approximate and detail coefficients are obtained by using the following equations:

$$CD(j) = \sum_i s(i)g(2j - i) \qquad (6.1)$$

$$CA(j) = \sum_i s(i)h(2j-i) \qquad (6.2)$$

Where s(i) is the original speech signal, $1 \leq i \leq N$, where N is the number of sample values, CD(j) is the detail coefficients and CA(j) is the approximate coefficients calculated using high pass(g) and low pass(h) filter. The approximate coefficients are the low frequency components, which are highly sensitive in nature, therefore, it can be easily detected by the human auditory system. The detail coefficients are the representation of the high frequency components, which are not very sensitive in nature therefore it is difficult to detect. The decomposition up to the third level of DWT of the audio signal is shown in Figure 6.10.

In the Figure 6.10 we have seen that DWT decomposes the sample values of audio into the low frequency components (approximate coefficients CA1) and the high frequency components (detailed coefficients CD1). After that approximate coefficients are further decomposed with the help of DWT into approximate coefficients (CAA1) and detail coefficients (CAD1). The number of decomposition level depends on the application. After nth level decomposition the number of approximate and detail coefficients are 2n. Then original signal is reconstructed using inverse discrete wavelet transform (IDWT). As shown in Figure 6.12 DWT has multi resolution property which means that at higher frequency, DWT provides good time resolution and poor frequency resolution on the contrary, at lower frequencies it provides a good frequency resolution and poor time resolution. Therefore DWT analyses the complete signal in both time and frequency domain efficiently. Above two properties enable DWT good perceptual transparency for an audio signal.

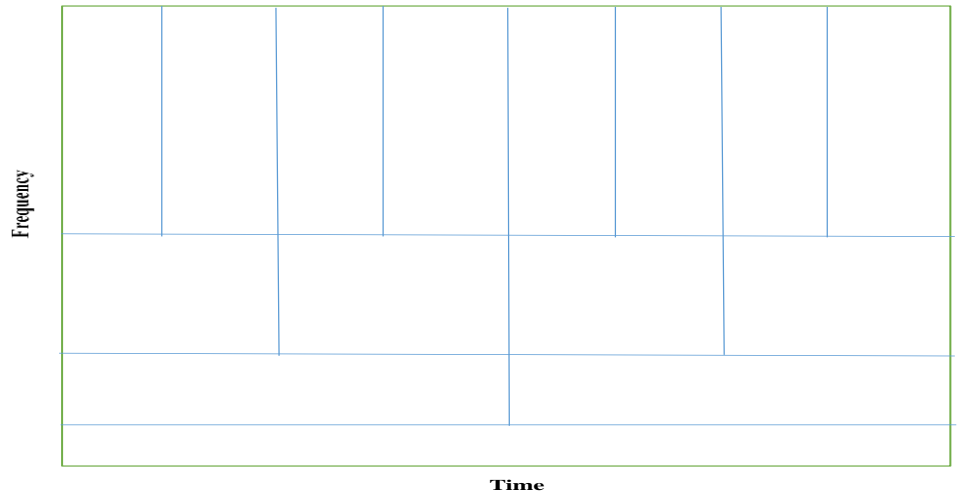

Figure 6.12: Multiresolution property of DWT

In this approach, we have applied a (normalized) Daubechies filter [20] which has been explained by assuming the signal having 8 sample values.

$$c = (c(0), c(1), c(2), c(3)) = (\frac{1+\sqrt{3}}{4\sqrt{2}}, \frac{3+\sqrt{3}}{4\sqrt{2}}, \frac{3-\sqrt{3}}{4\sqrt{2}}, \frac{1-\sqrt{3}}{4\sqrt{2}})$$

as a low pass filter and

$$d = (d(0), d(1), d(2), d(3)) = (c(3), -c(2), -c(0))$$

as a high pass filter. A normalized Daubechies filter is used for providing smoothness in original speech signal. Here smoothness means the number of non-zero coefficients are much higher than the number of zeros so that the signal becomes continuous and easy to handle. It looks very similar to sin function. $\frac{1}{\sqrt{2}}$ is a normalization factor, so that $||c2|| = ||d2|| = 1$. Here we get four high pass and four low pass filter coefficients because of filter length of size four.

$$S = (s(0), s(1), s(2), s(3), s(4), s(5), s(6), s(7)),$$

Where s is the number of sample values.

The steps of DWT process are as follows:

Step1: Number of sample values of the original signal are increased periodically by applying padding process. We have applied a periodic extension to perform padding in speech signal because most of the speech signal are periodic in nature. Extended signal after padding has been denoted by tilde ($\tilde{S}$).

$$\tilde{S} = (s(6), s(7), s(0), s(1), s(2), s(3), s(4), s(5), s(6), s(7))$$

Low and High pass filter coefficients are added and shifted two steps down:

$$s_1 = \begin{pmatrix} c(0)s(6) + c(1)s(7) + c(2)s(0) + c(3)s(1) \\ c(0)s(0) + c(1)s(1) + c(2)s(2) + c(3)s(3) \\ c(0)s(2) + c(1)s(3) + c(2)s(4) + c(3)s(5) \\ c(0)s(4) + c(1)s(5) + c(2)s(6) + c(3)s(7) \\ d(0)s(6) + d(1)s(7) + d(2)s(0) + d(3)s(1) \\ d(0)s(0) + d(1)s(1) + d(2)s(2) + d(3)s(3) \\ d(0)s(2) + d(1)s(3) + d(2)s(4) + d(3)s(5) \\ d(0)s(4) + d(1)s(5) + d(2)s(6) + d(3)s(7) \end{pmatrix}$$

Step2: In this step last four coefficients are fixed and unchanged while first four coefficients are written in a linear combination of coefficients obtained from low and high pass filters and vector $s_1$. To perform this step it is necessary to extend the first half of the vector $s_1$ periodically because throughout this process we assume that our signal are periodic in nature.

$$\widetilde{s_1} = (s_1(2), s_1(3), s_1(0), s_1(1), s_1(2), s_1(3), s_1(4), s_1(5), s_1(6), s_1(7))$$

$$s_2 = \begin{pmatrix} c(0)s_1(2) + c(1)s_1(3) + c(2)s_1(0) + c(3)s_1(1) \\ c(0)s_1(0) + c(1)s_1(1) + c(2)s_1(2) + c(3)s_1(3) \\ d(0)s_1(2) + d(1)s_1(3) + d(2)s_1(0) + d(3)s_1(1) \\ d(0)s_1(0) + d(1)s_1(1) + d(2)s_1(2) + d(3)s_1(3) \\ s_1(4) \\ s_1(5) \\ s_1(6) \\ s_1(7) \end{pmatrix}$$

Step3: For calculating first two rows of $s_3$, vectors need to be extended by padding first two elements. Here in vector $s_3$, last six coefficients have been fixed and unchanged; only first two rows of $s_3$ are calculated.

$$\widetilde{s_2} = (s_2(0), s_2(1), s_2(0), s_2(1), s_2(2), s_2(3), s_2(4), s_2(5), s_2(6), s_2(7))$$

As the final step we act with the filter on the first four elements of this vector:

$$s_3 = \begin{pmatrix} c(0)s_2(0) + c(1)s_2(1) + c(2)s_2(0) + c(3)s_2(1) \\ d(0)s_2(0) + d(1)s_2(1) + d(2)s_2(0) + d(3)s_2(1) \\ s_2(2) \\ s_2(3) \\ s_2(4) \\ s_2(5) \\ s_2(6) \\ s_2(7) \end{pmatrix}$$

s3 is final coefficients obtained after DWT transform. It shows the third level decomposition of the speech signal where approximate coefficients carries maximum information. Approximate coefficients are similar with the original human voice whereas details coefficients carries noise.

### 6.2.3 Feature Extraction:

Different types of features like MFCC, HFCC have been calculated from wavelet coefficients of original audio signals. HFCC is a modification of MFCC feature extraction technique [150] [204]. HFCC features are decoupled in nature which provides more appropriate features of the speech signal.

The steps for extracting HFCC coefficients are:

i) Wavelet coefficients obtained after DWT decomposition is divided into the frames. The number of sample values in each frame is sn = ⌊num/len⌋, where num is the total number of DWT coefficients and len is the number of frames. For removing side ripples and maintaining continuity in the signal, hamming window is applied to the framed signal.

$$hm(sn) = 0.54 - 0.46 \cos\left(\frac{2\pi sn}{N-1}\right), 0 \leq sn \leq num - 1 \quad (6.3)$$

$$y(sn) = s_3(sn) * hm(sn) \quad (6.4)$$

Where hm (n) is the window signal, $s_3(sn)$ is the wavelet coefficients, y(n) is the output signal and N is the number of samples in each frame.

ii) After that Fourier coefficients are calculated from the windowed signal using Discrete Fourier Transform (DFT).

$$Y(sn) = FFT(y(sn)) \quad (6.5)$$

Where $s_3(w)$, hm(w) and y(w) are the Fourier Transforms.

iii) DFT coefficients are mapped into the mel scale which has been expressed as:

$$Mel(f) = 2595*\log 10\ (1+f/700) \quad (6.6)$$

$$f(k) = sn*fs/num \quad, k = 0\ldots\ldots sn-1$$

where fs is the frequency range used in filter bank. The filter bank is designed as:

$$H(k, b) = \begin{cases} 0, & \text{if } f(k) < fc(b-1) \\ f(k) - fc(b-1)/(fc(b) - fc(b-1)), \\ \quad \text{if } fc(b-1) \leq f(k) < fc(b) \\ f(k) - fc(b+1))/fc(b) - fc(b-1), \\ \quad \text{if } fc(b) \leq f(k) < fc(b+1) \\ 0 \\ \quad \text{if } f(k) < fc(b-1) \end{cases} \quad (6.7)$$

Where fc (b) is the central frequency of the filter and b is the number of filters used in filter bank.

iv) Find the log energy output of each of the mel frequencies.

$$S(b) = \sum_{k=0}^{sn-1} H(k, b) * Y(sn) \quad (6.8)$$

Where $b = 1, 2, \ldots m$ and m is the number of filters

v) Coefficients me1, me2…..are generated by applying Discrete Cosine Transform (DCT) on theses mel frequencies.

$$Me(sn) = DCT(Sb) \quad (6.9)$$

Where Sk is the Mel spectrum and k is the number of cepstral coefficients.

vi) HFCC features are obtained by replacing filter bank H(k, b) by the filter bank designed by Mark D. Skowronski and John G. Harris [204]. The HFCC feature is very similar to MFCC feature, the only difference between these two methods is that, the filter bandwidth is coupled with the other filter bank parameters (frequency range, number of filters). In HFCC filter center frequencies are equally spaced in mel frequency scale, as in the MFCC method, but filter bandwidth is a design parameter measured in equivalent rectangular bandwidth (ERB):

$$ERB(fc) = 6.23\, fc^2 + 93.39\, fc + 28.52 Hz \quad (6.10)$$

Where fc is a central frequency.

The HFCC and MFCC filter bank of SAWAN speech signal is shown in Figure 6.13 and 6.14.

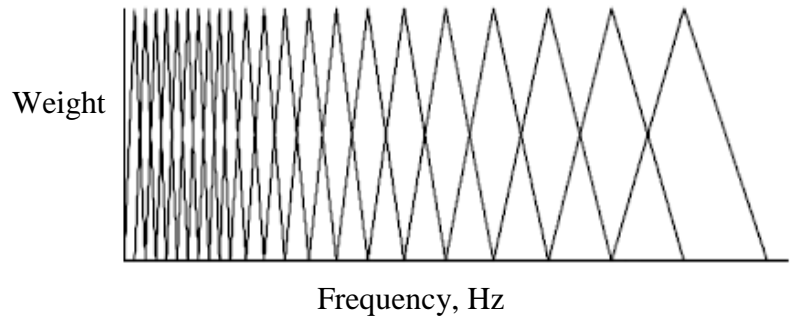

Figure 6.13: Triangular MFCC filter bank

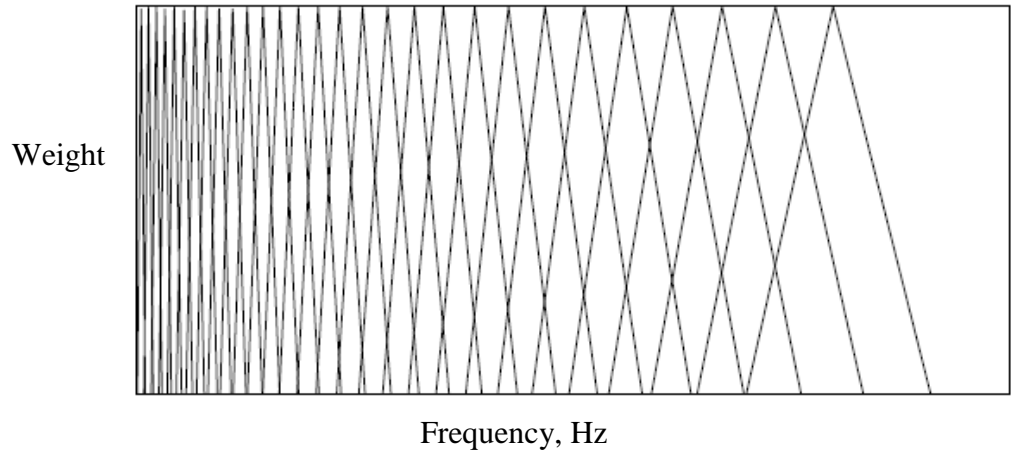

Figure 6.14: Triangular HFCC filter bank

### 6.2.4 Dimension Reduction:

Dimension of HFCC features is reduced using PCA [205]. The steps of PCA are:

i) HFCC features of raw data sets are taken.

ii) The mean m of HFCC features are calculated. Then it is subtracted from each data value.

$$S(n) = me(sn) - m \qquad (6.11)$$

iii) The covariance matrix of the data set has been calculated.

$$C(n) = cov(S(n)) \qquad (6.12)$$

Where S (n) is the subtracted mean and C (n) is the covariance matrix.

iv) Eigen value and Eigen vector of the covariance matrix are calculated. Those Eigen vectors which are having highest Eigenvalues, known as the principal components, are selected.

$$Q(n) = eig(C(n)) \qquad (6.13)$$

v) The final data set has been generated by multiplying transpose of Eigen vectors with the mean adjusted original dataset.

$$tr(n) = QT\ S(n) \qquad (6.14)$$

### 6.2.5 Classification:

Classification is done using Bayesian classifier [206]. Here the test speech signal is classified by calculating maximum probability density function.

**Bayesian Algorithm:**

- Let there be C classes w1, w2 ….wC with prior probabilities $P_1, P_2… P_C$ = 1/C and class conditional density functions:

    $p_1, p_2,………. p_C.$ Where $0<=P_i<=1$ (for every i) and $\sum_{i=1}^{C} P_i = 1$.

- Reduced features obtained after PCA are considered as a training data set $Y_i = (tr1, tr2……trn)$.

- Let X=(x1, x2, xi) be the test vector put in a $i^{th}$ class. Where $X \in \pi_i$, $\pi$ is a work space. $\Pi \subseteq |R^m$, $x \in |R^m$ set of all possible observations and m is the number of components.

- Here the probability density function (PDF) is calculated using the Gaussian multivariate distribution function which is expressed as:

$$P_i(X) = \text{agrmax}\left(\frac{1}{(\sqrt{2\pi})^m |\Sigma_i|^{1/2}} exp\left\{-\frac{1}{2}(X-\mu_i)' \Sigma_i^{-1}(X-\mu_i)\right\}\right) \qquad (6.15)$$

Where Pi(X) is the PDF, μ=Mean, Σ= Co-variance, |Σ| =determinant of Co-variance, m is the dimension of given data. Test vector belongs to the class having highest PDF value. Here we consider C=12.

### 6.2.6 Experimental Results and Performance Analysis

Experiments have been performed on twelve Hindi isolated words (Hindi months). Each word is spoken six times by six different persons. Words are recorded in .wav format using audacity software. Each word is recorded at a bit-rate of 256 kbps and 16 KHz frequency. Recorded words are shown in Table 6.1.

Table 6.1: List of Words

| Serial No. | Words | Serial No. | Words |
|---|---|---|---|
| 1 | Chatra | 7 | Kwar |
| 2 | Baisakh | 8 | Kartik |
| 3 | Jeyesth | 9 | Aghan |
| 4 | Aasadh | 10 | Paush |
| 5 | Sawan | 11 | Magh |
| 6 | Bhdrapad | 12 | Falgun |

This database has also been recorded in different modes like slow speech means a word speaks by a people very slowly (8 KHz frequency) and average recording time of each word is 5 to 10 seconds. Medium speech means a word is recorded with medium frequency and the recording time is 4 to 7 seconds. Last one is the high speech where speech signals have been spoken very fastly. The recording time lies between 1 to 2 seconds.

Experiments have been performed to test the recognition accuracy associated with the speech signals done in two environments, with DWT and without DWT. We have taken two features of speech signal one is MFCC and the other one is HFCC. HFCC features are calculated at 12.5 KHz sampling frequency. Here we used 29 triangular filters for calculating HFCC coefficients. Performances of the experiments are analyzed in various environments like DWT with HFCC, DWT with MFCC etc. The implementation is performed using MATLAB 10. Here speech signals have been decomposed up to $1^{st}$ $2^{nd}$, $3^{rd}$ and $4^{th}$ level of decomposition using Daubechies-6 Wavelet Transform. Experimentally we found that $3^{rd}$ level decomposition reduces maximum noise with minimum loss of information. Robustness of Hindi speech recognition system has been tested by calculating the misclassification error rate which has been shown in Table 6.2, 6.3 and 6.4.

misclassification rate
$= (\text{no. of misclassified words}/\text{Total no. of test samples}) * 100$ (6.16)

Table 6.2. Table of Testing Details Using MFCC with Bayesian and HFCC, DWT with Bayesian

| Total no.of samples | Test Samples | Misclassification Rate (After MFCC and Bayesian)(%) | Avg. Processing Time(After MFCC and Bayesian)(Sec) | Misclassification Rate (After DWT, HFCC and Bayesian)(%) | Avg. Processing Time(After DWT, HFCC and Bayesian)(Sec) |
|---|---|---|---|---|---|
| 1-50 | 20 | 25 | 0.56 | 5 | 0.66 |
| 1-100 | 30 | 35 | 0.74 | 9 | 0.79 |
| 1-150 | 50 | 39 | 0.96 | 14 | 0.99 |
| 1-200 | 70 | 48 | 1.89 | 19 | 2.01 |
| 1-250 | 90 | 54 | 2.99 | 24 | 3.11 |
| 1-300 | 100 | 56 | 4.76 | 28 | 4.99 |
| 1-400 | 140 | 61 | 5.89 | 32 | 6.78 |

Table 6.3. Table of testing details using PCA and PCA with DWT

| Total no.of samples | Test Samples | Misclassification Rate(After PCA)(%) | Avg. Processing Time(After PCA)(Sec) | Misclassification Rate(After DWT and PCA)(%) | Avg. Processing Time( sec) (After DWT and PCA) |
|---|---|---|---|---|---|
| 1-50 | 20 | 27 | 0.16 | 25 | 0.56 |
| 1-100 | 30 | 40 | 0.36 | 35 | 0.74 |
| 1-150 | 50 | 48 | 0.41 | 39 | 0.96 |
| 1-200 | 70 | 59 | 0.49 | 48 | 1.89 |
| 1-250 | 90 | 63 | 0.52 | 54 | 2.99 |
| 1-300 | 100 | 69 | 0.56 | 56 | 4.76 |
| 1-400 | 140 | 74 | 0.63 | 59 | 5.87 |

Table 6.4. Table of testing details using MFCC and PCA and MFCC, DWT and PCA

| Total no.of samples | TestSamples | Misclassification Rate ( After MFCC and PCA)(%) | Avg. Processing Time(sec) After MFCC and PCA(Sec) | Misclassification Rate ( After MFCC, DWT and PCA) (%) | Avg. Processing Time(sec) After MFCC, DWT and PCA(Sec) |
|---|---|---|---|---|---|
| 1-50 | 20 | 25 | 0.32 | 20 | 0.45 |
| 1-100 | 30 | 27 | 0.70 | 26 | 1.10 |
| 1-150 | 50 | 38 | 0.91 | 35 | 2.12 |
| 1-200 | 70 | 46 | 1.76 | 33 | 2.84 |
| 1-250 | 90 | 52 | 2.98 | 39 | 3.86 |
| 1-300 | 100 | 53 | 4.56 | 37 | 5.03 |
| 1-400 | 140 | 58 | 5.21 | 55 | 6.23 |

From Table 6.2, 6.3, 6.4 we observe that HFCC, DWT with Bayesian provides minimum misclassification error rate in comparison to other techniques because of multiresolution property of DWT and ERB property of HFCC technique. In this chapter DWT is used for reducing the noise present in the original signal and Bayesian is used for classifying an unknown speech using a multivariate probability density function. Here we observe that the computational complexity of HFCC, DWT with Bayesian has not been very higher than the other techniques. This is because the decomposition property of DWT which reduces the size of the signal makes processing faster. All the techniques explained above are speaker independent, means it does not depend on the speaker (who is speaking, weather it is a male or female). From Table 6.2, 6.3 and 6.4 we see that, as the number of test samples increases the misclassification error rate also increases. Also we have compared the classification results with various other classifiers which is shown in Table 6.5. And found that HMM gives highest classification accuracy than other classifiers (distance classifier, Bayesian and KNN). Because HMM handles the time series data very effectively than others.

Table 6.5 Classification results at different classifiers

| Classifiers | Classification rate (%) |
|---|---|
| Euclidean Distance | 90 |
| Bhattacharya Distance | 93 |
| Manhattan Distance | 87 |
| KNN (K=2) | 88 |
| Bayesian Classifier | 92 |
| HMM | 95 |

Table 6.6. Classification results obtained at different modes of speech

|  | Classification rate (%) |
|---|---|
| Slow Speech | 95 |
| Medium Speech | 94.3 |
| High Speech | 89.25 |

Results shown in Table 6.6 shows that the slow speech and medium speech gives highest classification rate than high speech. After that the experiments are performed on dataset recorded at different distances like 0.5 meter, 1 meter, 2m, 3m and 4 m. We also extended our dataset upto 5000 samples. The Results are dipicted in Table 6.7.

Table 6.7. Classification results obtained at different distances

| Total no. of samples | 0.5meter | 1 meter | 2 meter | 3 meter | 4 meter |
|---|---|---|---|---|---|
| 1-500 | 100 | 100 | 100 | 96 | 93 |
| 1-1000 | 100 | 100 | 100 | 96 | 93 |
| 1-1500 | 100 | 100 | 99 | 95 | 89 |
| 1-2000 | 100 | 100 | 98 | 94 | 86 |
| 1-3000 | 100 | 100 | 98 | 94 | 86 |
| 1-4000 | 98 | 99 | 96 | 92.8 | 85.1 |
| 1-5000 | 98 | 97 | 95 | 90 | 80 |

After carefully analyzing the data presented in Table 6.7, it has been observed that as the distance between the speaker and human increases the accuracy decreases. This happens because of the involvement of external factors like environmental noise, human error etc. The Proposed algorithm has also been tested on large number of English words and found that, it gives sufficient accuracy (93%) on English words.

### 6.2.7 Conclusion

A novel DWT and HFCC based speaker invariant speech recognition system has been developed and tested with Hindi as well as English words. This system has brought a tremendous appeal in real world speech biometric applications. It is indeed a challenging job to process those speech signals in order to extract intrinsic speech signatures. We have applied two fold process of feature extraction. One is a DWT decomposition technique, applied on speech signal and another is the HFCC feature extraction method. This extracted feature is prone to reducing maximum noises so that irrelevant information can be removed from the speech signal. A probabilistic approach based on naïve Bayes' rule and HMM has been used in classification purposes with the strength of handling nonlinearity in the speech signals.

Comparative analysis of various existing methods have been performed with respect to the proposed methods using twelve Hindi words with six speakers. Features of speech signals are extracted using Daubechies Wavelet Transform-6. Than the dimensions of features are reduced using PCA and classified using Bayesian and HMM classifier. Among all the results, we have seen that HFCC, DWT and HMM based method provide high recognition rate in comparison to the other methods due to the decoupling nature of HFCC features, which is very helpful for reducing noise of a speech signal. From experimental results we see that the recognition accuracy decreases as the number of test samples increases because the Bayes' decision rule classifies only a limited number of samples, as the size of sample increases it fails to classify because it deals only with univariate datasets whereas our dataset are multivariate in nature and therefore, we have applied HMM as a classifier and found better accuracy than other classifiers (distance based classifier, KNN etc.). Experiments have also been performed at various distances and it has been observed that as the distance increases the misclassification rate is also increases. Misclassification rate has also been increased with increasing number of speakers. These are the limitation of this methodology.

## 6.3  Multimodal Decision based Human Robot Interaction

A single mode (gesture or speech) is not sufficient to establish interactions between human and robot. Because each mode has its own limitations like for gesture mode of interaction, the light condition, skin color variation are the major issues. Similarly in speech mode, the effect of background noise, variations in human voice etc. are the problems arises during interaction.  These limitation have been minimized by combining these two modes. First question which comes into mind that how we combine these two modes? Because these two modes are completely different in nature. Gesture mode of interaction is based on image processing and speech mode is based on the signal processing. These two modes are combined using feature level fusion as well as decision level fusion. In feature level fusion the information coming from the different traits are combined before classification but in the decision level fusion the information is combined after the classification. In this chapter we have used a decision level fusion for combining the results coming from the different individualities shown in Figure 6.15. For combining the results obtained from different modalities, we define some test cases at the decision level and get the final result.

### 6.3.1  Decision Level Fusion

"It is a strategy to combine the decisions obtained from different systems to produce the final decision." In decision level fusion, the fusion has been performed in various ways like using AND, OR, weighted average, decision tree etc. Here the hard decisions obtained from various modes are merged to obtain the global result.

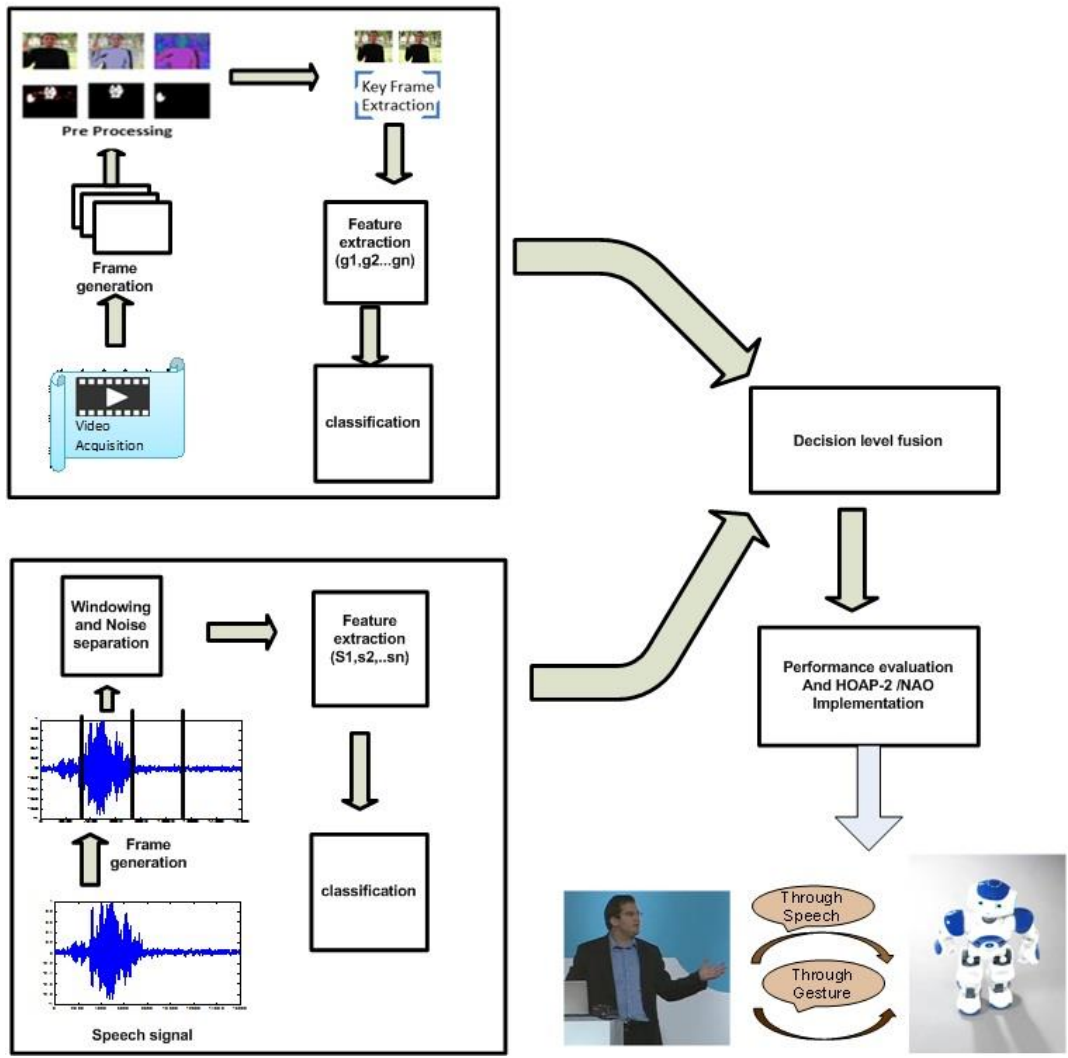

Figure 6.15 Framework of multimodal system

### 6.3.2  Fusion Strategies

There are various strategies available for getting the combined results obtained from the local results. These local results are obtained from different classifiers used in each mode. Here we used Support Vector Machine (SVM) [183], Hidden Markov Model (HMM) [196] and Bayesian as a classifier. Each classifier gives an individual result specific to each mode. Fusing two different modes (gesture and speech) at the decision level we define some test cases. These test cases have been defined on the basis of likelihood values obtained after applying HMM as a classifier. The description of symbols used in the test cases are defined in Table 6.8.

Table 6.8 Description of symbols

| | |
|---|---|
| Source 1($\lambda_1$) = Gesture | Source 2 ($\lambda_2$) = Speech |
| Likelihood of ($\lambda_1$) = $\alpha_1$ | Likelihood of ($\lambda_2$) = $\alpha_2$ |
| Threshold of ($\lambda_1$) = $\theta_1$ | Threshold of ($\lambda_2$) = $\theta_2$ |
| Class Label of ($\lambda_1$)=$C_1$ | Class Label of ($\lambda_1$)=$C_2$ |

Table 6.9 Test cases for decision level fusion

| IF | THEN |
|---|---|
| ($\alpha_1 \geq \theta_1$ AND $\alpha_2 \geq \theta_2$) | IF ($\alpha_2 \leq \alpha_1$) <br> Then **FinalClassLabel**=$C_1$ <br> ELSE **FinalClassLabel**=$C_2$. |
| ($\alpha_1 \geq \theta_1$ AND $\alpha_2 < \theta_2$) | Then **FinalClassLabel**=$C_1$ |
| ($\alpha_1 < \theta_1$ AND $\alpha_2 \geq \theta_2$) | Then **FinalClassLabel**=$C_2$ |
| ($\alpha_1 < \theta_1$ AND $\alpha_2 < \theta_2$) | Then $\alpha_1 = w_1 \times \alpha_1$ and $\alpha_2 = w_2 \times \alpha_2$ <br> IF ($\alpha_2 \leq \alpha_1$) <br> Then FinalClassLabel=$C_1$ <br> ELSE FinalClassLabel=$C_2$. |

Table 6.9 shows all the test cases defined on the basis of likelihood values obtained in each mode. The likelihood value at each test sample of each of the mode is shown in Figure 6.16 and 6.17. In both the figures triangle shows the likelihood value of classified samples and circle shows the likelihood value of misclassified samples. By seeing the likelihood values of classified samples we define a threshold for each of the mode. The threshold value for gesture is 0.8 and the threshold value for speech mode is 0.85.

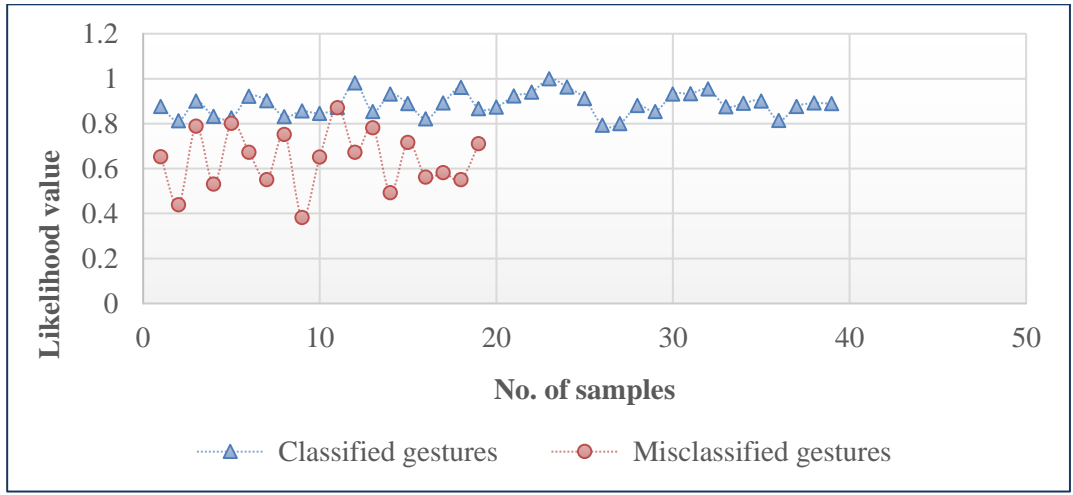

Figure 6.16 Likelihood values of classified/ misclassified gestures

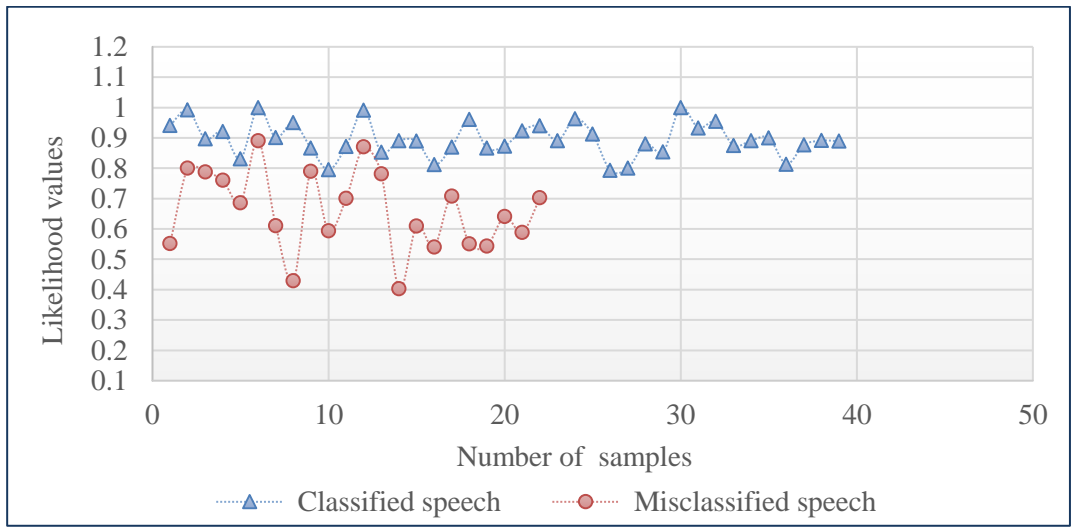

Figure 6.17 Likelihood values of classified/ misclassified speech

### 6.3.3 Classification Results

Experiments are performed on the basis of these test cases. Here 840 gestures are used as a probe gestures and 2000 speech samples are used as a testing speech. Out of 840, 798 gestures are correctly classified therefore the accuracy of individual gesture mode is 95%. Similarly for speech out of 2000 samples 1800 samples are classified correctly, the accuracy is 90% which is shown in Figure 6.18 and 6.19. This is the accuracy of each individual mode.

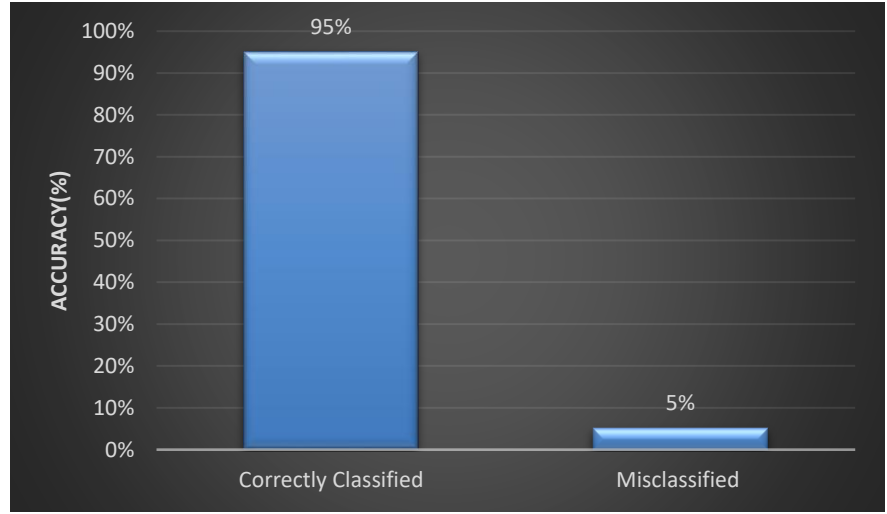

Figure 6.18 Classification results for gestures

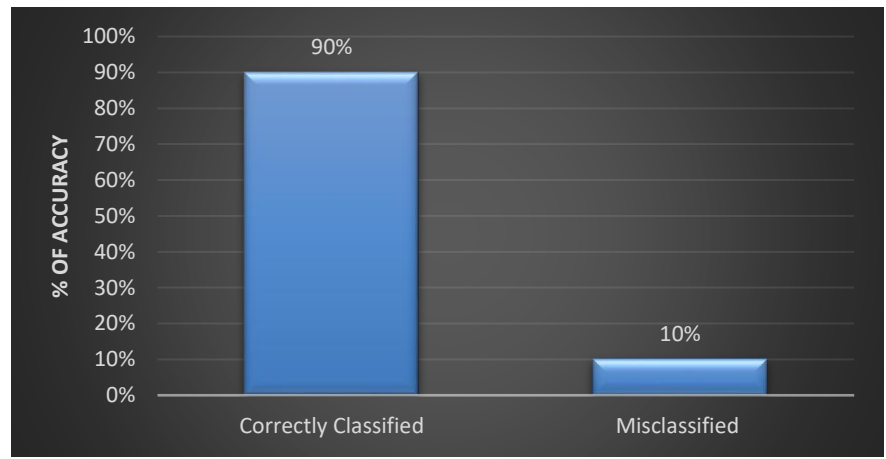

Figure 6.19 Classification results for speech

In a multimodal system we fuse a decision obtained from each of the mode. This fusion is done on the basis of the test cases defined in Table 6.10. This fusion method increases the overall accuracy by Five percent. Which makes the communication between human and robot more effective. The results of multimodal classification is shown in Table 6.10.

Table 6.10 Multimodal classification accuracy

| Gesture | Classified/ misclassified | Speech | Classified/ Misclassified | Multimodal classification |
|---|---|---|---|---|
| How | Classified | How | Classified | Classified |
| Agree | Classified | Agree | Misclassified | Classified |
| Close | Classified | Close | Misclassified | Classified |
| One | Misclassified | One | Classified | Classified |
| Two | Misclassified | Two | Classified | Classified |
| Three | Classified | Three | Classified | Classified |
| Four | Classified | Four | Classified | Classified |
| Five | Classified | Five | Classified | Classified |
| Seven | Misclassified | Seven | Misclassified | Misclassified |
| Room | Misclassified | Room | Classified | Misclassified |
| Study | Classified | Study | Misclassified | Classified |
| Coffee | Misclassified | Coffee | Classified | Classified |
| Namaste | Classified | Namaste | Misclassified | Classified |

If a command performs by both gesture as well as speech mode simultaneously and both the mode classified it correctly then overall status is classified. If one mode classified it correctly and another mode misclassified it then the decision takes place one the basis of higher likelihood value and the weights are assigned according to that. The overall classification result is shown in Table 6.7.

### 6.3.4   Implementation on HOAP-2 Robot

Lastly the results obtained through multimodal decision level fusion is implemented on HOAP-2 robot shown in Figure 6.20. The complete description of HOAP-2 robot is given in chapter 3.

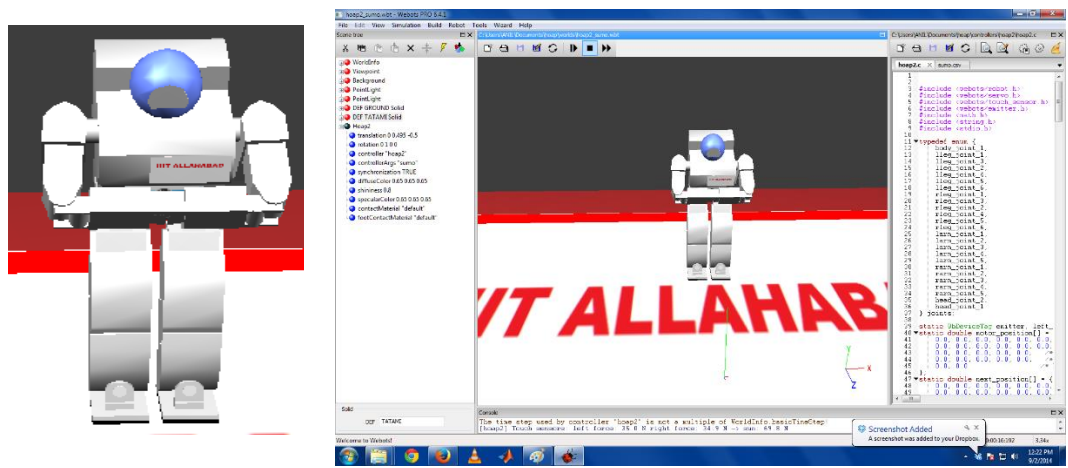

Figure 6.20 Namaste gesture performed by HOAP-2 Robot

### 6.3.5 Conclusion

A decision level fusion strategy is used for combining the results obtained from two different modes gesture and speech. A test cases have been defined by considering the likihood values for this fusion. These likihood values are calculated for each probe gesture and speech using HMM. Fusion on the decision-level improves the performance of the multimodal (gesture and speech) system significantly. Fusion seems only beneficial for classification with unequal false-negative and false-positive rates.

This method does not work well in the circumstances in which both the modes misclassify a particular word or command. This limitations have been overcome by applying various other techniques like artificial neural network, fuzzy algorithm, Dempster-Shefer theory etc. for fusion which increases the system performance.

# Chapter 7

# Conclusion, Recommendation for Future Work

## 7.1 Conclusion

Throughout the research we tried to establish that the gesture and speech are the most common biometric trait which can be used for human robot interactions (HRI). Understanding of HRI through speech and gesture makes the robot social and are easy to include in our day today activities. Robots have been used for entertainment, learning, as a helper and so many other purposes. Gesture and speech are usually used by physically challenged persons for communication establishment with the machine and/or humans. Among these two modes, a single one has not been sufficient because gesture fails when light is not good enough and speech fails when the voice of a person is not clear, therefore, fusing both the modes in our framework we can take the advantage of both. This is our concept of hybrid framework design.

This dissertation has focused on proposing new feature extraction techniques for increasing gesture and speech recognition performance for HRI. Methodology for integrating these two modes have been investigated and experimented over the human robot communication system. Various methods related to such multimodal communication are introduced using computer vision and machine learning techniques. Following are some of the contributions of our investigation.

- A novel hand tracking algorithm has been proposed with proper justification which generates a good quality of silhouette hand sequences.

- A framework for HRI using isolated ISL gestures has been established where DWT with MFCC feature extraction technique have been applied. This technique has been shown to be robust for acquiring person invariant features with small changes in hand shape with respect to the small variations of gesture types.
- A concept of possibility theory based HMM has been addressed where all the three problems of classical HMM have been redefined and solved. Theoretically as well as experimentally, we have validated that the time complexity of PTBHMM is N time less than the time complexity of existing HMM which is $N^2T$. We have tested our novel PTBHMM algorithm on real time continuous ISL gesture recognition and observed that the system responds very quickly in comparison to classical HMM.
- A NAO based framework for continuous ISL gesture recognition has been proposed. Three different features (speed, shape and orientation) of the hand have been extracted and combined in a Cartesian coordinate system. These hybrid features comprehends maximum information of hand gestures and then it has been classified using possibility theory based HMM.
- For speaker invariant Hindi speech recognition system DWT with Human Factor Cepstral Coefficient (HFCC) has been compared with DWT with MFCC. DWT with HFCC has been found to be more efficient in finding appropriate features.
- A multimodal framework for HRI has been proposed where the fusion of gesture and speech have been done at the decision level. The results obtained from different modes have been combined using weighted decision rules. This makes the human robot communication system more accurate with minimum false positive.

We have applied an effective and efficient approach for background modelling and tracking of hand and also tried to reduce the computational complexity. Although this approach is efficient and helpful in improving the performance, it cannot avoid all the errors introduced due to the luminance and assimilation of person's cloth colour with the background colour. This approach also induces limitation for tracking algorithm as it depends highly on above stated background modelling algorithm with limited range. We have prescribed a new method for extracting the cepstral features of hand of isolated gestures. Dimensionality of the cepstral features is reduced using discrete wavelet transform when DWT decomposition is performed up to the 4$^{th}$ level.

A gradient based method has been addressed for overlapping frame extraction from a sequence of continuous ISL gestures. These long gestures are the combination of different isolated gestures including grammar. In a video sequence of continuous gesture, if a change in the gesture appears frequently then it is difficult to identify the breaking points. Complexity is a big issue in an NAO based real time multimodal system which has been solved by proposing PTBHMM. It deals with spatiotemporal dataset. Shape, velocity and orientation are the key features of any sign language gesture which are performed through the hand. These features are extracted using wavelet descriptor, frame differencing and atan2 function applied on the LH (low, high) and HL (high low) coefficients obtained after 3rd level decomposition of DWT. We employed a slew of distance metric classifiers together with K-nearest neighbour, hidden Markov model and PTBHMM in order to examine the best classification for the projected features. In house dataset of isolated as well as continuous ISL gesture have been created at varying distances with full black sleeve dresses. The proposed algorithm has also been tested on benchmark Sheffield Kinect gesture dataset [177].

Subsequently, a speaker invariant speech recognition system has been addressed for HRI where Hindi words are used for the ease of communication. These Hindi words have been recorded at 16000 Hz frequency of mono signals. All words are recorded in close environment as well as outdoor environment, having environmental noise using audacity software with various numbers of speakers. Pre-emphasis, utterance detection, feature extraction are solved using DWT and human factor cepstral coefficients (HFCC). It has been observed that DWT separates the complete signal into two parts, one is low frequency known as approximate coefficient and the second one is high frequency known as detail coefficient. Approximate coefficient contains maximum information with minimum disturbances. The validity of the system has been tested using HMM, and Bay's decision rule using HOAP-2 humanoid robot framework. The accuracy of the system decreases as the distance between human and robot increases. The noise of the system highly affects the recognition accuracy and have one should use appropriate noise reduction technique.

A decision based multimodal fusion strategy has been implemented for making the communication system robust because a single medium is not enough for a good communication. We define rules on the basis of likelihood values obtained after each mode of classification. These rules have been defined on the priority basis, which may fail when both of them have the same likelihood values. To avoid this situation we need to perform appropriate likelihood value refinement, if still the problem persists, we can give higher weightage to the mode whose classification rate is high.

All our experience noted down in the thesis may lead to a successful development of human-machine interaction system which can remove the hurdles of communication among physically challenged persons.

## 7.2    Recommendation for Future Work

Based on our research experience in this exciting area of human-robot-interactions, following recommendations are made which may be helpful for the future researchers to carry forward the task of developing an effective multimodal HRI system:

- Vision based analysis of the silhouette images has been addressed, but the features obtained through sensors are totally untouched. The human robot interaction could be improved through appropriate sensor development.
- The current work is based on the geometrical features of the hand. For better biometric identification using point cloud, depth map and detector such as Kinect, a 3D volumetric model may be explored for better understanding of gesture data.

- All the experiments are performed on fixed black coloured cloth with full sleeves. It can be extended to situations carrying different colour clothes, half sleeves or having full hand.
- Emotions and age of a person in speech dataset have not been considered in this thesis. Considering all such features of speech would certainly make the HRI system more effective.
- In this thesis fusion has been addressed using rule based system which can further be extended with other techniques like ANFIS architecture, decision tree etc. which may make the system more reliable.

gesture recognition. Vol. 12. 1995.

[117] Rady, Engy R., et al. "Speech Recognition System Based on Wavelet Transform and Artificial Neural Network."

[118] Kulkarni, Vaishali S., and S. D. Lokhande. "Appearance based recognition of american sign language using gesture segmentation." International Journal on Computer Science and Engineering 2.03 (2010): 560-565.

[119] Nandy, Anup, et al. "Recognizing & interpreting indian sign language gesture for human robot interaction." Computer and Communication Technology (ICCCT), 2010 International Conference on. IEEE, 2010.

[120] Nandy, Anup, et al. "Recognition of isolated indian sign language gesture in real time." Information Processing and Management. Springer Berlin Heidelberg, 2010. 102-107.

[121] Nandy, Anup, et al. "Classification of Indian Sign Language In Real Time." International Journal on Computer Engineering and Information Technology (IJCEIT) 10.15 (2010): 52-57.

[122] Kishore, P. V. V., et al. "Video Audio Interface for Recognizing Gestures of Indian Sign." International Journal of Image Processing (IJIP) 5.4 (2011): 479.

[123] Prasad, Jay Shankar, and Gora Chand Nandi. "Clustering method evaluation for hidden Markov model based real-time gesture recognition." 2009 International Conference on Advances in Recent Technologies in Communication and Computing. IEEE, 2009.

[124] Kishore, P. V. V., and P. Rajesh Kumar. "A video based Indian sign language recognition system (INSLR) using wavelet transform and fuzzy logic." International Journal of Engineering and Technology 4.5 (2012): 537.

[125] Bhuyan, M. K., Debashis Ghosh, and P. K. Bora. "A framework for hand gesture recognition with applications to sign language." India Conference, 2006 Annual IEEE. IEEE, 2006.

[126] Bhuyan, M. K., Debanga Raj Neog, and Mithun Kumar Kar. "Hand pose
*IIIT-A, Robotics and AI Lab* 170

Machine Learning and Cybernetics, 1-19.

[223] Singh, A. K., Baranwal, N., & Nandi, G. C. (2015, August). Human perception based criminal identification through human robot interaction. In Contemporary Computing (IC3), 2015 Eighth International Conference on (pp. 196-201). IEEE.

[224] Singh, A. K., Joshi, P., & Nandi, G. C. (2014, July). Face recognition with liveness detection using eye and mouth movement. In Signal Propagation and Computer Technology (ICSPCT), 2014 International Conference on. IEEE (pp. 592-597).

[225] Singh, A. K., & Nandi, G. C. (2012, October). Face recognition using facial symmetry. In Proceedings of the Second International Conference on Computational Science, Engineering and Information Technology (pp. 550-554). ACM.

[226] Singh, A. K., Chakraborty, P., & Nandi, G. C. (2015, November). Sketch drawing by NAO humanoid robot. In TENCON 2015-2015 IEEE Region 10 Conference (pp. 1-6). IEEE.

[227] Singh, A. K., & Nandi, G. C. (2016). NAO humanoid robot: Analysis of calibration techniques for robot sketch drawing. Robotics and Autonomous Systems, 79, 108-121.

[228] Singh, A. K., Joshi, P., & Nandi, G. C. (2014). Face liveness detection through face structure analysis. International Journal of Applied Pattern Recognition, 1(4), 338-360.

[229] Singh, A. K., Kumar, A., Nandi, G. C., & Chakroborty, P. (2014, July). Expression invariant fragmented face recognition. In Signal Propagation and Computer Technology (ICSPCT), 2014 International Conference on (pp. 184-189). IEEE.

[230] Singh, A. K., & Nandi, G. C. (2017). Visual perception-based criminal identification: a query-based approach. Journal of Experimental & Theoretical Artificial Intelligence, 29(1), 175-196.

[231] Singh, A. K., Joshi, P., & Nandi, G. C. (2016). Development of a Fuzzy Expert System based Liveliness Detection Scheme for Biometric
*IIIT-A, Robotics and AI Lab* 181

Authentication. arXiv preprint arXiv:1609.05296.